\documentclass[11pt]{article}
\usepackage{graphicx} 
\usepackage{natbib}
\usepackage[a4paper, top=0.8in, bottom=0.8in, left=0.9in, right=0.9in]
{geometry}
\usepackage{setspace}
\usepackage{amsmath} 
\usepackage{longtable}
\usepackage{booktabs}
\usepackage{array}
\usepackage{threeparttable}
\usepackage{changepage}
\usepackage{hyperref}
\usepackage{float}
\usepackage{multirow}
\usepackage{pdflscape}
\usepackage{float}

\usepackage{subcaption}

\newcolumntype{C}[1]{>{\centering\arraybackslash}p{#1}}

\title{Yield Curve Prediction with Machine Learning: Forecasting Approaches and the Role of Macroeconomic Predictors}

\vspace{1em}

\author{
Jeron Tan Kang\thanks{I am grateful to Denis Tkachenko for his guidance, insightful feedback, and support throughout this paper. Any remaining errors are my own.}
}
\date{27 July 2026}

\begin{document}

\pagenumbering{roman}  
\setcounter{page}{1}

\maketitle


\begin{abstract}
This paper compares direct-yield and factor-based approaches to U.S.\ Treasury yield curve forecasting using a common high-dimensional macroeconomic information set. Forecasts are evaluated on monthly zero-coupon yields over the 2015--2025 out-of-sample period. Gains over the random walk are concentrated at short maturities and in slope forecasts, and decline with the forecast horizon. Direct-yield models perform best for slope forecasts and are relatively stronger at short horizons, while factor-based models become more competitive at longer horizons. Macroeconomic predictors provide clear incremental predictive power, strongest for slope-related movements. A trading simulation reinforces that macro-augmented models perform best in slope trades. The simulation also highlights a gap between statistical and economic performance, as the random walk is a strong benchmark under statistical loss but performs poorly as a trading signal.
\end{abstract}


\clearpage

\setcounter{tocdepth}{2}
\tableofcontents

\newpage

\pagenumbering{arabic}  
\setcounter{page}{1}

\section{Introduction}

Forecasting the yield curve is important for both macroeconomics and finance. From a macroeconomic perspective, the shape of the yield curve is widely regarded as an informative summary of market expectations about future growth, inflation, and monetary policy. In particular, inversions of the yield curve have often been interpreted as signals of future economic slowdown or recession. For finance practitioners, anticipated yield curve movements shape portfolio allocation, duration management, hedging decisions, and relative-value trading strategies. Active fixed income investors take positions not only on the level of interest rates, but also on changes in the slope and curvature of the curve. Accurate forecasts of the yield curve at specific maturities and curve shape therefore have direct economic value.

Yield curve forecasting is challenging because bond yields at most maturities are highly persistent and close to non-stationary. As a result, the random walk no-change forecast has proven difficult to outperform out of sample, and much of the term-structure forecasting literature is motivated by that benchmark. One strand of the literature compares forecasting approaches, including factor-based models, maturity-specific direct forecasts, and functional methods, typically evaluated against the random walk benchmark. Another strand studies whether macroeconomic information contains predictive content beyond the yield curve itself. However, these two questions are often studied separately.

This paper brings these two questions together by comparing two forecasting approaches under the same macro-augmented forecasting design. The first is a maturity-specific direct forecasting approach, which I refer to as the \emph{direct-yield approach}. It estimates separate models for individual maturities and therefore does not impose cross-maturity restrictions. The second is a factor-structured approach based on the Dynamic Nelson--Siegel framework and its Svensson extension, which I refer to as the \emph{factor-based approach}. This approach first summarizes the curve using latent factors and then forecasts those factors forward. The central objective of the paper is to assess how these two approaches perform relative to each other once both are allowed to exploit a rich macroeconomic information set.

The role of macroeconomic predictors in yield curve forecasting is closely related to the macro spanning hypothesis. If the yield curve spans the relevant state variables for future interest-rate movements, then macroeconomic variables should add little or no incremental forecasting power once current yields are observed. In that case, practitioners concerned with forecasting rates would have little reason to move beyond parsimonious yield-based models, especially given the well-known low-rank structure of the U.S.\ yield curve. A further question is whether any such gains are concentrated in particular forecasting approaches or curve segments.

The paper also studies a separate but related issue: whether more flexible calibration of the Dynamic Nelson--Siegel--Svensson (DNSS) model improves forecasting performance. A large calibration literature evaluates alternative optimization methods mainly through in-sample fit. In this paper, the emphasis is instead on whether they improve out-of-sample forecasts. To study this, I estimate time varying DNSS decay parameters using a hybrid Particle Swarm Optimization and limited-memory Broyden--Fletcher--Goldfarb--Shanno algorithm with box constraints, and compare this against a baseline in which they are held fixed. I then test whether the estimated parameters that improve in-sample fit also translate into better out-of-sample forecast performance.

An additional contribution of the paper is to evaluate forecasts through a simple trading exercise rather than through RMSE alone. While the yield-forecasting literature remains focused mainly on statistical forecast accuracy, there is very limited work examining the trading value of yield-curve forecasts in a comparable bond-trading setting. Accordingly, in addition to RMSE, I construct duration and slope trading exercises aligned with the single-maturity and slope forecast targets. Although these trading rules are stylized and rely on simplifying assumptions, they provide a transparent way to assess whether statistically accurate forecasts also generate economic value. Performance is benchmarked against simple carry and roll-down and time-series-momentum style rules within the same trading environment.

Using monthly U.S.\ zero-coupon Treasury yields from \citet{FilipovicPelgerYe2024} and a broad set of macroeconomic predictors from FRED-MD, I evaluate out-of-sample forecasts for key maturities and slope measures over 2015--2025. The results show that direct-yield forecasts, especially when augmented with macroeconomic predictors, perform best at the short end of the curve and for slope forecasts, while factor-based forecasts are relatively more competitive in the belly and long end. The results also show that macroeconomic predictors provide significant incremental predictive power, with the strongest gains concentrated in slope-related movements. Although PSO-based DNSS specifications substantially improve in-sample fit, time-varying decay parameters estimated from this algorithm do not translate into better out-of-sample forecast performance. This indicates that stability of the factorization matters more for forecasting than maximizing cross-sectional fit. Finally, a trading exercise shows that the random walk is less compelling as an economic benchmark than as a statistical loss benchmark, and that macro-augmented models are particularly effective in slope-trading applications.

The remainder of the paper is organized as follows. Section 2 reviews the related literature. Section 3 describes the yield and macroeconomic data. Section 4 presents the forecasting methodology, including the direct-yield and factor-based models, the random forest specification, and the PSO-based DNSS estimation procedure. Section 5 reports the out-of-sample forecasting results and the trading exercise. Section 6 concludes.








\section{Literature Review}

\subsection{Yield Curve Forecasting Approaches}
\label{subsec:direct_vs_factor}

The \citet{NelsonSiegel1987} (NS) framework is an established method for yield curve modelling which gained popularity because of its parsimonious yet flexible representation of the term structure. It is capable of fitting monotonic, humped, and S-shaped curves, and explaining 96\% of the variation in bill yields across maturities. This framework was later extended by \citet{Svensson1994} into the Nelson--Siegel--Svensson (NSS) specification to better fit curvature. The NSS model later became the benchmark for producing fitted zero-coupon yields, with \citet{GurkaynakSackWright2007} fitting the NSS curve to Treasury prices to recover a smooth zero-coupon yield curve at all maturities.

Building on this, \citet{DieboldLi2006} proposed the Dynamic Nelson--Siegel (DNS) model as a forecasting framework. DNS fits the NS curve each period, interprets its coefficients as level, slope, and curvature factors, then forecasts yields by projecting these factors forward and reconstructing the curve using the NS loadings. In their original sample, DNS outperformed the random walk, and has since become a standard benchmark in the yield forecasting literature. However, that evidence was based on pre-2007 data, and subsequent work has revisited its relative performance in later samples and against more flexible alternatives.

One alternative is the direct forecasting approach, which estimates one predictive model per maturity without imposing cross-maturity restrictions. A drawback is that maturity level forecasts can yield an incoherent term structure when combined. However, many applications focus on a small set of maturities or yield spreads, such as curve steepener and flattener positions that trade the slope between two points on the curve (see, e.g., CME Group, 2013). In such instances, coherence of the full curve is less central than accuracy at the traded tenors. \citet{WangWangTu2025} study the U.S.\ 10-year yield and compare linear regression, decision trees, random forests, and a neural network using macroeconomic predictors, and find that the random forest performs the best out of sample.

Another approach forecasts the yield curve using functional methods, which treat the entire curve as a single observation and model its dynamics in function space rather than at selected maturities. This approach has shown promising empirical performance in several settings, including functional autoregressive forecasting of Eurodollar futures curves in \citet{kargin2008curve}, adaptive functional approaches to yield curve forecasting in \citet{chen2014adaptive}, and common functional principal component methods for international yield curves in \citet{zhang2017international}. However, this approach is not pursued in this paper because it typically models the yield curve as a single functional object driven mainly by its own lagged dynamics, making it less straightforward to incorporate the richer set of external predictors considered in this paper.

The yield forecasting literature commonly evaluates the DNS model against the random walk model, but few directly compare DNS with maturity specific direct forecasting. One such study is \citet{Rahimi2020}, who finds that direct methods can be competitive at the short end, while the DNS factor approach performs better at longer maturities and horizons. However, the analysis does not incorporate macroeconomic predictors, so the comparison is conducted under a relatively limited information set. 

\subsection{Macroeconomic Predictors in Yield Curve Forecasting}
\label{subsec:ml_macro}

A central question in term structure forecasting is whether macro variables add predictive power for future yields beyond what is contained in today’s yield curve. This is formalised by the macro spanning hypothesis, which posits that conditional on the yield curve, additional state variables should not improve forecasts of yields or bond returns.

An increasing number of studies show that macroeconomic variables can improve the prediction of future bond yields and bond returns. \citet{ang2003no} were the first to augment a standard three-factor affine model with macroeconomic variables. \citet{DieboldRudebuschAruoba2006} estimate a joint model of NS yield factors and key macro variables (real activity, inflation, and a monetary policy instrument) and find that macro shocks transmit primarily through the slope and level factors. In particular, policy rate, activity, and inflation shocks move the slope factor in the direction of a flatter curve, consistent with monetary policy raising the short end, while inflation surprises also raise the level factor more persistently, consistent with shifts in longer run inflation expectations. \citet{DePooterRavazzoloVanDijk2010} report that macro augmented models deliver larger gains in episodes of elevated interest rate path uncertainty (e.g., around the 2001 recession), whereas models without macro information perform relatively better when the term structure follows a more stable pattern or when the long short spread compresses. \citet{coroneo2016unspanned} use a joint dynamic factor model for yields and a large macro panel and identify two macroeconomic factors, related to economic growth and real interest rates, that are unspanned by the cross section of government bond yields, and have significant predictive power for future bond yields and excess returns. \citet{FreireRiva2025} use machine learning methods to show that macroeconomic variables are informative only for the shorter end of the yield curve, proxied by the short run NS factor which is empirically close to the slope of the yield curve, but with no evidence of improvement for the longer end.

So far, the literature has largely separated comparisons of forecasting approaches from studies that incorporate macroeconomic predictors into yield forecasts. \citet{DePooterRavazzoloVanDijk2010} is one of the few studies that jointly considers both questions in a single forecasting comparison. For maturity specific direct forecasting, they report that the best models at each forecast horizon are augmented with macro data. Specifically, they find that an autoregressive model for yields augmented with macro factors performs best at short horizons (1 and 3 months), while a vector autoregression for yields augmented with macro factors performs best at longer horizons (6 and 12 months). As for factor structured forecasting, they report that regardless of whether macroeconomic information is incorporated, they perform poorly across maturities and forecast horizons, contradicting the results of \citet{DieboldLi2006}. These comparisons are conducted within linear, low dimensional specifications with macro factors extracted from a panel of macro data instead of individual series. This helps navigate the curse of dimensionality and makes AR/VAR dynamics tractable. However, the factors were constructed using principal components analysis, which is unsupervised, so variables that are most relevant for forecasting may be underutilized if they do not load heavily on the first few principal components. This leaves open how the relative performance of forecast approaches change when modern machine learning methods are applied to a high dimensional information set augmented with macroeconomic information.

\subsection{Calibration of Nelson  Siegel  Svensson Decay Parameters}
\label{subsec:decay_calibration}

The NS and NSS frameworks represent the yield curve through a small set of latent factors and one or two decay parameters, respectively. These decay parameters determine the maturity location of the curvature loadings and therefore where curvature is concentrated along the maturity axis. As a result, they affect not only cross sectional fit, but also the interpretation and time series behaviour of the extracted factors. The parameter estimates of \citet{GurkaynakSackWright2007} are an established benchmark reference for fitted Treasury curves; they estimate parameters jointly by maximum likelihood, fitting the curve using duration weighted squared pricing errors.

However, the additional flexibility of the Svensson extension comes at the cost of a more difficult calibration problem. \citet{GilliGrosseSchumann2010} show that calibration with standard local optimizers can be fragile because the NSS objective is non convex, with multiple local solutions and strong interactions between the decay parameters and factor coefficients. \citet{gimeno2009genetic} likewise emphasize sensitivity to starting values and argue that global search procedures can reduce dependence on initial values. Together, these properties can lead to unstable parameter estimates across repeated runs.

Accordingly, a sizeable literature studies calibration strategies for the NSS framework, with many papers turning to global search algorithms to improve robustness. \citet{manousopoulos2009comparison} estimate the yield curve using Simulated Annealing and find that it outperforms gradient based methods in the more challenging Svensson setting. They recommend first using a global or direct search method to locate a good solution, and then refining that solution with a gradient based algorithm. \citet{GilliGrosseSchumann2010} propose differential evolution for NSS calibration and argue that it solves the problem reliably while delivering very good fits, with substantially greater stability across repeated runs than a restarted gradient based optimizer. Genetic algorithms, have also been reported to mitigate sensitivity to initial conditions and local optima, hence able to obtain more robust and stable fits than local search methods \citep{gimeno2009genetic,LakhanyPintarZhang2021, Ibanez2016}. Particle Swarm Optimization has also been reported to perform well in yield-curve calibration, \citet{QuirosGranadosTrejosZelaya2019} emphasize its ease of implementation, acceptable computational cost, and stable calibration outcomes, while \citet{AyoucheGharaibehAlQudah2016} show that a hybrid PSO–Nelder–Mead procedure delivers results that are reliably better than those obtained from a traditional derivative-based method.

While the preceding studies focus mainly on the in-sample calibration problem, the forecasting literature places a somewhat different emphasis. In forecasting-oriented implementations, the key question is not only how well the curve fits in sample, but also whether the resulting parameterization yields stable and useful factor dynamics out of sample. For this reason, a common practice in forecasting applications is to fix the decay parameters for stability and comparability over time. In the DNS approach, \citet{DieboldLi2006} fix the decay parameter at $\lambda=0.0609$, which places the peak of the curvature loading at roughly 30 months. They motivate this choice by noting that it makes estimation simpler and more numerically reliable, since the factor loadings become known and the factors can then be estimated each period by ordinary least squares rather than repeated nonlinear optimization.

There is comparatively limited literature that studies NSS calibration methods from an out-of-sample forecasting perspective where the focus is on yielding stable factors for better forecasts. One of the few papers to examine this link is \citet{Vela2013}, who finds that although differential evolution can improve calibration, the resulting parameter estimates may remain highly unstable and can adversely affect out-of-sample forecast performance.

\subsection{Yield Curve Construction as an Input to Forecasting}
\label{subsec:yield_curve_data}

The zero-coupon yield curve is not directly observed. Instead, it is inferred from a relatively sparse cross section of traded Treasury securities, which are coupon-bearing at all maturities except the 3-month bill, and whose prices can be noisy and affected by microstructure frictions and liquidity. Because yield forecasts are evaluated on the estimated curve, differences in curve construction can translate into different empirical conclusions.

The curve construction literature mainly differs in the restrictions imposed to recover discount factors, yields, or forward rates from observed coupon bond prices. Parametric approaches impose a low dimensional functional form and estimate a small set of parameters by minimizing pricing errors. A widely referenced benchmark is the U.S. Treasury yield curve estimated by \citet{GurkaynakSackWright2007} using the NSS specification. Nonparametric approaches instead prioritize flexibility in fitting the cross section, typically relying on local smoothing and explicit regularization. Classic examples include the \citet{fama1987information} piecewise constant forward rate construction and spline based estimators such as regression splines and smoothing splines. A prominent recent contribution is by \citet{LiuWu2021}, who construct a constant maturity zero-coupon curve using nonparametric kernel smoothing with an adaptive bandwidth tailored to Treasury issuance patterns. They show that replacing the commonly used G\"urkaynak  Sack  Wright curve with their estimates can change conclusions in influential applications such as \citet{CochranePiazzesi2005} and \citet{GiglioKelly2018} which highlights the economic significance of the underlying yield data.

One of the most recent contributions to zero-coupon yield curve construction is \citet{FilipovicPelgerYe2024}. They propose a robust closed form estimator of the discount curve that enforces smoothness through an explicit flexibility-smoothness trade off and is designed to remain stable in the presence of noisy Treasury prices. In extensive empirical comparisons against leading parametric and nonparametric benchmarks, including \citet{GurkaynakSackWright2007}, \citet{fama1987information}, \citet{Svensson1994}, and \citet{LiuWu2021}, they report uniformly smaller pricing and yield errors, greater robustness to outliers, and more stable implied forward rate series.

\subsection{Evaluation of Yield Forecasts based on Economic Value}
The yield curve forecasting literature focuses mainly on statistical forecast accuracy, most commonly through loss measures such as the root mean squared error. A smaller set of papers evaluates the economic value of forecasts, including portfolio allocation exercises in \citet{caldeira2016predicting} and trading exercises in \citet{guidolin2019forecasting}. To our knowledge, \citet{guidolin2019forecasting} is the only paper that forecasts yields using macro-related indicators and then evaluates those forecasts in a trading exercise that directly assesses their value by forming trading signals from predicted yield curve movements. However, its forecasting framework is more limited: it focuses on Dynamic Nelson--Siegel models with linear vector autoregression (VAR) and Markov-switching VAR dynamics, uses monetary policy indicators rather than a broad macro panel such as FRED-MD, and does not consider modern machine learning methods designed for high-dimensional predictors.

\subsection{Focus and Contribution of This Paper}
\label{subsec:lit_gaps}

Taken together, the literature points to three gaps that this paper addresses. First, existing work does not provide evidence comparing machine learning implementations of maturity-specific direct forecasts against DNS(S) factor-structured forecasts under a common macro-augmented information set. Second, while a sizeable literature uses global search or hybrid optimization methods to improve NSS calibration, there is limited evidence on whether these methods improve out-of-sample yield-curve forecasts beyond in-sample fit. This paper addresses this by testing whether optimizing decay parameters with such methods, rather than fixing them, yields measurable forecast gains. Third, to the best of our knowledge, since \citet{FilipovicPelgerYe2024} was released, their dataset has not been used for yield-curve forecasting, and this paper provides the first forecasting evaluation built on that data. In addition, this paper complements the statistical forecasting analysis with a simple trading exercise to assess whether the forecasts translate into economic value.

\section{Data}

\subsection{Yield Data}

This paper uses the U.S. zero-coupon Treasury yield data constructed by \citet{FilipovicPelgerYe2024}. The monthly zero-coupon yields are observed at month-end, that is, on the last day of each month. From the full estimated curve, I use the following key maturities:
\{3, 6, 12, 18, 24, 36, 60, 120, 240, 360\}\text{ months} which provide coverage of the short, medium, and long ends of the curve and are used as the forecasting targets throughout the paper. The models use a six-month information set based on lags 1--6 of the yield curve. In addition to forecasting yields at these maturities, I also consider slope movements of the yield curve through spread measures. For example, the slope measure \(24\_120\) is defined as $y_t(120)-y_t(24)$ which corresponds to the 2-year to 10-year slope.

\begin{figure}[H]
    \centering
    \makebox[\textwidth][c]{%
        \includegraphics[width=1.1\textwidth]{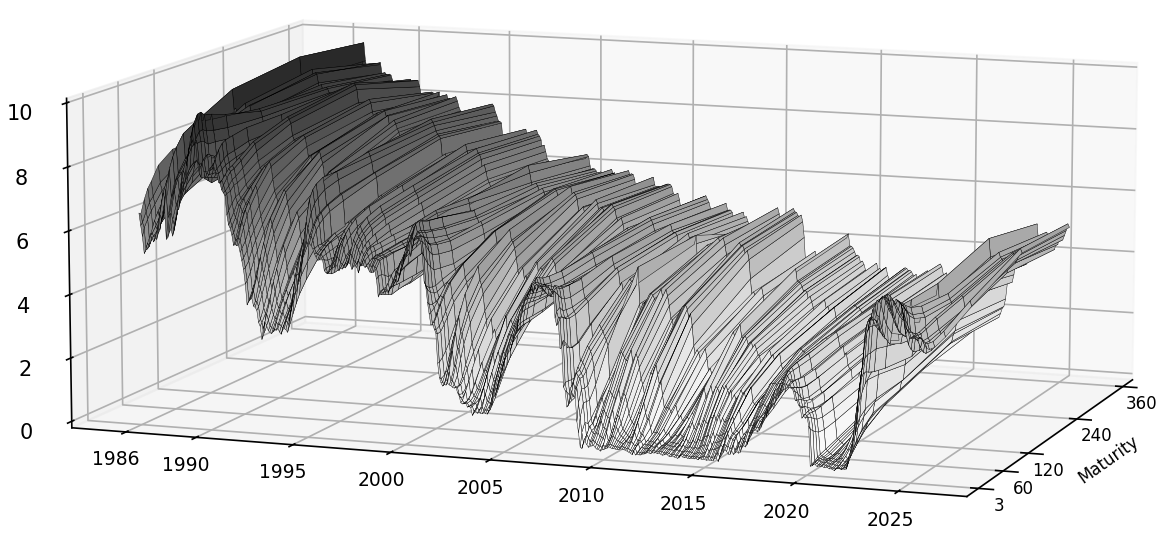}%
    }
    \caption{zero-coupon yield surface (\%) from 1986-05 to 2025-06.}
    \label{fig:yield_net}
\end{figure}

\autoref{fig:yield_net} shows the zero-coupon yield surface over the sample period considered in this paper, from 1986-05 to 2025-06. \autoref{tab:yield_summary} reports summary statistics for the selected maturities, together with the slope and curvature measures over the same period.

\begin{table}[htbp]
\centering
\caption{Summary statistics for zero-coupon yields at key maturities and spreads (\%) over 1986-05 to 2025-06}
\label{tab:yield_summary}
\begin{tabular}{lcccccc}
\hline
Maturity / Measure & Mean & Std. Dev. & Min & 25th pct. & Median & Max \\
\hline
3            & 3.136 & 2.483 & 0.002  & 0.378  & 3.130  & 9.154 \\
6            & 3.242 & 2.504 & 0.024  & 0.541  & 3.297  & 9.358 \\
12           & 3.403 & 2.533 & 0.062  & 0.798  & 3.504  & 9.608 \\
18           & 3.514 & 2.541 & 0.104  & 0.964  & 3.765  & 9.633 \\
24           & 3.604 & 2.520 & 0.121  & 1.113  & 3.854  & 9.504 \\
36           & 3.800 & 2.477 & 0.122  & 1.439  & 3.923  & 9.452 \\
60           & 4.148 & 2.371 & 0.233  & 1.957  & 4.109  & 9.289 \\
120          & 4.712 & 2.222 & 0.528  & 2.786  & 4.478  & 9.641 \\
240          & 5.236 & 2.138 & 1.042  & 3.339  & 4.963  & 9.933 \\
360          & 5.115 & 1.939 & 1.292  & 3.435  & 4.850  & 9.673 \\
24\_120      & 1.108 & 0.989 & -1.137 & 0.290  & 0.994  & 3.381 \\
24\_360      & 1.511 & 1.318 & -1.136 & 0.507  & 1.207  & 4.401 \\
60\_360      & 0.967 & 0.869 & -0.659 & 0.285  & 0.730  & 3.212 \\
\hline
\end{tabular}
\end{table}

\subsection{Macroeconomic Data}

The macroeconomic data used in this paper consist of the full panel of monthly U.S. indicators from FRED-MD \citet{mccracken2016fred}, which provides broad coverage of real activity, labor market conditions, housing, money, credit, and prices. Following the FRED-MD documentation, the corresponding transformation codes are applied to each series; see \autoref{tab:macro_vars_full}.

To avoid look-ahead bias, a conservative uniform publication lag of three months is imposed on all macro indicators. This choice is motivated by the slower release schedules of some series, especially housing-related indicators such as housing starts and permits, and ensures that all predictors are plausibly available at the forecast origin. The macro information set therefore consists of lags 4 to 6 of each monthly macro variable. 123 macro variables are used from FRED-MD hence this macro augmentation contributes 369 lagged predictors.
\section{Forecasting Methodology}

\subsection{Random Walk Benchmark}
Following the literature, the random walk is considered as the key benchmark. Under this specification, the forecast at horizon $h$ is simply the most recently observed yield, so that
\begin{equation}
\widehat{y}_{t+h\mid t}(\tau) = y_t(\tau).
\label{eq:rw_benchmark}
\end{equation}
where $y_t(\tau)$ denotes the zero-coupon yield at date $t$ and maturity $\tau$.

\subsection{Direct-Yield Forecast Approach}
For each forecast horizon $h$ and each yield maturity $\tau$ in the set of key maturities, a separate maturity-specific direct-yield forecast model is estimated. Rather than imposing a joint dynamic structure on the full yield curve, this approach treats each maturity as a separate prediction target.

The target variable is the $h$-step change in the yield,
\begin{equation}
\Delta_h y_t(\tau) \equiv y_{t+h}(\tau) - y_t(\tau).
\label{eq:direct_forecast_delta_def}
\end{equation}
Let $\mathbf{x}_t$ denote the predictor vector observed at time $t$. For each pair $(h,\tau)$, the forecasting equation is
\begin{equation}
\Delta_h y_t(\tau) = f_h^{(\tau)}(\mathbf{x}_t) + u_{t+h}^{(\tau)},
\label{eq:direct_forecast_delta}
\end{equation}
where $f_h^{(\tau)}(\cdot)$ is a maturity- and horizon-specific prediction function.

The corresponding forecast of the yield change is
\begin{equation}
\widehat{\Delta_h y}_{t+h\mid t}(\tau) = \widehat{f}_h^{(\tau)}(\mathbf{x}_t).
\label{eq:direct_forecast_delta_hat}
\end{equation}
The yield level forecast is then recovered as
\begin{equation}
\widehat{y}_{t+h\mid t}(\tau) = y_t(\tau) + \widehat{\Delta_h y}_{t+h\mid t}(\tau).
\label{eq:direct_forecast_delta_reconstruct}
\end{equation}

\subsection{Factor-Based Forecast Approach}

The Dynamic Nelson--Siegel (DNS) specification is
\begin{equation}
y_t(\tau)
= \beta_{0t}
+ \beta_{1t}\left(\frac{1-e^{-\lambda \tau}}{\lambda \tau}\right)
+ \beta_{2t}\left(\frac{1-e^{-\lambda \tau}}{\lambda \tau} - e^{-\lambda \tau}\right),
\label{eq:dns}
\end{equation}
where $\lambda>0$ controls the decay of the factor loadings. The decay parameter $\lambda$ determines how quickly the loadings decline with maturity and therefore where the curvature loading attains its peak. A smaller $\lambda$ implies slower decay and more weight at longer maturities, while a larger $\lambda$ implies faster decay and more weight at shorter maturities. The coefficients have natural economic interpretations. The coefficient $\beta_{0t}$ is a level factor because its loading is constant across maturities, so changes in $\beta_{0t}$ shift the entire curve. The coefficient $\beta_{1t}$ is a slope factor because its loading starts near one at short maturities and decays monotonically toward zero, so changes in $\beta_{1t}$ affect the short end more than the long end. The coefficient $\beta_{2t}$ is a curvature factor because its loading is hump-shaped: it is near zero at very short and very long maturities, but largest at intermediate maturities, so it primarily governs the medium-term shape of the yield curve.

This paper follows the factorization in \citet{DieboldLi2006}, which differs slightly from the original specification in \citet{NelsonSiegel1987}. As emphasized by \citet{DieboldLi2006}, this reparameterization is preferable because it yields more distinct factor loadings and hence clearer economic interpretations of the latent factors. In the original Nelson--Siegel form, two of the loadings are both monotonically decreasing and can be quite similar, which makes interpretation less transparent and can introduce multicollinearity in estimation.

The Dynamic Nelson--Siegel--Svensson (DNSS) extension adds a second curvature term:
\begin{equation}
y_t(\tau)
= \beta_{0t}
+ \beta_{1t}\left(\frac{1-e^{-\lambda_1 \tau}}{\lambda_1 \tau}\right)
+ \beta_{2t}\left(\frac{1-e^{-\lambda_1 \tau}}{\lambda_1 \tau} - e^{-\lambda_1 \tau}\right)
+ \beta_{3t}\left(\frac{1-e^{-\lambda_2 \tau}}{\lambda_2 \tau} - e^{-\lambda_2 \tau}\right),
\label{eq:dnss}
\end{equation}
where $\lambda_1,\lambda_2>0$ determine the locations of the two hump shaped
curvature loadings.

In this extension, the additional factor $\beta_{3t}$ introduces a second hump shaped loading, governed by $\lambda_2$, which increases flexibility in fitting medium to long maturity curvature while preserving the level-slope-curvature interpretation of the first three factors. DNS is nested within DNSS as a restricted case with the second curvature term suppressed
(i.e., $\beta_{3t}=0$ under the notation in \eqref{eq:dnss}).

\paragraph{Forecasting Dynamics}\mbox{}\par
After extracting the factors, factor dynamics are forecast directly at the target horizon, rather than forecasting yields maturity by maturity. Let
\[
\boldsymbol{\beta}_t =
\begin{cases}
(\beta_{0t},\beta_{1t},\beta_{2t})', & \text{under DNS},\\[4pt]
(\beta_{0t},\beta_{1t},\beta_{2t},\beta_{3t})', & \text{under DNSS}.
\end{cases}
\]
The $h$-step-ahead factor forecast is
\begin{equation}
\widehat{\boldsymbol{\beta}}_{t+h\mid t} = \widehat{\Phi}_h(\mathcal{I}_t),
\label{eq:beta_forecast_generic}
\end{equation}
where $\mathcal{I}_t$ denotes the information set available at time $t$. The predicted yield curve is then obtained by substituting $\widehat{\boldsymbol{\beta}}_{t+h\mid t}$ into the DNS or DNSS loading equations.

In the baseline specification, factor dynamics are modeled using an AR(1) process. For each factor, a direct $h$-step forecasting equation is estimated using only its first lag as the predictor, following \citet{DieboldLi2006}. Under this specification, the factors are forecast in levels.

I next consider a macro-augmented Random Forest specification. In this case, the information set is expanded to include lagged term structure factors, lagged yields, and lagged macroeconomic predictors,
\[
\mathcal{I}_t
=
\left\{
\boldsymbol{\beta}_{t-1}, \boldsymbol{\beta}_{t-2}, \ldots,\,
\mathbf{y}_{t-1}, \mathbf{y}_{t-2}, \ldots,\,
\mathbf{x}_{t-4}, \mathbf{x}_{t-5}, \ldots
\right\},
\]
where \(\boldsymbol{\beta}_t\) denotes the vector of term structure factors, \(\mathbf{y}_t\) denotes the vector of yields across the selected maturities, and \(\mathbf{x}_t\) denotes the macro predictor vector. Yield and factor information enters with lags starting at \(t-1\), while macro predictors enter only from \(t-4\) onward to reflect the imposed 3-month publication lag. For the Random Forest model, the target variable is defined as the $h$-step change in each factor,
\begin{equation}
\Delta_h \beta_{i,t} \equiv \beta_{i,t+h} - \beta_{i,t}.
\label{eq:beta_diff_target}
\end{equation}
For each factor $i$, the forecast is written as
\begin{equation}
\widehat{\Delta_h \beta}_{i,t+h\mid t} = \widehat{f}^{(\beta_i)}_h(\mathcal{I}_t),
\label{eq:beta_rf_diff_forecast}
\end{equation}
where $\widehat{f}^{(\beta_i)}_h(\cdot)$ denotes the factor- and horizon-specific Random Forest prediction function. The implied factor level forecast is then recovered as
\begin{equation}
\widehat{\beta}_{i,t+h\mid t} = \beta_{i,t} + \widehat{\Delta_h \beta}_{i,t+h\mid t}.
\label{eq:beta_diff_reconstruct}
\end{equation}

\paragraph{Time Varying Decay Parameters}\mbox{}\par

In addition, I consider a specification with time varying decay parameters. Here, the cross sectional estimation at each date $t$ produces not only the factor vector $\boldsymbol{\beta}_t$, but also date specific decay parameters $\lambda_{1t}$ and $\lambda_{2t}$. However, because these parameters determine the basis functions of the yield curve, forecasting them jointly with the factor coefficients would make the forecasting problem considerably more complex. Hence, I use a random walk model for the decay parameters. That is, at forecast origin $t$, I set the future decay parameters equal to their most recently estimated values. 
\begin{equation}
\widehat{\lambda}_{1,t+h\mid t} = \lambda_{1t}, 
\qquad
\widehat{\lambda}_{2,t+h\mid t} = \lambda_{2t}.
\label{eq:lambda_random_walk}
\end{equation}

The estimation of time varying decay parameters are covered in the next sub-section.

\subsubsection{Estimation of Time Varying NSS Decay Parameters}
\label{sec:nss_pso_calibration}

The estimation of decay parameters for the NSS framework involves solving a non-convex objective function which is difficult to optimize with local methods alone \citet{GilliGrosseSchumann2010}. I implement a hybrid optimization algorithm that uses Particle Swarm Optimization (PSO), a global optimization algorithm, followed by the limited-memory Broyden--Fletcher--Goldfarb--Shanno algorithm with box constraints (L-BFGS-B) as a local refinement step \citet{byrd1995limited}. In this hybrid optimization procedure, PSO is first run for a fixed number of iterations to explore the parameter space, after which L-BFGS-B is applied to each particle in the resulting swarm and the best local optimum is retained.

This hybridization is useful because PSO can become inefficient near the optimum, where particles may overshoot the solution and converge slowly. In contrast, L-BFGS-B is well suited for accurate local refinement once PSO has already identified a promising region of the parameter space. After the PSO stage ends, I initialize L-BFGS-B from the leading particles in the swarm and retain the solution with the lowest final objective value.

\paragraph{Particle Swarm Optimization Algorithm}\mbox{}\par

Particle Swarm Optimization is a stochastic global optimization method inspired by the collective behaviour of socially organized groups such as bird flocks, fish schools, and animal herds \citep{parsopoulos2010particle}. Its core idea is that candidate solutions move through the search space while sharing information about previously discovered promising regions.

I begin with the baseline constrained specification
\begin{equation}
\begin{aligned}
\min_{\theta_t} \quad & \sum_{j=1}^{M} \left(y_t(\tau_j) - \hat{y}_t(\tau_j; \theta_t)\right)^2 \\
\text{s.t.} \quad 
& \lambda_{1,t} \geq \lambda_{2,t} \geq 0 \quad,
\end{aligned}
\label{eq:nss_pso_problem_baseline}
\end{equation}

where $y_t(\tau_j)$ is the observed yield at maturity $\tau_j$ and $\hat{y}_t(\tau_j; \theta_t)$ is the NSS fitted yield. The condition $\lambda_{1,t} \geq \lambda_{2,t} \geq 0$ ensures that both decay parameters are nonnegative, so that the exponential loading functions decay with maturity, while also imposing ordering on the two decay parameters. The parameter vector is
\[
\theta_t = \left(\beta_{0,t}, \beta_{1,t}, \beta_{2,t}, \beta_{3,t}, \lambda_{1,t}, \lambda_{2,t}\right)'.
\]

\begin{figure}[H]
    \centering
    \setlength{\abovecaptionskip}{3pt}
    \setlength{\belowcaptionskip}{0pt}

    \begin{subfigure}[t]{0.44\textwidth}
        \centering
        \includegraphics[width=\linewidth]{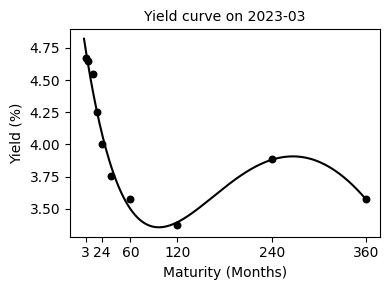}
    \end{subfigure}
    \hfill
    \begin{subfigure}[t]{0.44\textwidth}
        \centering
        \includegraphics[width=\linewidth]{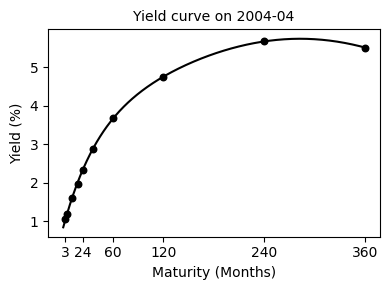}
    \end{subfigure}

    \vspace{0.05em}

    \begin{subfigure}[t]{0.44\textwidth}
        \centering
        \includegraphics[width=\linewidth]{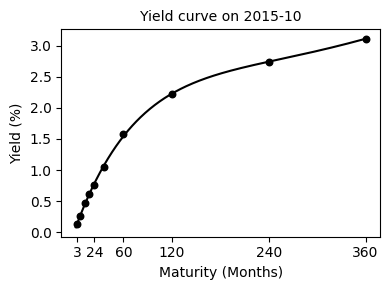}
    \end{subfigure}
    \hfill
    \begin{subfigure}[t]{0.44\textwidth}
        \centering
        \includegraphics[width=\linewidth]{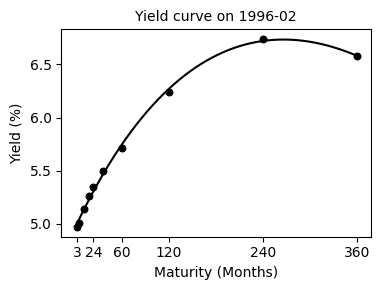}
    \end{subfigure}

    \caption{DNSS-PSO yield curves for the four dates with the largest negative $\beta_0$ estimates. The dots denote the observed yields.}
    \label{fig:selected_yield_curves_negative_beta0}
\end{figure}

Under this baseline formulation, the fitted curves are often visually satisfactory, but the resulting factor coefficients are not always economically meaningful. In particular, some dates produce large negative spikes in the level factor $\beta_{0,t}$. These episodes reflect weak identification rather than genuinely negative long-run rates. The issue is that, on dates where the cross section is very flat over a substantial portion of the curve and exhibits little curvature, the objective function is nearly unchanged across certain combinations of NSS parameters. In such cases, the optimizer cannot clearly distinguish between offsetting values of $\beta_{0,t}$ and $\beta_{1,t}$, so the level and slope factors become effectively interchangeable. The model therefore fits the cross section mechanically, with good fit, but the decomposition into economically interpretable factors breaks down. This is illustrated in Figure~\ref{fig:selected_yield_curves_negative_beta0}, which plots the fitted and observed yield curves for dates corresponding to the four largest negative spikes in the estimated level factor. More broadly, these episodes suggest that NSS-type models are more difficult to identify when the yield curve is unusually flat and contains little curvature, even though they tend to perform better when the cross section exhibits richer shape variation.

To address this issue, I also consider a more restrictive specification that imposes a zero lower bound on both the long-end and short-end limits of the fitted yield curve:
\begin{equation}
\begin{aligned}
\min_{\theta_t} \quad & \sum_{j=1}^{M} \left(y_t(\tau_j) - \hat{y}_t(\tau_j; \theta_t)\right)^2 \\
\text{s.t.} \quad 
& \beta_{0,t} \geq 0, \\
& \beta_{0,t} + \beta_{1,t} \geq 0, \\
& \lambda_{1,t} \geq \lambda_{2,t} \geq 0 \quad.
\end{aligned}
\label{eq:nss_pso_problem_zlb}
\end{equation}

The restriction $\beta_{0,t} \geq 0$ imposes a zero lower bound on the long-end limit of the fitted yield curve, while $\beta_{0,t} + \beta_{1,t} \geq 0$ imposes a zero lower bound on its short-end limit. These additional constraints improve the economic interpretability of the estimated factors by ruling out the most extreme weak-identification outcomes, although this comes at the cost of slightly weaker out-of-sample forecast performance.

With the optimization problem defined, I move on to describe the swarm-based search procedure used to solve it. Each individual in the swarm is a particle, which represents one candidate NSS parameter vector. The swarm is initialized by drawing particle positions randomly over the search region, while initial velocities are also randomly assigned and then bounded to prevent excessively large moves. Let
\[
x_i^{(k)} = \left(\beta_{0,i}^{(k)}, \beta_{1,i}^{(k)}, \beta_{2,i}^{(k)}, \beta_{3,i}^{(k)}, \lambda_{1,i}^{(k)}, \lambda_{2,i}^{(k)}\right)'
\]
denote the position of particle $i$ at iteration $k$.

At the initial step, the objective function is evaluated at each particle's starting position. These initial evaluations determine each particle's personal best and its neighborhood best. At each subsequent iteration, every particle updates its velocity and then its position according to
\[
v_i^{(k+1)} = \omega_k v_i^{(k)}
+ c_1 r_{1,i}^{(k)} \left(p_i^{(k)} - x_i^{(k)}\right)
+ c_2 r_{2,i}^{(k)} \left(n_i^{(k)} - x_i^{(k)}\right),
\]
\[
x_i^{(k+1)} = x_i^{(k)} + v_i^{(k+1)},
\]
where $\omega_k$ is the inertia parameter, $c_1$ and $c_2$ are acceleration coefficients, and $r_{1,i}^{(k)}$ and $r_{2,i}^{(k)}$ are random draws from a uniform distribution on $(0,1)$,  $p_i^{(k)}$ denotes the particle's personal best position up to iteration $k$, $n_i^{(k)}$ denotes the best position found within particle $i$'s neighborhood and $v_i^{(k)}$ denotes its associated velocity. In this implementation, the inertia weight decreases linearly from $w_{\text{start}}$ to $w_{\text{end}}$ over the PSO run, so that the swarm places greater weight on exploration in early iterations and more weight on local refinement in later iterations.

In a global-best PSO, every particle is attracted to the best solution found by the entire swarm, whereas in a local-best PSO each particle is attracted only to the best solution found within its neighborhood. Rather than using a single global-best particle throughout, I adopt an adaptive neighborhood structure based on a ring topology, which \citet{parsopoulos2010particle} describe as a standard and computationally simple choice for local PSO variants. They note that the local-best variant has better exploration properties because information about good solutions diffuses gradually through neighboring particles rather than being transmitted immediately to the whole swarm, which gives the algorithm more scope to avoid suboptimal solutions. They also emphasize that neighborhood topology and size affect the exploration--exploitation trade-off. Accordingly, I initialize the swarm with a small ring neighborhood so that information spreads gradually in the early iterations, helping preserve swarm diversity. I then increase the neighborhood radius over the PSO run, allowing information to diffuse more broadly across particles as the search progresses.

Hence, the algorithm consists of four key elements: the particle's current position, which represents its candidate parameter vector; the particle's personal best position, which records the best solution that particle has found so far; the neighborhood-best position, which summarizes the best solution available within the particle's current ring neighborhood; and the particle's velocity, which governs movement through the search space by combining inertia, individual learning, and social learning from the neighborhood.

To improve stability across adjacent dates, I initialize one particle at the previous period's optimal parameter vector, which acts as a warm start. I also include a fixed anchor particle with baseline decay values $\lambda_1 = 0.0609$ and $\lambda_2 = 0.01$ and the associated factor coefficients initialized by OLS applied to the yield cross section for that date. The remaining particles are initialized randomly, so the swarm retains broad exploratory coverage of the admissible parameter space.

\paragraph{L-BFGS-B Algorithm}\mbox{}\par

After the PSO stage, the local refinement step uses the limited-memory Broyden--Fletcher--Goldfarb--Shanno algorithm with box constraints (L-BFGS-B). L-BFGS-B is a quasi-Newton method for smooth nonlinear optimization. Rather than relying only on first order updates in gradient descent, it uses gradient information to build an approximation to the inverse Hessian, which generally improves the speed and accuracy of local convergence. The limited memory version stores only a small amount of recent information, and the simple box constraints ensure that parameters remain within pre specified bounds.

\paragraph{PSO--L-BFGS-B Hyperparameters}\mbox{}\par
I use a swarm size of 300 particles and set a maximum of 1{,}000 PSO iterations, with early termination determined by a stall criterion based on an objective improvement tolerance of \(10^{-8}\). The initial neighborhood size is set to 10\% of the swarm, and post-PSO local refinement is performed using L-BFGS-B on the top 10 particles, with a maximum of 500 iterations, function tolerance \(10^{-12}\), and gradient tolerance \(10^{-8}\). The PSO search domain is given by \(\lambda_1,\lambda_2 \in [10^{-9},\,0.4]\) \footnote{The upper bound of 0.4 for the decay parameters is chosen with reference to the range of estimated NSS decay parameters reported by \citet{GurkaynakSackWright2007}.} and \(\beta_j \in [-1,1]\) for \(j=0,1,2,3\). Under these settings, the PSO stage typically reaches a stable neighborhood of the optimum, so the subsequent L-BFGS-B step mainly acts as a local polishing step.

\paragraph{Estimated Time-Varying DNSS Parameters}\mbox{}\par
Three figures summarize the parameter estimates; the reported \(\lambda\) values are expressed as monthly decay parameters. Figure~\ref{fig:nss_params_uncon} reports the estimates under the baseline specification. Consistent with the identification issues discussed above in the PSO methodology, the baseline estimates contain several episodes in which \(\beta_{0,t}\) becomes sharply negative, reflecting weak identification rather than genuinely negative long-run rates. Figure~\ref{fig:selected_yield_curves_negative_beta0} examines the four most extreme episodes in greater detail and shows that they are associated with unusually flat yield curves with little curvature over much of the cross section. In these cases, the model continues to fit the observed yields closely, but the decomposition into economically interpretable level and slope factors becomes unstable. Figure~\ref{fig:nss_params_zlb} reports the parameter estimates under the zero lower bound specification in equation~\eqref{eq:nss_pso_problem_zlb}. Under this specification, \(\beta_{0,t}\) is no longer allowed to enter the negative region, and the resulting \(\beta_{1,t}\) estimates appear closer to those reported by \citet{GurkaynakSackWright2007}. For comparison, Figure~\ref{fig:nss_params_gsw} reports the estimates of \citet{GurkaynakSackWright2007}.

In this PSO algorithm to estimate time-varying decay parameters, I work with a six-parameter PSO particle consisting of two decay parameters and four factor coefficients. However, for in-sample fitted values, I do not use the four factor coefficients from PSO estimation and instead extract factor coefficients using OLS from equation~\eqref{eq:dnss}, conditional on the decay parameters estimated by PSO. In the out-of-sample exercise, factor coefficients are forecasted using AR or RF models, while the decay parameters are modelled as random walk. Forecasted yields are then reconstructed using OLS from equation~\eqref{eq:dnss} once again. This keeps the mapping between yields, factor coefficients, and decay parameters consistent across the in-sample and out-of-sample stages, which supports a stable forecasting framework.


\subsubsection{Evaluating Yield Curve Fit}
To evaluate yield curve fit of DNSS specifications using decay parameters estimated from PSO, the cross-sectional fitting error is computed for each date $t$ as
\[
\mathrm{RMSE}_t = \sqrt{\frac{1}{M}\sum_{m=1}^{M}\left(y_t(\tau_m)-\hat{y}_t(\tau_m)\right)^2},
\]
where $M$ is the number of maturities, $y_t(\tau_m)$ is the observed zero-coupon yield, and $\hat{y}_t(\tau_m)$ is the fitted yield. Table~\ref{tab:factor_fit_rmse} reports the average of $\mathrm{RMSE}_t$ over all dates in the sample. This comparison is carried out over the full 1986-05 to 2025-06 period because the objective here is to assess cross-sectional fit rather than forecast performance. For reference, the table also includes the fit from a baseline DNS model ($\lambda = 0.0609$) and a baseline DNSS model ($\lambda_1 = 0.0609,\ \lambda_2 = 0.01$). The DNSS-PSO and DNSS-PSO with zero lower bound (ZLB) specifications allow the decay parameters to vary over time. Among all specifications, the DNSS-PSO achieves the lowest RMSE. For the remainder of the fit evaluation, attention is restricted to the DNSS-PSO specification. As noted earlier, the ZLB specification is not pursued further because it delivers weaker out-of-sample forecasting performance.

\begin{figure}[H]
    \centering
    \setlength{\abovecaptionskip}{3pt}
    \setlength{\belowcaptionskip}{0pt}

    \begin{subfigure}[t]{0.90\textwidth}
        \centering
        \includegraphics[width=\linewidth,height=0.25\textheight,keepaspectratio]{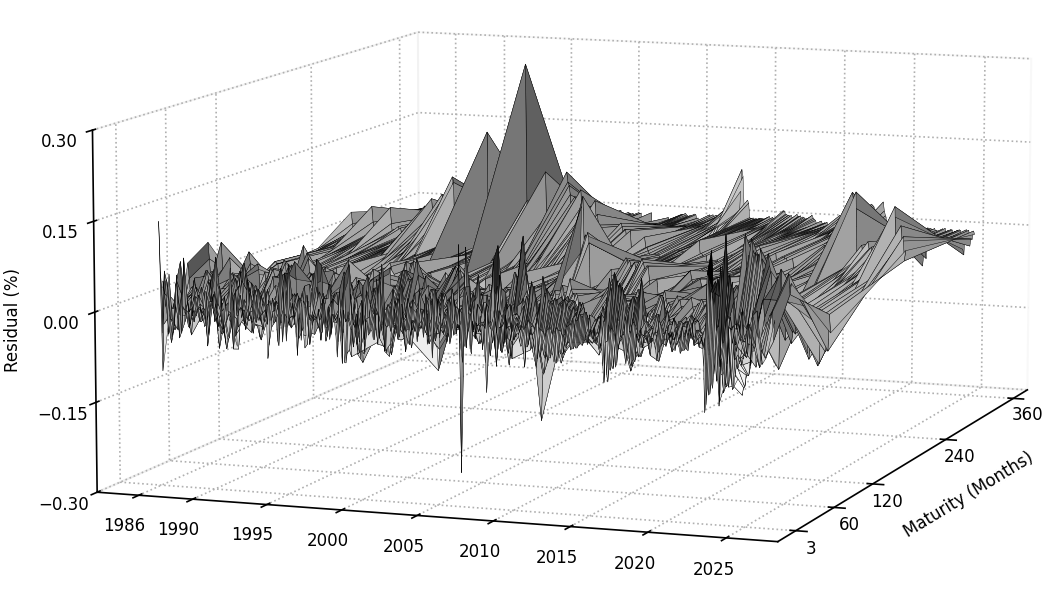}
    \end{subfigure}

    \vspace{0.1em}

    \begin{subfigure}[t]{0.90\textwidth}
        \centering
        \includegraphics[width=\linewidth,height=0.25\textheight,keepaspectratio]{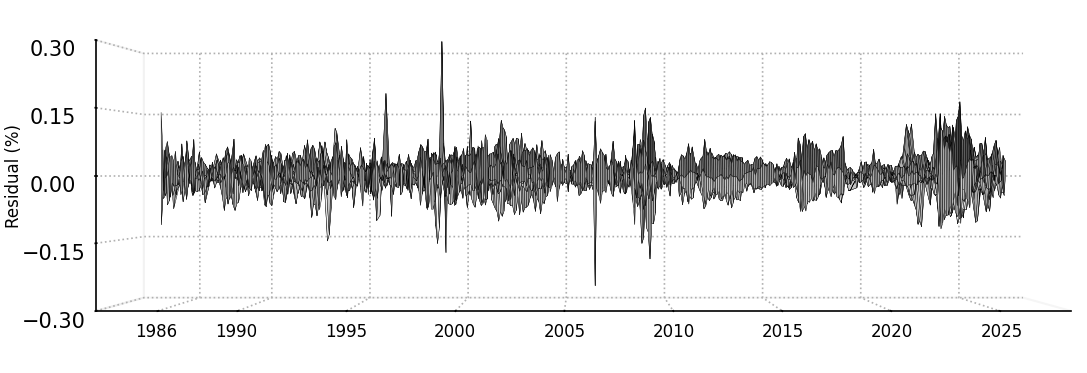}
    \end{subfigure}

    \caption{Yield curve residuals from the DNSS-PSO yield curves fitted month-by-month from 1986-05 to 2025-06. The same residual surface is shown under two viewing angles.}
    \label{fig:resid_surfaces_azim}
\end{figure}

\begin{figure}[H]
    \centering

    \begin{subfigure}[t]{0.44\textwidth}
        \centering
        \includegraphics[width=\linewidth]{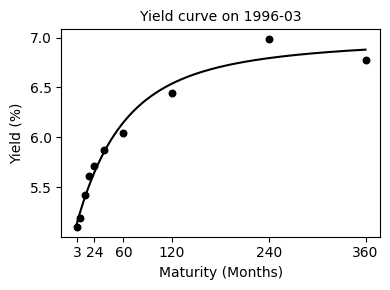}
    \end{subfigure}
    \hfill
    \begin{subfigure}[t]{0.44\textwidth}
        \centering
        \includegraphics[width=\linewidth]{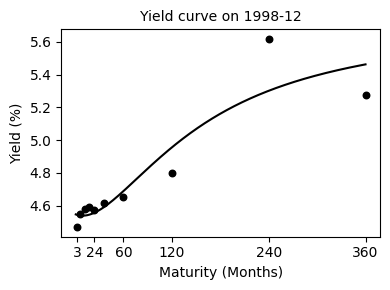}
    \end{subfigure}

    \vspace{0.4em}

    \begin{subfigure}[t]{0.44\textwidth}
        \centering
        \includegraphics[width=\linewidth]{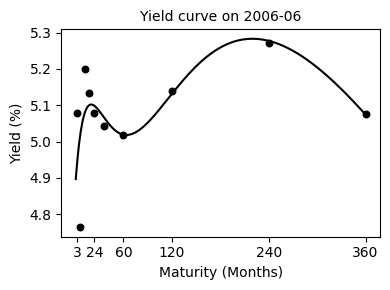}
    \end{subfigure}
    \hfill
    \begin{subfigure}[t]{0.44\textwidth}
        \centering
        \includegraphics[width=\linewidth]{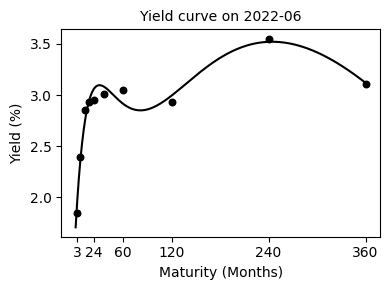}
    \end{subfigure}

    \caption{DNSS-PSO yield curves for the four dates with the highest total RMSE across maturities. The dots denote the observed yields.}
    \label{fig:selected_yield_curves_highest_rmse}
\end{figure}

\begin{figure}[H]
    \centering

    \begin{subfigure}[t]{0.44\textwidth}
        \centering
        \includegraphics[width=\linewidth]{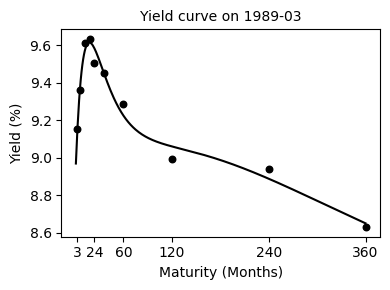}
    \end{subfigure}
    \hfill
    \begin{subfigure}[t]{0.44\textwidth}
        \centering
        \includegraphics[width=\linewidth]{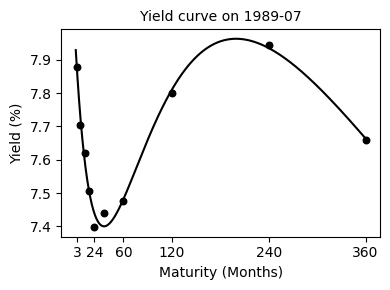}
    \end{subfigure}

    \vspace{0.4em}

    \begin{subfigure}[t]{0.44\textwidth}
        \centering
        \includegraphics[width=\linewidth]{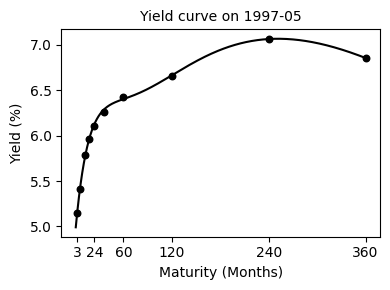}
    \end{subfigure}
    \hfill
    \begin{subfigure}[t]{0.44\textwidth}
        \centering
        \includegraphics[width=\linewidth]{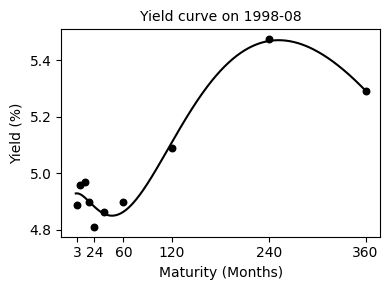}
    \end{subfigure}

    \caption{DNSS-PSO yield curves for the same four dates shown in \citet{DieboldLi2006}. The dots denote the observed yields.}
    \label{fig:selected_yield_curves_same_as_diebold}
\end{figure}

Focusing on the DNSS-PSO specification in more detail, the residual surface in Figure~\ref{fig:resid_surfaces_azim} indicates a close fit, with the magnitude of residuals remaining within 0.15\% for nearly all dates. The four dates with the largest cross-sectional residuals, corresponding to the largest spikes visible in the side view of Figure~\ref{fig:resid_surfaces_azim}, such as 2006-06 and 1998-12, are plotted in Figure~\ref{fig:selected_yield_curves_highest_rmse}. For additional illustration, Figure~\ref{fig:selected_yield_curves_same_as_diebold} plots the fitted and observed yield curves for four selected dates, chosen to match those examined in \citet{DieboldLi2006}.

\begin{table}[htbp]
\centering
\caption{Comparison of fit across baseline DNS(S) and specifications with decay parameters estimated from PSO}
\label{tab:factor_fit_rmse}
\renewcommand{\arraystretch}{1.15}
\begin{tabular}{lc}
\toprule
Model specification & RMSE (\% yield) \\
\midrule
DNS ($\lambda = 0.0609$) & 0.1140 \\
DNSS ($\lambda_1 = 0.0609,\ \lambda_2 = 0.01$) & 0.0923 \\
DNSS-PSO & 0.0284 \\
DNSS-PSO with ZLB & 0.0331 \\
\bottomrule
\end{tabular}

\vspace{0.35em}
\begin{minipage}{0.95\textwidth}
\footnotesize
Notes: The table reports the average date-level cross-sectional RMSE over the 1986-05 to 2025-06 sample period. RMSE is reported in percentage-yield units.
\end{minipage}
\end{table}

\subsection{Random Forest Model}
The Random Forest (RF) is the machine learning model used in both the direct-yield and factor-based forecasting approaches. Introduced by \citet{breiman2001random}, it is well suited to handling a large number of potentially correlated predictors, which makes it a natural choice for the high-dimensional macro-augmented predictor set. It also captures nonlinearities and interaction effects automatically, without requiring these features to be imposed in the model specification. This choice is also consistent with \citet{WangWangTu2025}, who find that RF performs best among the machine learning models considered in their yield forecasting exercise.

The RF is built from regression trees. A regression tree partitions the predictor space into \(M\) disjoint regions, \(\{R_1,\dots,R_M\}\), and assigns a constant prediction within each region. Its fitted value is written as
\[
f(x)=\sum_{m=1}^{M} c_m \, I(x \in R_m),
\]
where \(I(\cdot)\) is an indicator function and
\[
c_m = \frac{1}{N_m}\sum_{i:x_i\in R_m} y_i
\]
is the average of the target variable within region \(R_m\), with \(N_m\) denoting the number of observations in that region. In this way, the tree approximates the conditional mean of the target variable by allowing the relationship between predictors and the forecasted yield or factor to differ across regions of the predictor space. Splits within each tree are chosen to reduce squared prediction error, so that terminal nodes group together observations with similar target values. However, a single tree is often unstable and sensitive to the particular sample on which it is estimated. Using many trees instead of one allows predictions to be averaged across trees, which reduces prediction variance. This idea underlies bootstrap aggregation, or bagging, in which many regression trees are estimated on bootstrap resamples of the training data and their predictions are averaged.

A limitation of plain bagging is that the trees can remain highly correlated, especially when a few predictors dominate the splitting decisions. Random Forest addresses this by restricting each split to a random subset of predictors of size \(m < P\), where \(P\) is the total number of candidate predictors. This decorrelates the trees, improves the gains from averaging, and is particularly useful when predictors are numerous and highly correlated. The RF forecast is then obtained by averaging the predictions across the \(B\) trees,
\[
\hat f_{\mathrm{RF}}(x)=\frac{1}{B}\sum_{b=1}^{B} T_b(x),
\]
where \(T_b(x)\) denotes the prediction from tree \(b\).

In implementation, each RF consists of 500 regression trees, with a minimum leaf size of 5 so that every terminal node contains at least 5 observations. The hyperparameter \(m\) is selected using recursive cross-validation. A validation period from 2010-01 to 2014-12 is used, corresponding to the final 60 targets of the in-sample period. Candidate \(m\) values are based on the predictor dimension \(P\):
\[
\left\{
\sqrt{P},\; \frac{P}{10},\; \frac{P}{6},\; \frac{P}{4},\; \frac{P}{3},\; \frac{P}{2},\; \frac{2P}{3}
\right\},
\]
with values rounded to the nearest integer. The \(m\) value with the lowest validation RMSE is selected. The resulting optimal choices for each model and forecast horizon are reported in Table~\ref{tab:chosen_mtry_all_models}.

\section{Results}

The results are organized to evaluate the forecasting system systematically. I begin with a broad overview of forecast performance relative to the random walk in Subsection~\ref{subsec:results_rw}. I then examine the results of each key aspect of the forecasting methodology: the comparison of forecasting approaches in Subsection~\ref{subsec:role_of_approach}, the role of macroeconomic predictors in Subsection~\ref{subsec:role_of_macro}, the link between in-sample fit and out-of-sample forecast performance in Subsection~\ref{subsec:role_of_pso}, and, finally, the trading exercise in Subsection~\ref{subsec:results_trading}.

\subsection{Forecast Evaluation Setup}

The forecasting exercise is conducted using a fixed size rolling window. The in-sample period runs from 1986-05 to 2014-12, while the out-of-sample evaluation period runs from 2015-01 to 2025-05. At each forecast origin, the model is re-estimated using the rolling estimation window, and forecasts are generated for horizons of 1, 3, 6, and 12 months ahead.

Forecast accuracy is evaluated out of sample by comparing predicted and realized yields over the evaluation period. Forecast accuracy is measured using the root mean squared error (RMSE), computed separately for each maturity and forecast horizon $h$:
\begin{equation}
RMSE_h = \sqrt{\frac{1}{T_h}\sum_{t=1}^{T_h} \left(y_{t+h} - \hat{y}_{t+h \mid t}\right)^2 }.
\end{equation}
Relative RMSE is also reported, defined as the ratio of the RMSE of the competing model to that of the random walk benchmark:
\begin{equation}
\text{Relative RMSE} = \frac{RMSE_{\text{Competing Model}}}{RMSE_{\text{Random Walk Model}}}.
\end{equation}
A value below one indicates that the competing model outperforms the random walk benchmark.

To compare predictive accuracy across models, I follow \citet{diebold1995comparing} and implement the Diebold--Mariano (DM) test. The DM test evaluates whether the mean difference in forecast loss between two models is statistically different from zero. In this paper, it is used to test whether one model delivers significantly lower out-of-sample forecast errors than another at a given maturity and forecast horizon.

Let $e_{1,t+h}$ denote the forecast error from the benchmark model and $e_{2,t+h}$ the forecast error from the competing model. Define the loss differential as
\begin{equation}
d_t = L(e_{1,t+h}) - L(e_{2,t+h}),
\end{equation}
where $L(\cdot)$ denotes the forecast loss function, taken here to be squared error loss. Let
\begin{equation}
\bar d = \frac{1}{T}\sum_{t=1}^T d_t.
\end{equation}
The Diebold--Mariano test statistic is then given by
\begin{equation}
DM = \frac{\bar d}{\hat{\sigma}_{\bar d}},
\qquad
\hat{\sigma}_{\bar d} = \sqrt{\frac{\widehat{S}_d(0)}{T}},
\end{equation}
where $\widehat{S}_d(0)$ denotes a heteroskedasticity and autocorrelation consistent (HAC) estimate of the standard error of $\bar d$ to account for serial correlation of forecast loss differentials. To select the HAC lag length, I examine the sample autocorrelation function of the loss differential series up to lag $\lfloor T^{1/3}\rfloor$. The selected HAC lag length is the largest lag up to $\lfloor T^{1/3}\rfloor$ for which the sample autocorrelation is statistically significant at the 5\% level using Bartlett confidence bands. If no lag is significant, the selected HAC lag length is set to zero. Under the null of equal predictive accuracy, \(E[d_t]=0\). A positive \(\bar d\) indicates that the competing model has lower forecast loss than the benchmark, while a negative \(\bar d\) indicates the opposite. The corresponding table of selected HAC lags is reported in the Appendix (Table~\ref{tab:hac_sig_lags}).

\subsection{Forecast Results Relative to the Random Walk}
\label{subsec:results_rw}

The forecast results are separated into the short end (3- to 24-month maturities), the belly (36- to 60-month maturities), the long end (120- to 360-month maturities), and slope forecasts. A summary of the competing models, including their descriptions and predictor sets, is provided in the Appendix; see Table~\ref{tab:model_predictor_summary}.

Tables~\ref{tab:rrmse_h1}, \ref{tab:rrmse_h3}, \ref{tab:rrmse_h6}, and \ref{tab:rrmse_h12} report the RMSE of each model specification relative to the random walk benchmark at the 1-, 3-, 6-, and 12-month forecast horizons, respectively. Panel~A reports yields-only specifications, while Panel~B reports macro-augmented specifications. Statistical significance relative to the random walk benchmark is denoted by stars: *** \(1\%\), ** \(5\%\), and * \(10\%\).

Overall, improvements over the random walk are concentrated mainly at the short end of the curve and, to a lesser extent, in slope forecasts. At the 1- and 3-month horizons, several models achieve relative RMSE below one at short maturities, with the strongest and most consistent gains coming from the direct-yield and fixed-PSO DNSS specifications, especially when macroeconomic predictors are included. For slope forecasts, the strongest performance comes mainly from the macro-augmented direct-yield approach. The direct-yield approach is compared to the factor-based approach in greater detail in section ~\ref{subsec:role_of_approach}. The belly and long end are harder to beat, relative RMSEs in these segments are usually close to or above one, and statistically significant improvements are less frequent.


For the macro-augmented models, a clear pattern across forecast horizons emerges. Forecast gains are strongest at short horizons, weaken at the 6-month horizon, and become much more limited by 12 months ahead. At the 12-month horizon, most specifications cluster close to one, indicating performance similar to the random walk benchmark, with only a few selective gains at particular long-end maturities or slope measures. This suggests that, even after incorporating macroeconomic information, the random walk remains a particularly strong benchmark at longer horizons.

By contrast, among the yield-only specifications with AR dynamics, the pattern across forecast horizons is more consistent with Diebold and Li (2006), with predictive improvements becoming more apparent at longer horizons. In fact, for single-maturity forecasting, these yield-only AR specifications broadly outperform the corresponding macro-augmented specifications at longer horizons. This indicates that the incremental predictive power of macroeconomic variables diminishes as the forecast horizon increases, while the information embedded in the yield curve itself remains relatively more useful for longer-horizon forecasts within the AR-based setting.

Macro augmentation helps, though not uniformly across the curve. Its gains are most visible at the short end and in slope-related forecasts, especially at the 1-, 3-, and 6-month horizons. In contrast, adding macroeconomic predictors does not systematically improve belly and long-end forecasts, and in some factor-based PSO specifications is associated with substantially worse performance relative to the random walk. The role of macroeconomic predictors is examined in greater detail in Subsection ~\ref{subsec:role_of_macro}.

Taken together, the results point to three broad conclusions. First, beating the random walk is feasible mainly for short-end and slope forecasts rather than for the curve as a whole. Second, forecasting performance relative to the random walk declines materially with the horizon. Third, the strongest performers are models that combine either direct forecasting or relatively stable factor structures with macroeconomic information, whereas the more flexible time-varying PSO specifications generally fail to improve on the random walk and often underperform it by a wide margin. The effect of using decay parameters estimated by PSO is examined in greater detail in Subsection ~\ref{subsec:role_of_pso}.

\subsection{Direct-Yield vs Factor-Based Approach}
\label{subsec:role_of_approach}

I next use the DM test to compare the direct-yield approach against the factor-based approach. Table~\ref{tab:approach_relative_rmse} reports the RMSE of each factor-based model relative to the corresponding direct-yield benchmark. Hence, a value above one indicates that the direct-yield model has lower RMSE and therefore more accurate forecasts. Panel~A reports the yields-only specifications, so the factor-based models are scaled relative to Direct-RF. Panel~B reports the macro-augmented specifications, so the factor-based models are scaled relative to Direct-RF-X.

The clearest relative advantage of the direct-yield approach is in slope forecasts. Relative RMSE is most often above one for the 24--120, 24--360, and 60--360 slopes, especially at short and medium horizons. By contrast, for single-maturity forecasts the relative performance across the curve is more mixed, with many relative RMSEs close to one. That said, when attention is restricted to the DNSS specifications rather than the DNS models, the factor-based approach tends to perform better than the direct-yield benchmark at the short end, where relative RMSE is more often below one.

A further pattern is that the factor-based approach steadily gains relative to the direct-yield approach as the forecast horizon increases. This is visible across all the factor-based specifications. Relative RMSEs generally move closer to one, and in many cases below one, as the horizon rises from $h=1$ to $h=12$. In the yields-only specifications, this effect becomes especially clear at the later forecast horizons. By $h=12$, in the yields-only specifications, the factor-based models consistently outperform the direct-yield benchmark for single-maturity forecasts across much of the curve. In the macro-augmented specifications, the same pattern is present but less pronounced. Even at longer horizons, however, the direct-yield benchmark tends to retain a clearer edge for slope-related targets.

Overall, the results indicate that the direct-yield approach is most useful at shorter forecast horizons, especially for forecasting slopes. By contrast, for single-maturity forecasts, the factor-based approach becomes relatively more attractive as the forecast horizon lengthens.

\subsection{The Role of Macroeconomic Predictors}
\label{subsec:role_of_macro}

\subsubsection{Test of Incremental Predictive Power}


To measure the incremental predictive power of macroeconomic variables, I use the DM test to compare the forecast accuracy of yield-only models against their macro-augmented counterparts. The results are reported in Table~\ref{tab:relative_rmse_macro_models}, where each entry gives the RMSE of the macro-augmented model relative to that of the corresponding yield-only specification. Values below one therefore indicate that macro augmentation improves forecast accuracy. For example, for the Direct-RF-X model at the 12-month-ahead horizon and the 3-month maturity, the reported value of 0.87 means that the RMSE of Direct-RF-X is 87\% of the RMSE of the Direct-RF model, corresponding to a 13\% reduction in forecast error.

From Table~\ref{tab:relative_rmse_macro_models}, it is clear that for Direct-RF-X, DNS-RF-X, and DNSS-RF-X, macro augmentation delivers statistically significant improvements at the slope targets at nearly all forecast horizons and slope measures. There are also improvements in other segments of the curve, but these are much more scattered across maturities and horizons. By contrast, the evidence for the slope segment is considerably more uniform, indicating that the incremental predictive ability of macroeconomic variables is relatively stronger for slope-related movements in the yield curve.


This differs somewhat from the sharper pattern reported in \citet{FreireRiva2025}, who find that macroeconomic variables provide valuable information \emph{only} for the short-run Nelson--Siegel factor, which is empirically close to the slope of the yield curve. Instead, the evidence here is more consistent with a setting in which macroeconomic variables provide the most incremental predictive value for the short-run Nelson--Siegel factor, while still providing some, though weaker incremental predictive value for other segments of the curve. It is notable that this similar pattern is present for the Direct-RF-X specification, which suggests that the stronger role of macroeconomic variables in forecasting slope-related movements is not confined to factor-based models, but also appears in the direct-yield approach.

A further result is that the direct-yield approach benefits more from macro augmentation than the factor-based approach. This is especially visible at the long end, where macro augmentation delivers a relatively uniform improvement for Direct-RF-X across forecast horizons, whereas the corresponding gains for DNS-RF-X and DNSS-RF-X are more mixed and often remain close to one. More broadly, the incremental predictive contribution of macro variables appears to diminish as more factor structure is imposed: moving from Direct-RF-X to DNS-RF-X and then to DNSS-RF-X, the gains from macro augmentation become less pronounced. This is consistent with residual traces of the spanning hypothesis, in the sense that as more factors are used to capture variation in the yield curve, macroeconomic predictors appear to add less incremental forecasting value.

\subsubsection{Feature Importance}

Feature importance is measured using impurity-based importance, also referred to as mean decrease in impurity. In regression trees, each split is chosen to reduce mean squared error. The importance of feature $j$ is therefore the total reduction in node impurity generated by all splits on that feature, weighted by the number of observations reaching the split, and then averaged across trees. Formally, for a single tree, the importance of feature $j$ can be written as
$$
I_j = \sum_{s \in \mathcal{S}_j} \frac{N_s}{N}\left( Q_s - \frac{N_{s,L}}{N_s}Q_{s,L} - \frac{N_{s,R}}{N_s}Q_{s,R} \right),
$$
where $\mathcal{S}_j$ is the set of splits using feature $j$, $N_s$ is the number of observations at node $s$, $N$ is the total number of observations, and $Q_s$ denotes node impurity, measured here by mean squared error. The forest-level importance is then obtained by averaging $I_j$ across all trees. In the figures below, these importances are further rescaled to sum to 100, so they can be interpreted as shares of total importance attributable to each predictor group. Because features that produce larger reductions in impurity or prediction loss get higher importance, the share of importance provides a summary of how useful each predictor was for forecasting the target.


Figure~\ref{fig:varimp_bigpicture} reports feature-importance shares for the DNS-RF-X, DNSS-RF-X, and Direct-RF-X models, with predictors collapsed into two groups: ``Macro'' and ``Yields''. From the feature importance measure $I_j$ defined above, I average feature importances over all trees, over the out-of-sample period from 2015-01 to 2025-06, and then across the four forecast horizons $h \in \{1,3,6,12\}$. Since the two bars for each target sum to 100, they can be interpreted as the shares of total importance attributable to macroeconomic information and yield-curve-based information, respectively. In Figures~\ref{fig:varimp_bigpicture}a and \ref{fig:varimp_bigpicture}b, the ``Yields'' share is lowest for the $\beta_1$ target, while in Figure~\ref{fig:varimp_bigpicture}c it is lowest for the 24\_120 target, although in each case only by a modest margin relative to the other targets. This is consistent with the incremental predictive-power results, which indicate that macroeconomic variables are most important for slope-related forecasts.

This also differs from the sharper pattern reported in \citet{FreireRiva2025}. In their feature-importance analysis, the short-run $\beta_1$ factor shows a much more pronounced reduction in the importance share of yield-curve information, suggesting that macroeconomic variables are especially relevant only for that segment. By contrast, the pattern here suggests that macroeconomic information matters across targets more broadly, but is relatively more important for slope-related forecasts than for other parts of the curve.

The present findings from the test of incremental predictive power and feature importance differ somewhat from \citet{FreireRiva2025}, who conclude that macroeconomic variables provide valuable information \emph{only} for the short-run Nelson--Siegel factor. Part of this may reflect differences in empirical design, since \citet{FreireRiva2025} study an out-of-sample period from 1990 to 2021, whereas the present analysis focuses on 2015 to 2025, and the two studies also use different yield datasets. Nonetheless, both studies point to the same broad conclusion that macroeconomic variables are most important for forecasting slope-related movements in the yield curve, despite the different forecasting setups, yield datasets, and the use of an additional direct-yield approach in this paper.

\subsubsection{Granular Analysis of Macroeconomic Predictors}

\begin{figure}[!htbp]
\centering
\begin{tabular}{cc}
\begin{tabular}{c}
\includegraphics[width=0.4\textwidth]{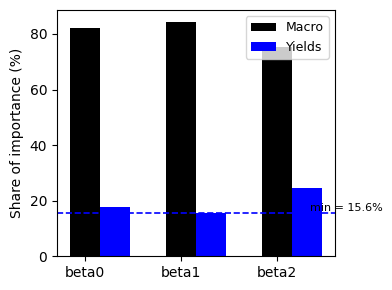} \\
\small (a) DNS-RF-X
\end{tabular}
&
\begin{tabular}{c}
\includegraphics[width=0.5\textwidth]{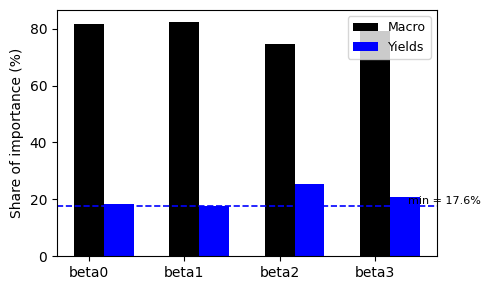} \\
\small (b) DNSS-RF-X
\end{tabular}
\\[0.8em]
\multicolumn{2}{c}{
\begin{tabular}{c}
\includegraphics[width=0.9\textwidth]{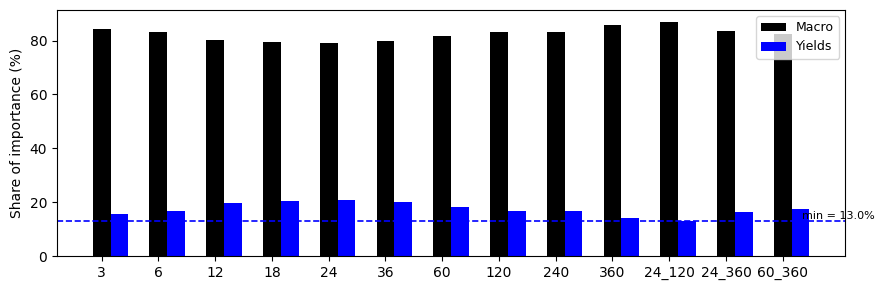} \\
\small (c) Direct-RF-X
\end{tabular}
}
\end{tabular}
\caption{Variable-importance composition across macroeconomic categories for the DNS-RF-X, DNSS-RF-X, and Direct-RF-X models.}
\label{fig:varimp_bigpicture}
\end{figure}

More broadly, Figure~\ref{fig:varimp_bigpicture} shows that macroeconomic predictors account for roughly 80\% of total importance across targets. Taken together with the DM test results for incremental predictive power of macro predictors, this provides clear evidence that macroeconomic information plays a meaningful role in yield-curve prediction. The discussion therefore now turns to a more granular examination of which macroeconomic variables matter most.

For the direct-yield specification, Figure~\ref{fig:varimp_direct_rf_x} first disaggregates macroeconomic predictors into eight broad categories (see Appendix~\ref{app:macro_variables} for the category mappings), while Figure~\ref{fig:macro_indiv_direct_rf_x} goes one step further and reports the importance of individual macro series. These feature importance measures are averaged across the four forecast horizons, over 2015-01 to 2025-06, and over the trees in the random forest. Together, these show a shift in the composition of macroeconomic importance as tenor increases along the curve. As maturity increases, the share of importance attributed to the Prices category and the Output and Income category rises, while the share attributed to Interest and Exchange Rates declines. Within the Prices category, the variables that receive the largest importance shares are real personal consumption expenditures (PCE) on durable goods, Consumer Price Index (CPI) for durables, CPI for services, and CPI for medical care. Within the Output and Income category, the variables that receive the largest importance shares are Capacity Utilization: Manufacturing, IP: Durable Materials, and IP: Manufacturing (SIC). Within the Interest and Exchange Rates category, the variables that receive the largest importance shares are 6-Month Treasury Constant Maturity (C) Minus FEDFUNDS, 1-Year Treasury C Minus FEDFUNDS, 5-Year Treasury C Minus FEDFUNDS, and 10-Year Treasury C Minus FEDFUNDS. This is consistent with \citet{DieboldRudebuschAruoba2006}, who link monetary policy most closely to slope and short-end adjustments, while associating inflation expectations and broader macroeconomic conditions more strongly with level and longer-end movements in the yield curve. Likewise, the rising importance of the Output and Income category with tenor is intuitive, since variables such as capacity utilization and industrial production capture the strength and persistence of economic activity, which should matter more for yields that embed expectations over longer horizons.

The same logic appears within the slope measures. For the 24--120 slope, the Interest and Exchange Rates category has the highest importance share, with the same key variables receiving the largest shares. Since these variables measure Treasury yields at different maturities relative to the Fed Funds rate, they summarize how the curve is priced relative to the policy anchor. This is consistent with the view that the 24--120 slope is shaped by expectations about the future path of monetary policy, inflation, and term premia. By contrast, for the longer-span slopes 24--360 and 60--360, the largest macroeconomic share comes from the Labor Market category, particularly All Employees: Goods-Producing Industries, All Employees: Wholesale Trade, All Employees: Durable Goods, and All Employees: Trade, Transportation \& Utilities, which is consistent with these farther-out slope measures embedding more information about longer-run growth conditions. This pattern is reinforced by the factor-based DNS-RF-X model (see Figure~\ref{fig:varimp_dns_rf_x}): for $\beta_0$, interpreted as the long-run factor, Prices accounts for a larger share of importance than Interest and Exchange Rates, whereas for $\beta_1$, interpreted as the slope or short-run factor, Interest and Exchange Rates is dominant by a clear margin. Taken together, the direct-yield and factor-based evidence point to a common interpretation that inflation-related variables matter relatively more at the long end, while policy-rate-related variables matter relatively more for slope and short-run movements.


Overall, the results support a standard macro-finance interpretation of yield curve dynamics. Variables related to inflation and real economic activity matter relatively more for level and longer-end movements, whereas policy-rate-related variables matter relatively more for slope and short-run movements. This aligns with \citet{DieboldRudebuschAruoba2006}, who link the slope factor closely to the funds rate and interpret the level factor as reflecting the bond market's perception of long-run inflation.

\subsection{Factor-Based Forecasting with PSO Optimized Decay Parameters}
\label{subsec:role_of_pso}

To assess whether DNSS specifications that use decay parameters estimated by the PSO algorithm improve out-of-sample forecast accuracy, I use the DM test to compare each PSO-based specification against its corresponding baseline DNSS benchmark. The baseline benchmark fixes the decay parameters at $\lambda_1 = 0.0609$ and $\lambda_2 = 0.01$. The DNSS-AR-PSO and DNSS-RF-PSO models allow the decay parameters to vary over time. The results in Table~\ref{tab:relative_rmse_pso_models} are reported relative to the corresponding baseline benchmark. That is, forecasts from the DNSS-AR-PSO model are evaluated against the forecasts from the DNSS-AR model using baseline decay parameters, while forecasts from the DNSS-RF-PSO model are evaluated against those from the DNSS-RF model using baseline decay parameters. Values below one therefore indicate that the PSO-based specification achieves lower RMSE and thus improves forecast accuracy relative to the baseline benchmark.

The results for the time-varying PSO specification are weak overall. For DNSS-AR-PSO, gains relative to the baseline DNSS-AR benchmark are limited and are concentrated mainly in slope forecasts, especially for the 24--120 spread at $h=1$ and $h=6$, while most maturity targets remain close to or above one. For DNSS-RF-PSO, performance is substantially worse at short horizons: at $h=1$ and $h=3$, relative RMSE is well above one across most maturity and slope targets, indicating clear deterioration in forecast accuracy. Some improvement appears at $h=6$, where relative RMSE falls below one at the short end and for the 24--120 slope, but these gains remain narrow. By $h=12$, improvements are confined mainly to the far long end, especially the 240- and 360-month maturities. Overall, the time-varying PSO variant does not deliver broad forecasting gains; its benefits are modest and selective for DNSS-AR-PSO and very limited for DNSS-RF-PSO, while its poor performance at short horizons is particularly notable.

A plausible explanation is that once the decay parameters are allowed to vary over time, the basis of the curve also changes over time. This makes the extracted factors less stable and less comparable across dates, which in turn makes forecasting future yields from those factors more difficult. The weaker performance of DNSS-RF-PSO relative to DNSS-AR-PSO is also consistent with this interpretation. Because the random forest is more flexible, it may fit noise generated by the unstable factor representation, whereas the AR model imposes more structure and is therefore somewhat less exposed to this problem. It is also important to note that the forecasting design used here remains relatively simple, since the time-varying decay parameters are forecast using a random walk, that is, a no-change benchmark. As such, these results should not be interpreted as definitive evidence against time-varying decay parameters. The conclusions may differ under forecasting approaches that model the dynamics of the decay parameters more explicitly, especially in ways that allow them to evolve jointly with the factor processes.

Overall, in the current implementation, the time-varying PSO variant improves in-sample yield-curve fit but does not deliver corresponding forecasting gains. Given the relatively poor forecast performance of the time-varying PSO variant, together with the weaker economic interpretation of the factors extracted from the PSO-based specifications, I do not include these models in the discussions of forecasting approaches in Subsection~\ref{subsec:role_of_approach} and the role of macroeconomic predictors in Subsection~\ref{subsec:role_of_macro}.

\subsection{Trading Exercise Setup}
\label{subsec:results_trading}
In the previous section, forecast performance was evaluated using root mean squared error (RMSE). While RMSE captures statistical accuracy, it does not necessarily indicate whether more accurate yield forecasts translate into economically meaningful gains. This section therefore considers an alternative loss function based on a trading backtest, in order to assess whether the model rankings persist once forecasts are mapped into investment decisions.

Given that the forecasts are generated at monthly frequency, the backtest is designed to reflect an active fixed-income trading setting. At each forecast origin, the predicted yield is converted into a trading signal for a given maturity. To evaluate realized payoffs in a way that is consistent with the zero-coupon yield data, yields are mapped into synthetic zero-coupon bond holding-period returns. For each maturity, a synthetic zero-coupon bond is priced at the forecast origin using the observed yield, and its holding-period return is computed as the bond ages into a shorter remaining maturity over the forecast horizon. The holding period is set equal to the forecast horizon so that the prediction horizon maps directly into the length of time the position is held. This provides a natural and intuitive way to align forecast horizons with trading outcomes in the backtest.

Let $P_t(\tau)$ denote the price at time $t$ of a zero-coupon bond with maturity $\tau$, and let $h$ denote the holding period. The total holding-period return is defined as
$$
R_{t,t+h}^{\mathrm{total}}(\tau)
=
\frac{P_{t+h}(\tau-h)-P_t(\tau)}{P_t(\tau)}.
$$

The roll-down component is defined as the holding-period return under an unchanged yield curve:
$$
R_{t,t+h}^{\mathrm{roll-down}}(\tau)
=
\frac{P_t(\tau-h)-P_t(\tau)}{P_t(\tau)}.
$$

The residual curve-move component is then given by
$$
R_{t,t+h}^{\mathrm{curve}}(\tau)
=
R_{t,t+h}^{\mathrm{total}}(\tau)
-
R_{t,t+h}^{\mathrm{roll-down}}(\tau).
$$

Hence, the total return can be decomposed as
$$
R_{t,t+h}^{\mathrm{total}}(\tau)
=
R_{t,t+h}^{\mathrm{roll-down}}(\tau)
+
R_{t,t+h}^{\mathrm{curve}}(\tau).
$$

The roll-down component represents the return that would be earned if the yield curve did not move after entry, while the curve-move component captures the additional gain or loss arising from changes in the yield curve between entry and exit.

\paragraph{Duration Trades from Single-Maturity Forecasts}\mbox{}\par

To evaluate forecasts at individual maturities, I focus on three key points on the curve: 24, 120, and 360 months. The trading strategy is designed to align closely with the forecast evaluation. At each month, the model’s predicted yield for a given maturity is compared against the current yield at that same maturity. If the forecasted yield is higher than the current yield, a short position is taken, since a rise in yield implies a fall in the price of the corresponding zero-coupon bond. Conversely, if the forecasted yield is lower than the current yield, a long position is taken. The position is then held for $h$ months, where $h$ matches the forecast horizon, and is closed at the end of the holding period using the price of the now-aged zero-coupon bond with remaining maturity $\tau-h$. Because forecasts are generated monthly, this process is repeated each month. For example, under the $h=12$ specification, if the 120-month yield is forecast to rise relative to its current level, the strategy enters a short position in the 120-month zero-coupon bond and closes that position 12 months later, when the bond has rolled down to 108 months.

For comparison of performance, I consider three simple benchmarks. The first is based on the random walk model. Since the random walk always predicts no change in yields, this benchmark takes a constant long position. The second is a roll-down benchmark, which takes a position in the direction of the highest unchanged-curve roll-down return. Although fixed-income practitioners often refer to such trades more broadly as carry-and-roll strategies, in the present zero-coupon setup this is more precisely a roll-down strategy. In other words, it goes long when the roll-down return under an unchanged yield curve is positive, and short when it is negative. The third benchmark is a time-series momentum strategy, which takes positions based on the trailing excess returns. I follow the strategy proposed in the AQR white paper by \citet{hurst2010understanding}, which has been shown to perform well in trading U.S. Treasuries under a broadly similar setup.

Let $\hat{y}_{t,t+h}(\tau)$ denote the forecast of the zero-coupon yield at maturity $\tau$ made at time $t$ for horizon $h$, and let $y_t(\tau)$ denote the current yield. Define the forecasted yield change as
$$
\Delta \hat{y}_{t,t+h}(\tau) = \hat{y}_{t,t+h}(\tau) - y_t(\tau).
$$

For the forecast-based models, the trading position is given by
$$
w_t^{\mathrm{model}}(\tau,h)
=
-\frac{\Delta \hat{y}_{t,t+h}(\tau)}{\sigma^{\mathrm{model}}_{\Delta \hat{y},\tau,h}}.
$$
where $\sigma^{\mathrm{model}}_{\Delta \hat{y},\tau,h}$ denotes the standard deviation of the forecasted yield-change signal for that model, maturity $\tau$, and horizon $h$.

For the roll-down benchmark, let
$$
R_{t,t+h}^{\mathrm{roll-down}}(\tau)
=
\frac{P_t(\tau-h)-P_t(\tau)}{P_t(\tau)}
$$
denote the unchanged-curve roll-down return. The corresponding volatility-scaled position is
$$
w_t^{\mathrm{roll\text{-}down}}(\tau,h)
=
\frac{R_{t,t+h}^{\mathrm{roll\text{-}down}}(\tau)}{\sigma^{\mathrm{roll\text{-}down}}_{\tau,h}}.
$$

For the time-series momentum benchmark, let $M_t(\tau)$ denote the cumulative excess return over the previous 12 months for maturity $\tau$, that is,
$$
M_t(\tau)=\sum_{j=1}^{12} r^{\mathrm{excess}}_{t-j}(\tau).
$$
The volatility-scaled position is then
$$
w_t^{\mathrm{tsmom}}(\tau,h)
=
\frac{M_t(\tau)}{\sigma_{M,\tau,h}}.
$$

Finally, positions are capped to prevent leverage:
$$
w_t^{*}(\tau,h)
=
\max\left\{-1,\,\min\left(w_t(\tau,h),\,1\right)\right\}.
$$

\paragraph{Slope Trades from Slope Forecasts}\mbox{}\par
To evaluate forecasts at slope targets, I focus on three key spreads along the zero-coupon curve: 24--120, 24--360, and 60--360 months. At each month, the model’s predicted slope is compared against the current slope. If the predicted slope is higher than the current slope, the strategy enters a steepener, taking a long position in the short-maturity zero-coupon bond and a short position in the long-maturity zero-coupon bond. If the predicted slope is lower than the current slope, the strategy enters a flattener, taking a short position in the short-maturity bond and a long position in the long-maturity bond. To isolate changes in the slope of the curve rather than outright duration exposure or parallel shifts in yields, the two legs are sized to be DV01-neutral at entry, where DV01 denotes the dollar value change in a bond’s price for a one-basis-point change in its yield. The position is then held for $h$ months, where $h$ matches the forecast horizon, and is closed using the prices of the two aged bonds at maturities $\tau_a-h$ and $\tau_b-h$.

For comparison of performance, a DV01-neutral roll-down benchmark is used, which takes the direction of the steepener or flattener with the higher expected return under an unchanged yield curve.
Let the forecasted slope change over horizon $h$ be denoted by
$$
\Delta \hat{S}_{t,t+h}(\tau_a,\tau_b)
=
\hat{S}_{t,t+h}(\tau_a,\tau_b)-S_t(\tau_a,\tau_b).
$$

For the forecast-based slope models, the volatility-scaled trading signal is given by
$$
s_t^{\mathrm{model}}(\tau_a,\tau_b,h)
=
\frac{\Delta \hat{S}_{t,t+h}(\tau_a,\tau_b)}
{\sigma^{\mathrm{model}}_{\Delta \hat{S},\tau_a,\tau_b,h}}.
$$

For the roll-down benchmark, let $R_{t,t+h}^{\mathrm{roll\text{-}down,\,steep}}(\tau_a,\tau_b)$ denote the unchanged-curve roll-down return of a DV01-neutral steepener. The corresponding volatility-scaled signal is
$$
s_t^{\mathrm{roll\text{-}down}}(\tau_a,\tau_b,h)
=
\frac{R_{t,t+h}^{\mathrm{roll\text{-}down,\,steep}}(\tau_a,\tau_b)}
{\sigma^{\mathrm{roll\text{-}down}}_{\tau_a,\tau_b,h}}.
$$

To isolate slope movements rather than outright duration exposure, the two legs are sized to be DV01-neutral at entry. Let $\mathrm{DV01}_t(\tau_a)$ and $\mathrm{DV01}_t(\tau_b)$ denote the dollar value change for a one-basis-point move in the short- and long-maturity zero-coupon bonds. The DV01-neutrality condition is
$$
|w_{a,t}|\,\mathrm{DV01}_t(\tau_a)
=
|w_{b,t}|\,\mathrm{DV01}_t(\tau_b).
$$

Normalizing the long-maturity leg to one unit in absolute value, the baseline hedge ratio is
$$
w_b^{0}=1,
\qquad
w_a^{0}=\frac{\mathrm{DV01}_t(\tau_b)}{\mathrm{DV01}_t(\tau_a)}.
$$

The final trade weights are then
$$
w_{a,t}=s_t\,w_a^{0},
\qquad
w_{b,t}=-s_t\,w_b^{0}.
$$
Hence, if $s_t>0$, the strategy enters a steepener by going long the short-maturity bond and short the long-maturity bond; if $s_t<0$, it enters a flattener by shorting the short-maturity bond and going long the long-maturity bond.

For all strategies across duration and slope trades, the scaling standard deviations are estimated using two years of signals from the in-sample period. In addition, a transaction cost of 1 basis point one way is applied to each trade. This trading setup is designed to mirror the forecast evaluation as closely as possible across the four forecast horizons and the selected maturity and slope targets. The intention is to keep the trading simulation simple and subject to minimal strategy tuning. This helps keep the comparison transparent and reduces scope for cherry-picking or results driven by ex-post design choices.


\subsection{Discussion of Trading Results}

For each duration target (4-, 120-, and 360-month maturities) and each slope target (24--120, 24--360, and 60--360), all strategies are evaluated at the 1-, 3-, 6-, and 12-month forecast horizons. For example, Table~\ref{tab:slope_table_24_120_h1} reports the performance metrics for the 24--120 slope trade at the one-month-ahead forecast horizon across all strategies. To provide a broad overview across these many target--horizon combinations, I summarize the performance of each strategy across all trading setups.

Tables~\ref{tab:duration_summary_total_return} and~\ref{tab:slope_summary_total_return} report a summary of strategy performance for single maturity trades and slope trades respectively, averaged across targets and forecast horizons. In total, there are 14 strategies based on yield forecast models with 3 benchmarks strategies for duration trades based on single-maturity forecasts and 1 benchmark strategy for slope trades based on slope forecasts. For each strategy, the summary tables report its average rank, the number of first-place finishes, and the number of top-3 finishes across all forecast targets and horizons. They also report the average annualized return, average annualized roll-down and the average annualized curve move, together with the average hit rate. Though reported elsewhere, Sharpe ratios are omitted from these summary tables because they are computed using holding-period returns, and the holding period differs across forecast horizons. The Sharpe ratios are not directly comparable across horizons since the corresponding return volatilities are not annualized because doing so requires a stronger assumption that the returns are independently and identically distributed, which is unlikely to hold. Hence, for these tables which summarize performance across different holding periods, the annualized return is used as the main performance metric.

\subsubsection{Trading Results Ranked by Total Returns}
First, I evaluate overall strategy performance using total annualized returns net of fees. Hence tables~\ref{tab:duration_summary_total_return} and~\ref{tab:slope_summary_total_return} are sorted in descending order of average rank based on the \emph{total annualized return}. Across both duration and slope trades, the factor-based AR specifications with PSO-estimated decay parameters rank the highest overall and outperform all benchmark strategies. This is notable given that these AR specifications are not macro-augmented. The factor-based AR specifications using the baseline decay parameters also perform strongly, although they are occasionally surpassed by the roll-down benchmark. By contrast, the direct-yields models perform relatively poorly. In the duration trades, they rank at the bottom, underperform all benchmarks, and deliver negative average returns. In the slope trades, they remain below the benchmark and below average relative to the other models, although their average returns are still positive. In both the duration and slope summary tables, the factor-based RF specifications underperform the roll-down benchmark and tend to rank in the middle of the pack.

\subsubsection{Trading Results Ranked by Curve-move Returns}
I next isolate the component of performance attributed purely to curve movements. Tables~\ref{tab:duration_summary_curve_return} and~\ref{tab:slope_summary_curve_return} re-rank the strategies using annualized \emph{curve-move returns} rather than total returns. This isolates the part of performance associated with correctly predicting yield-curve movements, with roll-down effects excluded.

The duration-trade rankings change noticeably under this criterion. When performance is ranked by returns attributable to curve moves, the random walk benchmark performs the worst. This is an important contrast with the earlier RMSE results, under which the random walk appears to be a difficult benchmark to beat. The trading exercise shows that strong forecast accuracy under statistical loss metrics does not necessarily translate into strong trading performance. Because the random walk forecasts no change, its implied trading signal is effectively a buy-and-hold position, which does not exploit information in the yield curve or macroeconomic variables. By contrast, models that embed richer information can generate materially better trading outcomes even when their RMSE improvements appear modest.

This distinction is especially clear in episodes of large yield-curve adjustment. The random walk performs relatively well when yields move sideways, but it performs poorly when yields move sharply. For example, during the 2022 bear flattening episode, the random walk incurred losses, whereas models using richer yield-curve and macroeconomic information were able to capitalize on the move. Figure~\ref{fig:24m_overtime} plots the 24-month yield over the backtest period. Figure~\ref{fig:rw_24_h1} shows the signals and P\&L attribution for the random walk strategy: although it performs adequately in quieter periods, much of its earlier gains are erased during the 2022 bear flattening. By contrast, Figure~\ref{fig:dnss_rf_x_24_h1} shows that the DNSS-RF-X model takes large short positions during this episode and earns an approximately 4\% net gain over that period.

This result is also robust more broadly. Even under the ranking based on total annualized returns, the DNS-AR specifications still outperform the random walk benchmark and generate substantially higher P\&L. Hence, while the random walk remains a strong statistical benchmark under RMSE, it is relatively weaker as an economic benchmark in trading applications.

The slope-trade rankings also change markedly under this criterion. All five macro-augmented strategies take the top 5 spots (see table ~\ref{tab:slope_summary_curve_return}). This further corroborates the results in this paper which indicate that macroeconomic predictors are especially important for forecasting slope dynamics, and is also in line with the findings of \citet{FreireRiva2025}. In particular, the Direct-RF-X model performs the strongest, ranking first in 7 out of 12 slope trading tasks. This also aligns with the earlier RMSE-based results relative to the random walk, where Direct-RF-X was the best performing model for slope prediction. By contrast, the factor-based AR models, which ranked most highly when performance was evaluated using total return, fall substantially in the rankings based on curve-move returns. This suggests a trade-off whereby strategies that generate stronger returns from accurate curve predictions may also incur larger losses from unfavorable roll-down. 

To illustrate this point, strategy diagnostic plots are shown for two models, which include the trade position sizing, total returns, and the decomposed returns attributed to curve move and roll-down over the out-of-sample period. Figure~\ref{fig:24_120_overtime} plots the 24-120 slope over the backtest period. The Direct-RF-X model in Figure~\ref{fig:direct_rf_x_24_120_h1} shows flattener positions taken from 2017 to 2019 and again from late 2021 through 2022 which generated substantial gains attributed to curve movements, but these were partly offset by negative roll-down. These dates correspond to periods of bear flattening which the model accurately predicted. However, because the yield curve at entry was relatively upward sloping, the negative roll-down from the short position at the shorter-maturity leg outweighed the positive roll-down from the longer-maturity leg. This highlights a structural challenge of slope strategies in the current trading setup whereby correctly forecasting the slope is not sufficient, since roll-down effects must also be managed. One possible refinement is to enter a flattener only when the longer-maturity bond also offers positive roll-down when held long. This would allow the positive roll-down on the long leg to partially offset the adverse roll-down on the short leg, though at the cost of relaxing DV01 neutrality and increasing exposure to other risks, such as outright level shifts in the yield curve. In practice, trade design would be even more complex, since investors must also account for coupon effects, convexity differences, among other considerations. Hence, the present trading exercise should be interpreted as a framework for comparing the economic value of forecasts under a common setup, rather than as a fully realistic implementation of slope trading in practice.

By contrast, the DNSS-AR model in Figure~\ref{fig:dnss_ar_24_120_h1} fails to capture the bear-flattening episodes from 2017 to 2019 and again from late 2021 through 2022, taking mostly neutral-to-steepener positions instead. The steepener positions in 2022 generated large losses from curve moves, but favorable roll-down more than offset these losses, so the strategy still outperformed the Direct-RF-X model in total return over the out-of-sample period.



Table~\ref{tab:slope_table_24_120_h1} reports the performance metrics for each model in the same 24--120 slope, one-month-ahead setting shown in Figures~\ref{fig:direct_rf_x_24_120_h1} and~\ref{fig:dnss_ar_24_120_h1}. This table is selectively included because this trading setting is discussed in greater detail, other tables are excluded for brevity.


Overall, the trading results reinforce three main points. First, strong performance under RMSE does not necessarily translate into strong trading performance: although the random walk is a difficult benchmark to beat under statistical loss because it forecasts no change, its implied buy-and-hold signal does not exploit information in the yield curve or macroeconomic variables, and is therefore economically outperformed by richer models. Second, the results further corroborate the importance of macroeconomic information for slope prediction, as the macro-augmented models generate the highest returns attributable purely to curve moves in slope trades. Third, for duration trades, the factor-based AR specifications broadly deliver the strongest performance despite their simplicity, while the time-varying PSO specifications perform more favorably than the RMSE results alone would suggest, indicating some promise if the joint forecasting of decay parameters and factors can be improved.


\section{Conclusion}

This paper studies U.S.\ yield curve forecasting by jointly comparing maturity-specific direct-yield forecasts and factor-based DNS(S) forecasts under a common high-dimensional macroeconomic information set. Using the zero-coupon yield data of \citet{FilipovicPelgerYe2024} and a fixed-window out-of-sample evaluation from 2015 to 2025, four main conclusions emerge.

First, relative to the random walk benchmark, forecasting gains are concentrated mainly at the short end of the curve and in slope forecasts rather than across the curve as a whole. These gains are strongest at short horizons and decline materially as the forecast horizon lengthens.

Second, the relative performance of the direct-yield and factor-based approaches depends on both the forecast target and the horizon. The direct-yield approach is most useful at shorter forecast horizons, especially for forecasting slopes. By contrast, the factor-based approach becomes relatively more attractive as the forecast horizon lengthens, though this advantage is limited to single-maturity forecasts.

Third, macroeconomic predictors provide clear incremental predictive power, strongest for the short-run Nelson--Siegel factor, which is closely related to the slope, while still contributing weaker gains for other parts of the curve. This conclusion is reinforced by the trading exercise, in which macro-augmented models generate the highest returns attributed to curve moves for slope trades. The feature-importance results also indicate that variables related to inflation and underlying economic activity matter relatively more at the long end, whereas policy-rate-related variables matter relatively more for short-run movements.

Fourth, better in-sample fit does not necessarily translate into better out-of-sample forecasting performance. Although the time-varying DNSS specifications materially improve in-sample fit, they do not deliver broad forecasting gains. These results suggest that in this setting, a more stable fixed calibration is more useful for forecasting than allowing decay parameters to vary over time.

Evaluating models through economic value in the trading exercise also reveals that strong performance under statistical loss measures such as RMSE does not necessarily translate into strong trading performance. In particular, the random walk, while a strong benchmark under statistical loss because it forecasts no change, performs much more poorly in trading terms because its implied buy-and-hold signal does not incorporate the richer cross-sectional and macroeconomic information exploited by the competing models. By contrast, richer models generate stronger trading outcomes, especially when yields move sharply. For duration trades, the factor-based AR specifications are broadly the strongest performers despite their simplicity. The trading results are also more favorable to the time-varying PSO specifications than the RMSE results, suggesting that their value may be understated by statistical forecast metrics alone and that the method may still have potential if the joint forecasting of decay parameters and factors can be improved.

Overall, the evidence points to a forecasting environment that is highly target- and horizon-dependent. Direct-yield models are most effective for short-horizon slope and front-end forecasts, factor-based models become more competitive for single-maturity forecasts at longer horizons, and macroeconomic information is most valuable for slope-related and short-run movements.

\clearpage

\bibliographystyle{apalike} 
\bibliography{refs}

\addtocontents{toc}{\protect\setcounter{tocdepth}{-1}}
\appendix

\section{Macroeconomic Variables}
\label{app:macro_variables}

This appendix documents the monthly macroeconomic variables used in the forecasting models.
I use the FRED-MD database of McCracken and Ng (2016) and apply the standard FRED-MD
transformation codes to construct stationary predictors where appropriate.\footnote{See
McCracken and Ng (2016) for the full database description and variable definitions.}

Table~\ref{tab:macro_vars_full} reports the full list of selected variables, including the
FRED code, category, description, and transformation code. In the forecasting exercise,
we additionally impose a conservative three month publication lag on all macro variables and
use a half year information set (lags 4--6) to ensure predictor availability at the forecast origin.

\subsection{Transformation Codes}
\label{app:macro_tcodes}

The transformation codes follow the FRED-MD convention:
\begin{itemize}
    \item Code 1: $z_t = x_t$
    \item Code 2: $z_t = x_t - x_{t-1}$
    \item Code 3: $z_t = x_t - x_{t-2}$
    \item Code 4: $z_t = \log(x_t)$
    \item Code 5: $z_t = \log(x_t/x_{t-1})$
    \item Code 6: $z_t = \log(x_t/x_{t-2})$
    \item Code 7: $z_t = \dfrac{x_t - x_{t-1}}{x_{t-1}}$
\end{itemize}

\subsection{Selected Macroeconomic Variables}
\label{app:macro_var_table}
{\scriptsize

\begin{longtable}{C{2.4cm} C{4.0cm} C{7.5cm} C{2.0cm}}
\caption{Full list of our macroeconomic variables} \label{tab:macro_vars_full} \\
\toprule
FRED Code & Category & Description & Transformation Code \\
\midrule
\endfirsthead

\multicolumn{4}{c}{\tablename\ \thetable\ (continued)} \\
\toprule
FRED Code & Category & Description & Transformation Code \\
\midrule
\endhead

\midrule
\multicolumn{4}{r}{Continued on next page} \\
\endfoot

\bottomrule
\endlastfoot

HOUST & Housing & Housing Starts: Total New Privately Owned & 4 \\
HOUSTMW & Housing & Housing Starts, Midwest & 4 \\
HOUSTNE & Housing & Housing Starts, Northeast & 4 \\
HOUSTS & Housing & Housing Starts, South & 4 \\
HOUSTW & Housing & Housing Starts, West & 4 \\
PERMIT & Housing & New Private Housing Permits (SAAR) & 4 \\
PERMITMW & Housing & New Private Housing Permits, Midwest (SAAR) & 4 \\
PERMITNE & Housing & New Private Housing Permits, Northeast (SAAR) & 4 \\
PERMITS & Housing & New Private Housing Permits, South (SAAR) & 4 \\
PERMITW & Housing & New Private Housing Permits, West (SAAR) & 4 \\

AAA & Interest and Exchange Rates & Moody's Seasoned Aaa Corporate Bond Yield & 2 \\
AAAFFM & Interest and Exchange Rates & Moody's Aaa Corporate Bond Minus FEDFUNDS & 1 \\
BAA & Interest and Exchange Rates & Moody's Seasoned Baa Corporate Bond Yield & 2 \\
BAAFFM & Interest and Exchange Rates & Moody's Baa Corporate Bond Minus FEDFUNDS & 1 \\
COMPAPFFx & Interest and Exchange Rates & 3-Month Commercial Paper Minus FEDFUNDS & 1 \\
CP3Mx & Interest and Exchange Rates & 3-Month AA Financial Commercial Paper Rate & 2 \\
EXCAUSx & Interest and Exchange Rates & Canada / U.S. Foreign Exchange Rate & 5 \\
EXJPUSx & Interest and Exchange Rates & Japan / U.S. Foreign Exchange Rate & 5 \\
EXSZUSx & Interest and Exchange Rates & Switzerland / U.S. Foreign Exchange Rate & 5 \\
EXUSUKx & Interest and Exchange Rates & U.S. / U.K. Foreign Exchange Rate & 5 \\
FEDFUNDS & Interest and Exchange Rates & Effective Federal Funds Rate & 2 \\
GS1 & Interest and Exchange Rates & 1-Year Treasury Rate & 2 \\
GS10 & Interest and Exchange Rates & 10-Year Treasury Rate & 2 \\
GS5 & Interest and Exchange Rates & 5-Year Treasury Rate & 2 \\
T10YFFM & Interest and Exchange Rates & 10-Year Treasury C Minus FEDFUNDS & 1 \\
T1YFFM & Interest and Exchange Rates & 1-Year Treasury C Minus FEDFUNDS & 1 \\
T5YFFM & Interest and Exchange Rates & 5-Year Treasury C Minus FEDFUNDS & 1 \\
TB3MS & Interest and Exchange Rates & 3-Month Treasury Bill & 2 \\
TB3SMFFM & Interest and Exchange Rates & 3-Month Treasury C Minus FEDFUNDS & 1 \\
TB6MS & Interest and Exchange Rates & 6-Month Treasury Bill & 2 \\
TB6SMFFM & Interest and Exchange Rates & 6-Month Treasury C Minus FEDFUNDS & 1 \\
TWEXAFEGSMTHx & Interest and Exchange Rates & Trade Weighted U.S. Dollar Index & 5 \\

AWHMAN & Labor Market & Avg Weekly Hours : Manufacturing & 1 \\
AWOTMAN & Labor Market & Avg Weekly Overtime Hours : Manufacturing & 2 \\
CE16OV & Labor Market & Civilian Employment & 5 \\
CES0600000007 & Labor Market & Avg Weekly Hours : Goods-Producing & 1 \\
CES0600000008 & Labor Market & Avg Hourly Earnings : Goods-Producing & 6 \\
CES1021000001 & Labor Market & All Employees: Mining and Logging: Mining & 5 \\
CES2000000007 & Labor Market & Avg Weekly Hours : Construction & 1 \\
CES3000000008 & Labor Market & Avg Hourly Earnings : Manufacturing & 6 \\
CLAIMSx & Labor Market & Initial Claims & 5 \\
CLF16OV & Labor Market & Civilian Labor Force & 5 \\
DMANEMP & Labor Market & All Employees: Durable goods & 5 \\
HWI & Labor Market & Help-Wanted Index for United States & 2 \\
HWIURATIO & Labor Market & Ratio of Help Wanted/No. Unemployed & 2 \\
MANEMP & Labor Market & All Employees: Manufacturing & 5 \\
NDMANEMP & Labor Market & All Employees: Nondurable goods & 5 \\
PAYEMS & Labor Market & All Employees: Total nonfarm & 5 \\
SRVPRD & Labor Market & All Employees: Service-Providing Industries & 5 \\
UEMP15OV & Labor Market & Civilians Unemployed - 15 Weeks \& Over & 5 \\
UEMP15T26 & Labor Market & Civilians Unemployed for 15-26 Weeks & 5 \\
UEMP27OV & Labor Market & Civilians Unemployed for 27 Weeks and Over & 5 \\
UEMP5TO14 & Labor Market & Civilians Unemployed for 5-14 Weeks & 5 \\
UEMPLT5 & Labor Market & Civilians Unemployed - Less Than 5 Weeks & 5 \\
UEMPMEAN & Labor Market & Average Duration of Unemployment (Weeks) & 2 \\
UNRATE & Labor Market & Civilian Unemployment Rate & 2 \\
USCONS & Labor Market & All Employees: Construction & 5 \\
USFIRE & Labor Market & All Employees: Financial Activities & 5 \\
USGOOD & Labor Market & All Employees: Goods-Producing Industries & 5 \\
USGOVT & Labor Market & All Employees: Government & 5 \\
USTPU & Labor Market & All Employees: Trade, Transportation \& Utilities & 5 \\
USTRADE & Labor Market & All Employees: Retail Trade & 5 \\
USWTRADE & Labor Market & All Employees: Wholesale Trade & 5 \\

BOGMBASE & Money and Credit & Monetary Base & 6 \\
BUSLOANS & Money and Credit & Commercial and Industrial Loans & 2 \\
CONSPI & Money and Credit & Nonrevolving Consumer Credit to Personal Income & 2 \\
DTCOLNVHFNM & Money and Credit & Consumer Motor Vehicle Loans Outstanding & 6 \\
DTCTHFNM & Money and Credit & Total Consumer Loans and Leases Outstanding & 6 \\
INVEST & Money and Credit & Securities in Bank Credit at All Commercial Banks & 6 \\
M1SL & Money and Credit & M1 Money Stock & 6 \\
M2REAL & Money and Credit & Real M2 Money Stock & 5 \\
M2SL & Money and Credit & M2 Money Stock & 6 \\
NONBORRES & Money and Credit & Reserves Of Depository Institutions & 7 \\
NONREVSL & Money and Credit & Total Nonrevolving Credit & 6 \\
REALLN & Money and Credit & Real Estate Loans at All Commercial Banks & 6 \\
TOTRESNS & Money and Credit & Total Reserves of Depository Institutions & 6 \\

ACOGNO & Orders and Inventories & New Orders for Consumer Goods & 5 \\
AMDMNOx & Orders and Inventories & New Orders for Durable Goods & 5 \\
AMDMUOx & Orders and Inventories & Unfilled Orders for Durable Goods & 5 \\
ANDENOx & Orders and Inventories & New Orders for Nondefense Capital Goods & 5 \\
BUSINVx & Orders and Inventories & Total Business Inventories & 5 \\
CMRMTSPLx & Orders and Inventories & Real Manuf. and Trade Industries Sales & 5 \\
ISRATIOx & Orders and Inventories & Total Business Inventories to Sales Ratio & 2 \\
RETAILx & Orders and Inventories & Retail and Food Services Sales & 5 \\
UMCSENTx & Orders and Inventories & Consumer Sentiment Index & 2 \\

CUMFNS & Output and Income & Capacity Utilization: Manufacturing & 2 \\
INDPRO & Output and Income & IP Index & 5 \\
IPBUSEQ & Output and Income & IP: Business Equipment & 5 \\
IPCONGD & Output and Income & IP: Consumer Goods & 5 \\
IPDCONGD & Output and Income & IP: Durable Consumer Goods & 5 \\
IPDMAT & Output and Income & IP: Durable Materials & 5 \\
IPFINAL & Output and Income & IP: Final products (Market Group) & 5 \\
IPFPNSS & Output and Income & IP: Final Products and Nonindustrial Supplies & 5 \\
IPFUELS & Output and Income & IP: Fuels & 5 \\
IPMANSICS & Output and Income & IP: Manufacturing (SIC) & 5 \\
IPMAT & Output and Income & IP: Materials & 5 \\
IPNCONGD & Output and Income & IP: Nondurable Consumer Goods & 5 \\
IPNMAT & Output and Income & IP: Nondurable Materials & 5 \\
RPI & Output and Income & Real Personal Income & 5 \\
W875RX1 & Output and Income & Real personal income ex transfer receipts & 5 \\

CPIAPPSL & Prices & CPI : Apparel & 6 \\
CPIAUCSL & Prices & CPI : All Items & 6 \\
CPIMEDSL & Prices & CPI : Medical Care & 6 \\
CPITRNSL & Prices & CPI : Transportation & 6 \\
CPIULFSL & Prices & CPI : All Items Less Food & 6 \\
CUSR0000SA0L2 & Prices & CPI : All items less shelter & 6 \\
CUSR0000SA0L5 & Prices & CPI : All items less medical care & 6 \\
CUSR0000SAC & Prices & CPI : Commodities & 6 \\
CUSR0000SAD & Prices & CPI : Durables & 6 \\
CUSR0000SAS & Prices & CPI : Services & 6 \\
DDURRG3M086SBEA & Prices & Personal Cons. Exp: Durable goods & 6 \\
DNDGRG3M086SBEA & Prices & Personal Cons. Exp: Nondurable goods & 6 \\
DSERRG3M086SBEA & Prices & Personal Cons. Exp: Services & 6 \\
OILPRICEx & Prices & Crude Oil, spliced WTI and Cushing & 6 \\
PCEPI & Prices & Personal Cons. Expend.: Chain Index & 6 \\
PPICMM & Prices & PPI: Metals and metal products & 6 \\
WPSFD49207 & Prices & PPI: Finished Goods & 6 \\
WPSFD49502 & Prices & PPI: Finished Consumer Goods & 6 \\
WPSID61 & Prices & PPI: Intermediate Materials & 6 \\
WPSID62 & Prices & PPI: Crude Materials & 6 \\

SP500 & Stock Market & S\&P's Common Stock Price Index: Composite & 5 \\
S\&P PE ratio & Stock Market & S\&P's Composite Common Stock: Price-Earnings Ratio & 5 \\
S\&P div yield & Stock Market & S\&P's Composite Common Stock: Dividend Yield & 2 \\
S\&P 500 & Stock Market & S\&P's Common Stock Price Index: Industrials & 5 \\
VIXCLSx & Stock Market & VIX & 1 \\

\end{longtable}

} 

\clearpage
\section{Random Forest Hyperparameter Tuning}

\begin{longtable}{llccccc}
\caption{Chosen \(m\) values from recursive cross-validation across all Random Forest specifications}
\label{tab:chosen_mtry_all_models}\\
\toprule
Model & \(P\) & Maturity/Factor & \(h=1\) & \(h=3\) & \(h=6\) & \(h=12\) \\
\midrule
\endfirsthead

\toprule
Model & \(P\) & Maturity/Factor & \(h=1\) & \(h=3\) & \(h=6\) & \(h=12\) \\
\midrule
\endhead

\midrule
\multicolumn{7}{r}{Continued on next page} \\
\endfoot

\bottomrule
\endlastfoot

\multirow[t]{15}{*}{Direct-RF}
& \multirow[t]{15}{*}{90}
& 3             & 9   & 9   & 9   & 9   \\
& & 6             & 9   & 9   & 9   & 9   \\
& & 12            & 9   & 9   & 9   & 9   \\
& & 18            & 15  & 22  & 15  & 9   \\
& & 24            & 9   & 22  & 15  & 9   \\
& & 36            & 9   & 45  & 30  & 9   \\
& & 60            & 9   & 45  & 45  & 9   \\
& & 120           & 9   & 22  & 22  & 9   \\
& & 240           & 9   & 30  & 30  & 60  \\
& & 360           & 9   & 9   & 45  & 22  \\
& & 24--120       & 9   & 60  & 60  & 30  \\
& & 24--360       & 15  & 30  & 45  & 30  \\
& & 60--360       & 9   & 30  & 30  & 30  \\
\midrule

\multirow[t]{15}{*}{Direct-RF-X}
& \multirow[t]{15}{*}{460}
& 3             & 21  & 21  & 21  & 21  \\
& & 6             & 76  & 21  & 21  & 21  \\
& & 12            & 46  & 21  & 21  & 76  \\
& & 18            & 153 & 21  & 21  & 46  \\
& & 24            & 153 & 21  & 21  & 46  \\
& & 36            & 21  & 306 & 46  & 46  \\
& & 60            & 21  & 46  & 230 & 306 \\
& & 120           & 76  & 21  & 76  & 153 \\
& & 240           & 21  & 21  & 21  & 306 \\
& & 360           & 153 & 21  & 21  & 306 \\
& & 24--120       & 76  & 230 & 115 & 46  \\
& & 24--360       & 76  & 115 & 21  & 46  \\
& & 60--360       & 21  & 230 & 21  & 21  \\
\midrule

\multirow[t]{3}{*}{DNS-RF}
& \multirow[t]{3}{*}{108}
& \(\beta_0\) & 11  & 36  & 72  & 27  \\
& & \(\beta_1\) & 10  & 10  & 72  & 27  \\
& & \(\beta_2\) & 11  & 11  & 72  & 27  \\
\midrule

\multirow[t]{3}{*}{DNS-RF-X}
& \multirow[t]{3}{*}{477}
& \(\beta_0\) & 119 & 48  & 48  & 238 \\
& & \(\beta_1\) & 80  & 22  & 80  & 318 \\
& & \(\beta_2\) & 22  & 318 & 22  & 48  \\
\midrule

\multirow[t]{4}{*}{DNSS-RF}
& \multirow[t]{4}{*}{114}
& \(\beta_0\) & 28  & 11  & 11  & 19  \\
& & \(\beta_1\) & 11  & 19  & 11  & 11  \\
& & \(\beta_2\) & 11  & 28  & 11  & 11  \\
& & \(\beta_3\) & 11  & 19  & 19  & 28  \\
\midrule

\multirow[t]{4}{*}{DNSS-RF-X}
& \multirow[t]{4}{*}{483}
& \(\beta_0\) & 48  & 22  & 48  & 161 \\
& & \(\beta_1\) & 48  & 22  & 22  & 322 \\
& & \(\beta_2\) & 161 & 22  & 48  & 22  \\
& & \(\beta_3\) & 22  & 22  & 80  & 80  \\
\midrule

\multirow[t]{4}{*}{DNSS-RF-PSO}
& \multirow[t]{4}{*}{114}
& \(\beta_0\) & 76  & 11  & 11  & 11  \\
& & \(\beta_1\) & 11  & 11  & 11  & 38  \\
& & \(\beta_2\) & 11  & 11  & 11  & 11  \\
& & \(\beta_3\) & 11  & 76  & 11  & 11  \\
\midrule

\multirow[t]{4}{*}{DNSS-RF-PSO-X}
& \multirow[t]{4}{*}{483}
& \(\beta_0\) & 48  & 48  & 80  & 121 \\
& & \(\beta_1\) & 242 & 48  & 48  & 161 \\
& & \(\beta_2\) & 48  & 121 & 22  & 22  \\
& & \(\beta_3\) & 48  & 22  & 322 & 322 \\
\midrule

\end{longtable}

\begin{minipage}{\textwidth}
\footnotesize
Notes: The table reports the selected number of predictors considered at each split (\(m\)) for each Random Forest specification, by maturity/factor and forecast horizon. For direct-yield models, the categories are maturities and slopes. For DNS and DNSS models, the categories are the NS(S) factors. \(P\) denotes the predictor dimension used in each specification.
\end{minipage}

\section{HAC Lag Selection}

\begin{longtable}{llccccccccccccc}
\caption{Number of significant HAC lags, capped at $\lfloor T^{1/3}\rfloor$, by model, forecast horizon, and maturity or slope category}
\label{tab:hac_sig_lags}\\
\toprule
\cmidrule(lr){3-7} \cmidrule(lr){8-9} \cmidrule(lr){10-12} \cmidrule(lr){13-15}
$h$ & Model & 3 & 6 & 12 & 18 & 24 & 36 & 60 & 120 & 240 & 360 & 24--120 & 24--360 & 60--360 \\
\midrule
\endfirsthead

\toprule
\cmidrule(lr){3-7} \cmidrule(lr){8-9} \cmidrule(lr){10-12} \cmidrule(lr){13-15}
$h$ & Model & 3 & 6 & 12 & 18 & 24 & 36 & 60 & 120 & 240 & 360 & 24--120 & 24--360 & 60--360 \\
\midrule
\endhead

\midrule
\multicolumn{15}{r}{Continued on next page} \\
\endfoot

\bottomrule
\endlastfoot

\multirow{14}{*}{1}
& Direct-RF           & 4 & 4 & 3 & 1 & 0 & 1 & 1 & 1 & 0 & 1 & 0 & 0 & 0 \\
& Direct-RF-X         & 4 & 4 & 2 & 0 & 0 & 0 & 0 & 1 & 0 & 0 & 1 & 1 & 0 \\
& DNS-AR              & 4 & 3 & 4 & 2 & 0 & 2 & 0 & 0 & 3 & 0 & 3 & 0 & 4 \\
& DNS-RF              & 3 & 3 & 1 & 0 & 0 & 1 & 0 & 1 & 3 & 0 & 1 & 0 & 4 \\
& DNS-RF-X            & 2 & 4 & 1 & 0 & 0 & 1 & 1 & 1 & 3 & 0 & 1 & 0 & 4 \\
& DNSS-AR             & 4 & 2 & 4 & 1 & 0 & 2 & 3 & 1 & 1 & 0 & 4 & 0 & 0 \\
& DNSS-AR-PSO         & 4 & 3 & 3 & 1 & 0 & 1 & 1 & 1 & 2 & 2 & 0 & 2 & 0 \\
& DNSS-RF             & 2 & 3 & 1 & 0 & 0 & 1 & 1 & 1 & 3 & 1 & 1 & 0 & 0 \\
& DNSS-RF-X           & 3 & 4 & 1 & 0 & 1 & 3 & 1 & 1 & 3 & 1 & 1 & 0 & 0 \\
& DNSS-RF-PSO         & 0 & 0 & 0 & 0 & 0 & 0 & 0 & 0 & 0 & 0 & 0 & 0 & 0 \\
& DNSS-RF-PSO-X       & 0 & 0 & 0 & 0 & 1 & 2 & 2 & 1 & 1 & 0 & 0 & 0 & 2 \\

\addlinespace
\midrule

\multirow{14}{*}{3}
& Direct-RF           & 4 & 4 & 3 & 2 & 2 & 2 & 2 & 1 & 1 & 1 & 1 & 2 & 1 \\
& Direct-RF-X         & 3 & 3 & 3 & 2 & 2 & 1 & 1 & 1 & 1 & 1 & 1 & 1 & 1 \\
& DNS-AR              & 3 & 4 & 4 & 2 & 2 & 2 & 2 & 2 & 2 & 2 & 2 & 2 & 4 \\
& DNS-RF              & 3 & 3 & 2 & 1 & 2 & 2 & 2 & 2 & 2 & 1 & 1 & 2 & 2 \\
& DNS-RF-X            & 3 & 3 & 2 & 2 & 2 & 1 & 1 & 2 & 1 & 1 & 3 & 1 & 2 \\
& DNSS-AR             & 4 & 3 & 2 & 2 & 2 & 4 & 4 & 2 & 1 & 2 & 2 & 2 & 1 \\
& DNSS-AR-PSO         & 3 & 3 & 3 & 2 & 2 & 2 & 2 & 2 & 2 & 2 & 1 & 1 & 1 \\
& DNSS-RF             & 4 & 4 & 2 & 1 & 2 & 3 & 2 & 1 & 2 & 1 & 3 & 3 & 2 \\
& DNSS-RF-X           & 4 & 4 & 3 & 2 & 2 & 1 & 1 & 2 & 1 & 2 & 3 & 2 & 1 \\
& DNSS-RF-PSO         & 3 & 0 & 0 & 0 & 0 & 0 & 0 & 1 & 2 & 2 & 0 & 0 & 0 \\
& DNSS-RF-PSO-X       & 1 & 0 & 0 & 0 & 0 & 1 & 3 & 2 & 1 & 1 & 0 & 0 & 2 \\

\addlinespace
\midrule

\multirow{14}{*}{6}
& Direct-RF           & 3 & 3 & 3 & 3 & 3 & 3 & 3 & 3 & 3 & 3 & 2 & 2 & 2 \\
& Direct-RF-X         & 4 & 4 & 4 & 3 & 3 & 3 & 3 & 3 & 3 & 2 & 2 & 2 & 2 \\
& DNS-AR              & 4 & 4 & 4 & 4 & 4 & 3 & 3 & 3 & 3 & 3 & 4 & 4 & 4 \\
& DNS-RF              & 4 & 4 & 4 & 3 & 3 & 4 & 4 & 4 & 4 & 2 & 2 & 2 & 4 \\
& DNS-RF-X            & 4 & 4 & 3 & 2 & 2 & 2 & 4 & 4 & 3 & 2 & 2 & 2 & 3 \\
& DNSS-AR             & 4 & 4 & 3 & 3 & 4 & 4 & 4 & 4 & 3 & 4 & 4 & 4 & 2 \\
& DNSS-AR-PSO         & 3 & 3 & 3 & 3 & 2 & 2 & 2 & 2 & 1 & 1 & 2 & 1 & 1 \\
& DNSS-RF             & 4 & 4 & 3 & 3 & 4 & 4 & 4 & 4 & 3 & 3 & 2 & 2 & 2 \\
& DNSS-RF-X           & 4 & 4 & 3 & 3 & 3 & 4 & 4 & 3 & 4 & 3 & 2 & 1 & 2 \\
& DNSS-RF-PSO         & 4 & 4 & 3 & 3 & 2 & 2 & 0 & 3 & 3 & 3 & 2 & 1 & 0 \\
& DNSS-RF-PSO-X       & 3 & 1 & 0 & 0 & 0 & 1 & 2 & 1 & 0 & 0 & 0 & 0 & 3 \\

\addlinespace
\midrule

\multirow{14}{*}{12}
& Direct-RF           & 4 & 4 & 4 & 4 & 4 & 4 & 4 & 4 & 4 & 4 & 4 & 3 & 2 \\
& Direct-RF-X         & 4 & 4 & 4 & 4 & 4 & 4 & 4 & 4 & 4 & 4 & 2 & 2 & 3 \\
& DNS-AR              & 4 & 4 & 4 & 4 & 4 & 4 & 4 & 4 & 4 & 4 & 4 & 4 & 3 \\
& DNS-RF              & 4 & 4 & 4 & 4 & 4 & 4 & 4 & 4 & 4 & 4 & 4 & 3 & 3 \\
& DNS-RF-X            & 4 & 4 & 4 & 4 & 4 & 4 & 4 & 4 & 4 & 4 & 4 & 3 & 2 \\
& DNSS-AR             & 4 & 4 & 4 & 4 & 4 & 4 & 4 & 4 & 4 & 4 & 4 & 4 & 3 \\
& DNSS-AR-PSO         & 3 & 3 & 3 & 3 & 3 & 2 & 1 & 2 & 2 & 2 & 3 & 2 & 2 \\
& DNSS-RF             & 4 & 4 & 4 & 4 & 4 & 4 & 4 & 4 & 4 & 4 & 2 & 2 & 2 \\
& DNSS-RF-X           & 4 & 4 & 4 & 4 & 4 & 4 & 4 & 4 & 4 & 3 & 2 & 2 & 3 \\
& DNSS-RF-PSO         & 1 & 1 & 1 & 1 & 1 & 1 & 1 & 3 & 3 & 3 & 1 & 1 & 1 \\
& DNSS-RF-PSO-X       & 4 & 0 & 0 & 0 & 0 & 0 & 0 & 1 & 1 & 1 & 0 & 0 & 0 \\

\end{longtable}

\begin{minipage}{\textwidth}
\footnotesize
Notes: The table reports the number of significant HAC lags, capped at $\lfloor T^{1/3}\rfloor$, for each model and forecast horizon for all maturities and slope categories. The number of available out-of-sample observations \(T\) differs across forecast horizons. Specifically, \(T=125\) for \(h=1\), \(T=123\) for \(h=3\), \(T=120\) for \(h=6\), and \(T=114\) for \(h=12\).
\end{minipage}

\section{Summary of Competing Models}

\begin{longtable}{>{\raggedright\arraybackslash}p{3.2cm} >{\raggedright\arraybackslash}p{5.7cm} >{\raggedright\arraybackslash}p{5.1cm}}
\caption{Summary of competing models}
\label{tab:model_predictor_summary}\\
\toprule
Model Name & Description & Predictor Set \\
\midrule
\endfirsthead

\toprule
Model Name & Description & Predictor Set \\
\midrule
\endhead

\midrule
\multicolumn{3}{r}{Continued on next page} \\
\endfoot

\bottomrule
\endlastfoot

Random Walk 
& Direct-yield prediction using a no-change random walk benchmark 
& Current yield at the same maturity, i.e. $y_t(\tau)$ \\ 
\midrule

Direct-RF
& Direct-yield prediction using Random Forest
& Lagged yields at all key maturities, i.e. $\{y_{t-\ell}(\tau_j)\}$ across the selected maturity set $\tau_j$ \\
\midrule

DNS-AR
& Factor-based prediction using AR(1) and baseline decay parameter $\lambda_1 = 0.0609$
& Own first lag only, i.e. $\beta_{i,t-1}$ \\
\midrule

DNSS-AR
& Factor-based prediction using AR(1) and baseline decay parameters $\lambda_1 = 0.0609$, $\lambda_2 = 0.01$
& Own first lag only, i.e. $\beta_{i,t-1}$ \\
\midrule

DNSS-AR-PSO
& Factor-based prediction using AR(1) and time-varying decay parameters from PSO estimation
& Own first lag only, i.e. $\beta_{i,t-1}$ \\
\midrule

DNS-RF
& Factor-based prediction using Random Forest and baseline decay parameter $\lambda_1 = 0.0609$
& Lags of all DNS factors, i.e. $\{\beta_{0,t-\ell}, \beta_{1,t-\ell}, \beta_{2,t-\ell}\}$ \\
\midrule

DNSS-RF
& Factor-based prediction using Random Forest and baseline decay parameters $\lambda_1 = 0.0609$, $\lambda_2 = 0.01$
& Lags of all DNSS factors, i.e. $\{\beta_{0,t-\ell}, \beta_{1,t-\ell}, \beta_{2,t-\ell}, \beta_{3,t-\ell}\}$ \\
\midrule

DNSS-RF-PSO
& Factor-based prediction using Random Forest and time-varying decay parameters from PSO estimation
& Lags of all DNSS factors, i.e. $\{\beta_{0,t-\ell}, \beta_{1,t-\ell}, \beta_{2,t-\ell}, \beta_{3,t-\ell}\}$ \\
\midrule

Direct-RF-X
& Macro-augmented direct-yield prediction using Random Forest
& Lagged yields at all key maturities together with lagged macro predictors, i.e. $\{y_{t-\ell}(\tau_j)\}$ and $\{x_{k,t-j}\}$ \\
\midrule

DNS-RF-X
& Macro-augmented factor-based prediction using Random Forest and baseline decay parameter $\lambda_1 = 0.0609$
& Lags of all DNS factors together with lagged macro predictors, i.e. $\{\beta_{0,t-\ell}, \beta_{1,t-\ell}, \beta_{2,t-\ell}\}$ and $\{x_{k,t-j}\}$ \\
\midrule

DNSS-RF-X
& Macro-augmented factor-based prediction using Random Forest and baseline decay parameters $\lambda_1 = 0.0609$, $\lambda_2 = 0.01$
& Lags of all DNSS factors together with lagged macro predictors, i.e. $\{\beta_{0,t-\ell}, \beta_{1,t-\ell}, \beta_{2,t-\ell}, \beta_{3,t-\ell}\}$ and $\{x_{k,t-j}\}$ \\
\midrule

DNSS-RF-PSO-X
& Macro-augmented factor-based prediction using Random Forest and time-varying decay parameters from PSO estimation
& Lags of all DNSS factors together with lagged macro predictors, i.e. $\{\beta_{0,t-\ell}, \beta_{1,t-\ell}, \beta_{2,t-\ell}, \beta_{3,t-\ell}\}$ and $\{x_{k,t-j}\}$ \\
\midrule

\midrule

\multicolumn{3}{p{14cm}}{\footnotesize Notes: $y_t(\tau)$ denotes the zero-coupon yield of maturity $\tau$ at time $t$, $\beta_{i,t}$ denotes the $i$th latent DNS or DNSS factor, and $x_{k,t}$ denotes the $k$th macroeconomic predictor. Models with suffix ``-X'' augment the predictor set with lagged macroeconomic variables. For yield and factor predictors, $\ell$ denotes lags 1 to 6. For macroeconomic predictors, $j$ denotes lags 4 to 6.}
\end{longtable}

\section{Estimated NSS Params from PSO}
\subsection{Time Varying PSO}
\begin{figure}[H]
    \centering
    
    \begin{subfigure}[t]{0.48\textwidth}
        \centering
        \includegraphics[width=\linewidth]{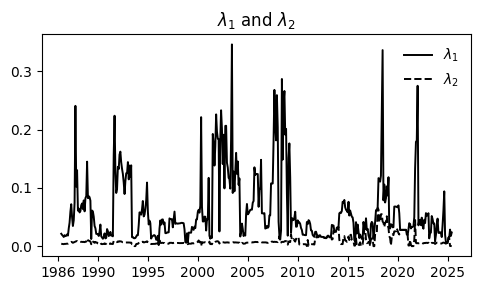}
    \end{subfigure}
    \hfill
    \begin{subfigure}[t]{0.48\textwidth}
        \centering
        \includegraphics[width=\linewidth]{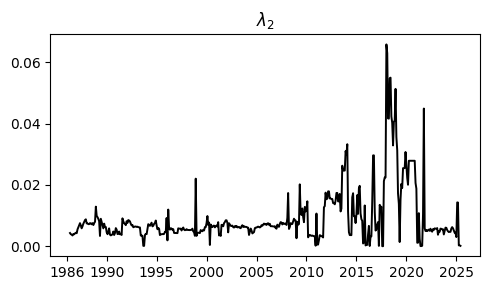}
    \end{subfigure}

    \vspace{0.4em}

    \begin{subfigure}[t]{0.48\textwidth}
        \centering
        \includegraphics[width=\linewidth]{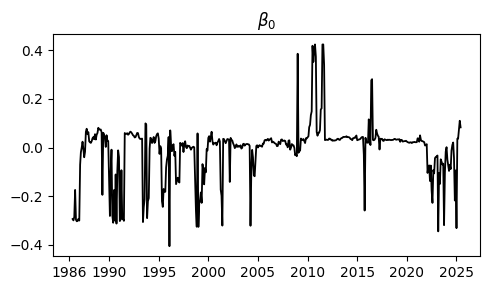}
    \end{subfigure}
    \hfill
    \begin{subfigure}[t]{0.48\textwidth}
        \centering
        \includegraphics[width=\linewidth]{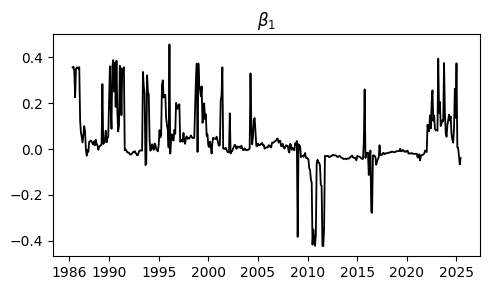}
    \end{subfigure}

    \vspace{0.4em}

    \begin{subfigure}[t]{0.48\textwidth}
        \centering
        \includegraphics[width=\linewidth]{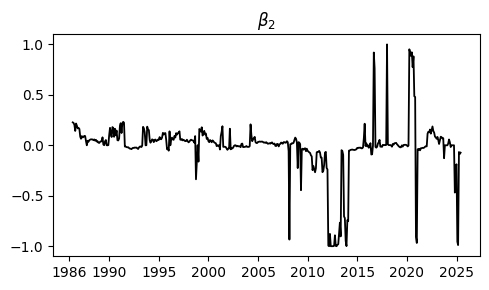}
    \end{subfigure}
    \hfill
    \begin{subfigure}[t]{0.48\textwidth}
        \centering
        \includegraphics[width=\linewidth]{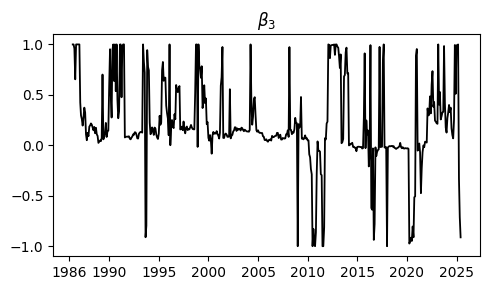}
    \end{subfigure}

    \caption{Estimated DNSS parameters under the PSO specification \eqref{eq:nss_pso_problem_baseline}.}
    \label{fig:nss_params_uncon}
\end{figure}

\begin{figure}[H]
    \centering
    
    \begin{subfigure}[t]{0.48\textwidth}
        \centering
        \includegraphics[width=\linewidth]{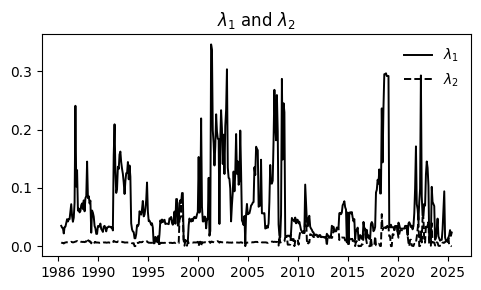}
    \end{subfigure}
    \hfill
    \begin{subfigure}[t]{0.48\textwidth}
        \centering
        \includegraphics[width=\linewidth]{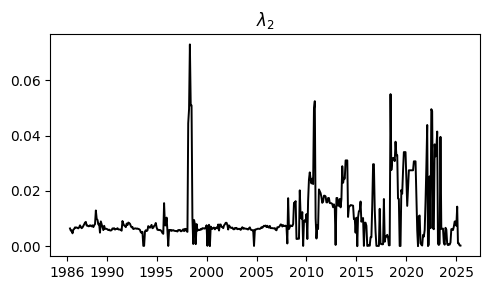}
    \end{subfigure}

    \vspace{0.4em}

    \begin{subfigure}[t]{0.48\textwidth}
        \centering
        \includegraphics[width=\linewidth]{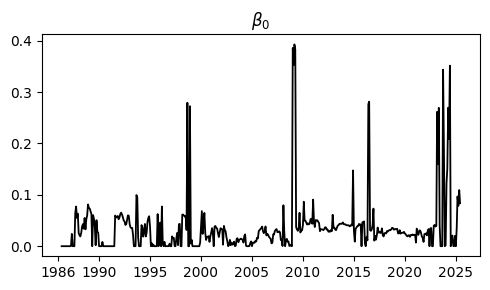}
    \end{subfigure}
    \hfill
    \begin{subfigure}[t]{0.48\textwidth}
        \centering
        \includegraphics[width=\linewidth]{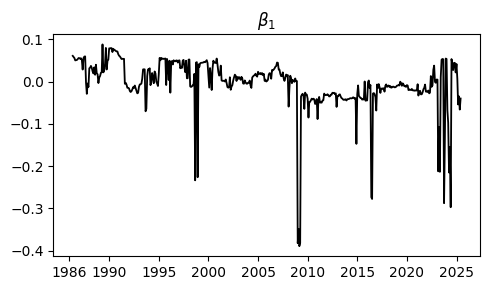}
    \end{subfigure}

    \vspace{0.4em}

    \begin{subfigure}[t]{0.48\textwidth}
        \centering
        \includegraphics[width=\linewidth]{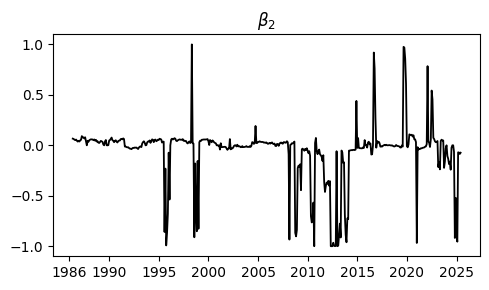}
    \end{subfigure}
    \hfill
    \begin{subfigure}[t]{0.48\textwidth}
        \centering
        \includegraphics[width=\linewidth]{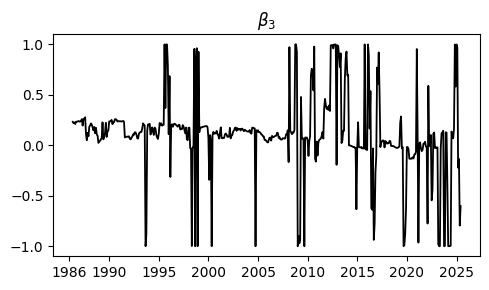}
    \end{subfigure}

    \caption{Estimated NSS parameters under the zero-lower-bound specification \eqref{eq:nss_pso_problem_zlb}.}
    \label{fig:nss_params_zlb}
\end{figure}

\begin{figure}[H]
    \centering
    
    \begin{subfigure}[t]{0.48\textwidth}
        \centering
        \includegraphics[width=\linewidth]{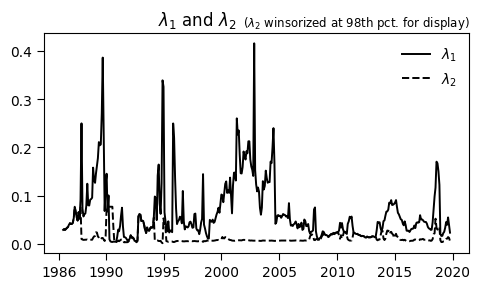}
    \end{subfigure}
    \hfill
    \begin{subfigure}[t]{0.48\textwidth}
        \centering
        \includegraphics[width=\linewidth]{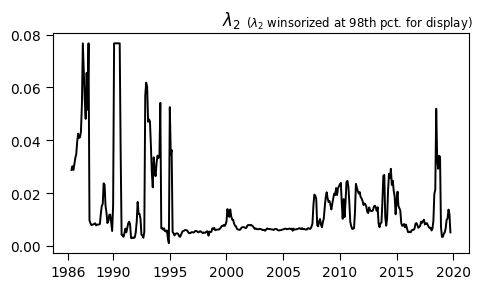}
        \caption{$\lambda_2$ winsorized at the 97.5th percentile}
    \end{subfigure}

    \vspace{0.4em}

    \begin{subfigure}[t]{0.48\textwidth}
        \centering
        \includegraphics[width=\linewidth]{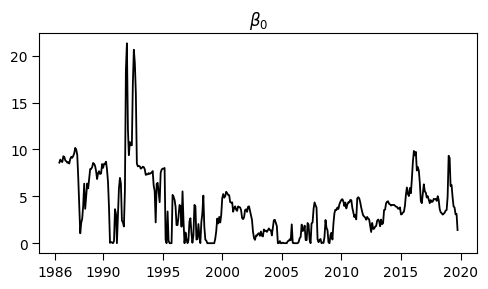}
    \end{subfigure}
    \hfill
    \begin{subfigure}[t]{0.48\textwidth}
        \centering
        \includegraphics[width=\linewidth]{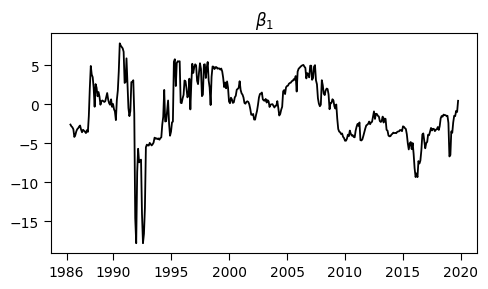}
    \end{subfigure}

    \vspace{0.4em}

    \begin{subfigure}[t]{0.48\textwidth}
        \centering
        \includegraphics[width=\linewidth]{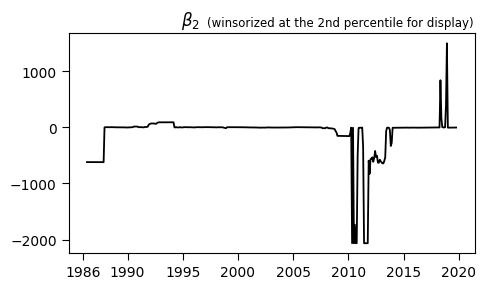}
    \end{subfigure}
    \hfill
    \begin{subfigure}[t]{0.48\textwidth}
        \centering
        \includegraphics[width=\linewidth]{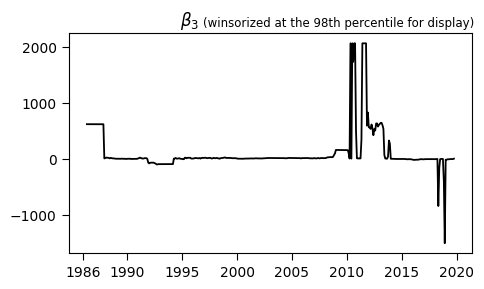}
    \end{subfigure}

    \caption{Estimated NSS parameters from \citet{GurkaynakSackWright2007}}
    \label{fig:nss_params_gsw}
\end{figure}

\subsection{Pooled PSO}
\begin{figure}[H]
\centering
\begin{tabular}{cc}
\includegraphics[width=0.48\textwidth]{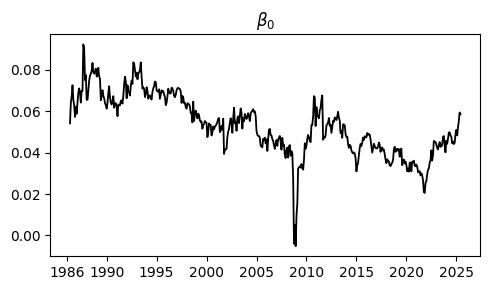} &
\includegraphics[width=0.48\textwidth]{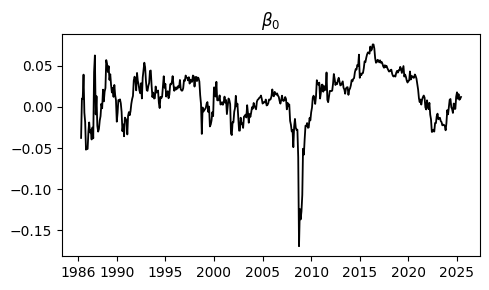} \\[0.6em]

\includegraphics[width=0.48\textwidth]{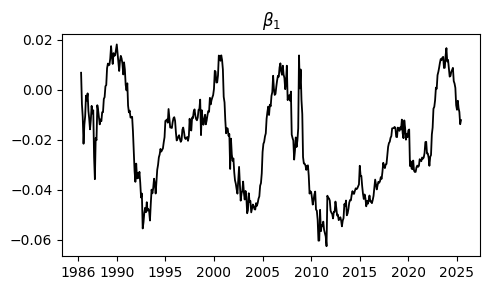} &
\includegraphics[width=0.48\textwidth]{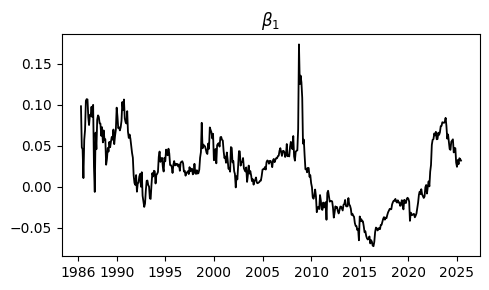} \\[0.6em]

\includegraphics[width=0.48\textwidth]{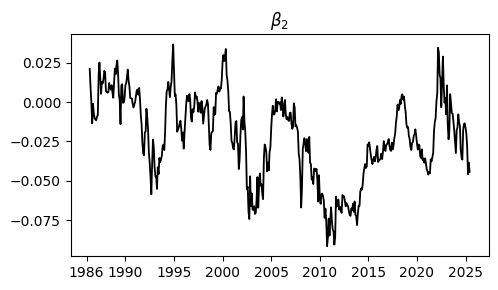} &
\includegraphics[width=0.48\textwidth]{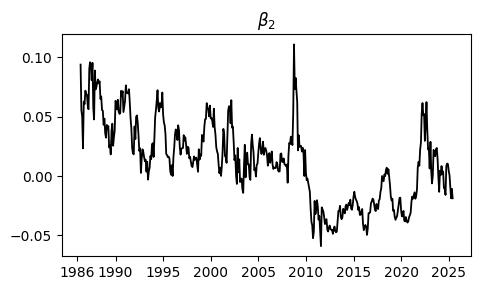} \\[0.6em]

\includegraphics[width=0.48\textwidth]{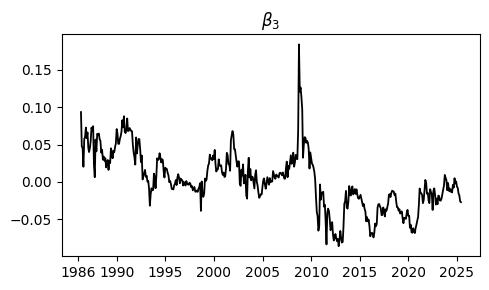} &
\includegraphics[width=0.48\textwidth]{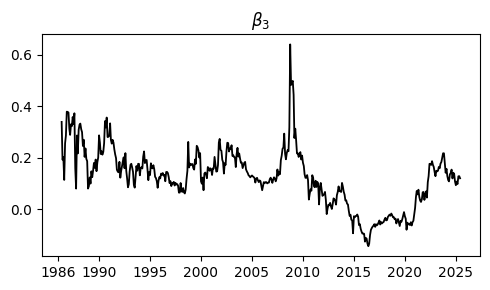} \\
\end{tabular}
\caption{Estimated DNSS factor series under the baseline and pooled-PSO specifications. The left column reports the baseline specification with $\lambda_1 = 0.0609$ and $\lambda_2 = 0.01$. The right column reports the pooled-PSO specification with $\lambda_1 = 0.035$ and $\lambda_2 = 0.006$.}
\label{fig:pooled_baseline_vs_pso_betas}
\end{figure}

\section{RMSE Tables}

\begin{table}[H]
\centering
\singlespacing
\caption{Relative RMSE at the 1-month forecast horizon}
\label{tab:rrmse_h1}
\small
\setlength{\tabcolsep}{4pt}
\renewcommand{\arraystretch}{1.15}
\makebox[\textwidth][c]{%
\begin{threeparttable}
\begin{tabular}{lccccc@{\hspace{0.6cm}}cc@{\hspace{0.6cm}}ccc@{\hspace{0.6cm}}ccc}
\toprule
& \multicolumn{5}{c}{Short-end} & \multicolumn{2}{c}{Belly} & \multicolumn{3}{c}{Long-end} & \multicolumn{3}{c}{Slope} \\
\cmidrule(lr){2-6} \cmidrule(lr){7-8} \cmidrule(lr){9-11} \cmidrule(lr){12-14}
Model & 3 & 6 & 12 & 18 & 24 & 36 & 60 & 120 & 240 & 360 & 24--120 & 24--360 & 60--360 \\
\midrule
\multicolumn{14}{l}{\textit{Panel A: Yields-only models}} \\
Direct-RF & 0.87** & 0.89** & 0.94* & 0.98 & 0.99 & 1.01 & 1.03 & 1.03 & 1.03 & 1.04 & 1.01 & 1.03 & 1.04 \\
DNS-AR & 0.98 & 1.05 & 1.24 & 1.14 & 1.04 & 1.01 & 1.06 & 1.15 & 1.33 & 1.15 & 1.83 & 1.11 & 1.78 \\
DNS-RF & 0.93 & 0.92* & 1.14 & 1.08 & 1.01 & 1.03 & 1.11 & 1.22 & 1.29 & 1.17 & 1.84 & 1.13 & 1.79 \\
DNSS-AR & 1.05 & 1.06 & 1.14 & 1.05 & 1.02 & 1.10 & 1.16 & 1.11 & 1.33 & 1.13 & 1.37 & 1.13 & 1.07 \\
DNSS-RF & 0.86 & 0.89* & 1.09 & 1.03 & 1.00 & 1.05 & 1.09 & 1.08 & 1.35 & 1.07 & 1.30 & 1.11 & 1.10 \\
DNSS-AR-PSO & 1.08 & 1.04 & 1.02 & 0.99 & 1.03 & 1.07 & 1.06 & 1.10 & 1.15 & 1.12 & 1.04 & 1.05 & 1.08 \\
DNSS-RF-PSO & 2.02 & 2.02 & 1.90 & 1.87 & 1.91 & 1.94 & 1.91 & 2.04 & 2.10 & 2.18 & 1.20 & 1.19 & 1.30 \\
\addlinespace
\multicolumn{14}{l}{\textit{Panel B: Macro-augmented models}} \\
Direct-RF-X & 0.86* & 0.86* & 0.91* & 0.96 & 0.98 & 0.98 & 1.00 & 1.02 & 1.02 & 1.02 & 0.97 & 0.97 & 0.98 \\
DNS-RF-X & 0.92 & 0.91 & 1.12 & 1.06 & 1.00 & 1.02 & 1.10 & 1.19 & 1.32 & 1.17 & 1.76 & 1.10 & 1.82 \\
DNSS-RF-X & 0.87 & 0.91* & 1.11 & 1.04 & 0.99 & 1.03 & 1.06 & 1.04 & 1.39 & 1.02 & 1.25 & 1.02 & 1.09 \\
DNSS-RF-PSO-X & 1.64 & 1.65 & 1.64 & 1.57 & 1.60 & 1.72 & 1.91 & 1.96 & 1.68 & 1.50 & 1.62 & 1.35 & 1.85 \\
\bottomrule
\end{tabular}
\begin{tablenotes}
\footnotesize
\item Notes: This table reports the RMSE at the 1-month horizon relative to that of the random walk benchmark. Values below 1 indicate an improvement over the random walk. Results are grouped by curve segment into the short end, belly, long end, and slope. Panel A reports yields-only specifications, while Panel B reports macro-augmented specifications. Statistical significance relative to the random walk benchmark is denoted by stars: *** \(1\%\), ** \(5\%\), and * \(10\%\).
\end{tablenotes}
\end{threeparttable}
}
\end{table}

\begin{table}[H]
\centering
\singlespacing
\caption{Relative RMSE at the 3-month forecast horizon}
\label{tab:rrmse_h3}
\small
\setlength{\tabcolsep}{4pt}
\renewcommand{\arraystretch}{1.15}
\makebox[\textwidth][c]{%
\begin{threeparttable}
\begin{tabular}{lccccc@{\hspace{0.6cm}}cc@{\hspace{0.6cm}}ccc@{\hspace{0.6cm}}ccc}
\toprule
& \multicolumn{5}{c}{Short-end} & \multicolumn{2}{c}{Belly} & \multicolumn{3}{c}{Long-end} & \multicolumn{3}{c}{Slope} \\
\cmidrule(lr){2-6} \cmidrule(lr){7-8} \cmidrule(lr){9-11} \cmidrule(lr){12-14}
Model & 3 & 6 & 12 & 18 & 24 & 36 & 60 & 120 & 240 & 360 & 24--120 & 24--360 & 60--360 \\
\midrule
\multicolumn{14}{l}{\textit{Panel A: Yields-only models}} \\
Direct-RF & 0.87** & 0.89** & 0.92** & 0.95 & 0.97 & 0.99 & 1.03 & 1.09 & 1.10 & 1.10 & 1.05 & 1.05 & 1.04 \\
DNS-AR & 0.97 & 1.05 & 1.13 & 1.09 & 1.03 & 1.00 & 1.00 & 0.99 & 1.17 & 1.03 & 1.35 & 1.04 & 1.27 \\
DNS-RF & 0.89 & 0.97 & 1.07 & 1.06 & 1.05 & 1.06 & 1.10 & 1.12 & 1.23 & 1.12 & 1.34 & 1.08 & 1.33 \\
DNSS-AR & 1.00 & 1.02 & 1.04 & 1.01 & 1.00 & 1.06 & 1.12 & 1.11 & 1.21 & 1.16 & 1.20 & 1.08 & 1.01 \\
DNSS-RF & 0.82* & 0.89* & 0.98 & 0.98 & 0.97 & 1.00 & 1.05 & 1.05 & 1.18 & 1.06 & 1.11 & 1.04 & 1.05 \\
DNSS-AR-PSO & 1.08 & 1.08 & 1.08 & 1.08 & 1.11 & 1.15 & 1.18 & 1.23 & 1.32 & 1.29 & 1.04 & 1.03 & 1.09 \\
DNSS-RF-PSO & 1.03 & 1.23 & 1.56 & 1.76 & 1.89 & 2.01 & 2.05 & 1.97 & 1.97 & 2.07 & 2.04 & 2.37 & 2.79 \\
\addlinespace
\multicolumn{14}{l}{\textit{Panel B: Macro-augmented models}} \\
Direct-RF-X & 0.81** & 0.82** & 0.85** & 0.90* & 0.93* & 0.97 & 0.98 & 1.01 & 1.04 & 1.05 & 0.96 & 0.94 & 0.98 \\
DNS-RF-X & 0.80* & 0.86* & 0.95 & 0.96 & 0.96 & 0.98 & 1.04 & 1.07 & 1.20 & 1.09 & 1.21 & 0.98 & 1.33 \\
DNSS-RF-X & 0.75* & 0.81* & 0.91 & 0.92* & 0.93* & 0.97 & 1.01 & 1.02 & 1.18 & 1.03 & 1.07 & 0.94 & 1.00 \\
DNSS-RF-PSO-X & 1.18 & 1.35 & 1.65 & 1.85 & 2.01 & 2.21 & 2.36 & 2.26 & 1.97 & 1.92 & 1.81 & 2.07 & 2.83 \\
\bottomrule
\end{tabular}
\begin{tablenotes}
\footnotesize
\item Notes: This table reports the RMSE at the 3-month horizon relative to that of the random walk benchmark. Values below 1 indicate an improvement over the random walk. Results are grouped by curve segment into the short end, belly, long end, and slope. Panel A reports yields-only specifications, while Panel B reports macro-augmented specifications. Statistical significance relative to the random walk benchmark is denoted by stars: *** \(1\%\), ** \(5\%\), and * \(10\%\).
\end{tablenotes}
\end{threeparttable}
}
\end{table}

\begin{table}[H]
\centering
\singlespacing
\caption{Relative RMSE at the 6-month forecast horizon}
\label{tab:rrmse_h6}
\small
\setlength{\tabcolsep}{4pt}
\renewcommand{\arraystretch}{1.15}
\makebox[\textwidth][c]{%
\begin{threeparttable}
\begin{tabular}{lccccc@{\hspace{0.6cm}}cc@{\hspace{0.6cm}}ccc@{\hspace{0.6cm}}ccc}
\toprule
& \multicolumn{5}{c}{Short-end} & \multicolumn{2}{c}{Belly} & \multicolumn{3}{c}{Long-end} & \multicolumn{3}{c}{Slope} \\
\cmidrule(lr){2-6} \cmidrule(lr){7-8} \cmidrule(lr){9-11} \cmidrule(lr){12-14}
Model & 3 & 6 & 12 & 18 & 24 & 36 & 60 & 120 & 240 & 360 & 24--120 & 24--360 & 60--360 \\
\midrule
\multicolumn{14}{l}{\textit{Panel A: Yields-only models}} \\
Direct-RF & 0.97 & 0.97 & 0.99 & 1.01 & 1.03 & 1.04 & 1.07 & 1.15 & 1.14 & 1.16 & 0.98 & 1.03 & 1.10 \\
DNS-AR & 1.00 & 1.05 & 1.10 & 1.08 & 1.05 & 1.01 & 1.00 & 0.97 & 1.13 & 1.00 & 1.26 & 1.04 & 1.06 \\
DNS-RF & 0.90 & 0.98 & 1.06 & 1.09 & 1.10 & 1.12 & 1.16 & 1.16 & 1.27 & 1.16 & 1.16 & 1.02 & 1.15 \\
DNSS-AR & 0.97 & 0.99 & 1.00 & 0.99 & 0.99 & 1.03 & 1.11 & 1.15 & 1.23 & 1.24 & 1.18 & 1.09 & 0.97 \\
DNSS-RF & 0.95 & 1.01 & 1.07 & 1.08 & 1.08 & 1.10 & 1.13 & 1.13 & 1.22 & 1.11 & 1.08 & 1.07 & 1.08 \\
DNSS-AR-PSO & 1.07 & 1.08 & 1.09 & 1.11 & 1.13 & 1.18 & 1.23 & 1.30 & 1.38 & 1.35 & 1.01 & 0.98 & 1.02 \\
DNSS-RF-PSO & 0.85 & 0.89 & 0.95 & 0.99 & 1.03 & 1.09 & 1.14 & 1.15 & 1.13 & 1.11 & 0.96 & 0.98 & 1.20 \\
\addlinespace
\multicolumn{14}{l}{\textit{Panel B: Macro-augmented models}} \\
Direct-RF-X & 0.87* & 0.88* & 0.90 & 0.92 & 0.95 & 0.98 & 1.02 & 1.07 & 1.07 & 1.08 & 0.87* & 0.88** & 0.98 \\
DNS-RF-X & 0.84** & 0.89* & 0.96 & 0.97 & 0.97 & 0.99 & 1.03 & 1.06 & 1.20 & 1.10 & 1.04 & 0.89* & 1.08 \\
DNSS-RF-X & 0.93 & 0.99 & 1.05 & 1.07 & 1.08 & 1.11 & 1.16 & 1.15 & 1.29 & 1.17 & 0.97 & 0.93 & 0.99 \\
DNSS-RF-PSO-X & 0.93 & 1.11 & 1.46 & 1.77 & 2.05 & 2.43 & 2.80 & 2.68 & 1.97 & 1.78 & 1.79 & 2.25 & 3.65 \\
\bottomrule
\end{tabular}
\begin{tablenotes}
\footnotesize
\item Notes: This table reports the RMSE at the 6-month horizon relative to that of the random walk benchmark. Values below 1 indicate an improvement over the random walk. Results are grouped by curve segment into the short end, belly, long end, and slope. Panel A reports yields-only specifications, while Panel B reports macro-augmented specifications. Statistical significance relative to the random walk benchmark is denoted by stars: *** \(1\%\), ** \(5\%\), and * \(10\%\).
\end{tablenotes}
\end{threeparttable}
}
\end{table}

\begin{table}[H]
\centering
\singlespacing
\caption{Relative RMSE at the 12-month forecast horizon}
\label{tab:rrmse_h12}
\small
\setlength{\tabcolsep}{4pt}
\renewcommand{\arraystretch}{1.15}
\makebox[\textwidth][c]{%
\begin{threeparttable}
\begin{tabular}{lccccc@{\hspace{0.6cm}}cc@{\hspace{0.6cm}}ccc@{\hspace{0.6cm}}ccc}
\toprule
& \multicolumn{5}{c}{Short-end} & \multicolumn{2}{c}{Belly} & \multicolumn{3}{c}{Long-end} & \multicolumn{3}{c}{Slope} \\
\cmidrule(lr){2-6} \cmidrule(lr){7-8} \cmidrule(lr){9-11} \cmidrule(lr){12-14}
Model & 3 & 6 & 12 & 18 & 24 & 36 & 60 & 120 & 240 & 360 & 24--120 & 24--360 & 60--360 \\
\midrule
\multicolumn{14}{l}{\textit{Panel A: Yields-only models}} \\
Direct-RF & 1.15 & 1.15 & 1.14 & 1.14 & 1.15 & 1.15 & 1.15 & 1.18 & 1.13 & 1.11 & 1.09 & 1.18 & 1.17 \\
DNS-AR & 1.03 & 1.06 & 1.08 & 1.06 & 1.04 & 1.00 & 0.98 & 0.94* & 1.10 & 0.99 & 1.19 & 1.00 & 0.88** \\
DNS-RF & 1.07 & 1.10 & 1.14 & 1.15 & 1.16 & 1.16 & 1.16 & 1.12 & 1.18 & 1.08 & 1.18 & 1.12 & 1.14 \\
DNSS-AR & 0.92 & 0.93 & 0.94 & 0.93 & 0.93 & 0.95 & 1.03 & 1.14 & 1.23 & 1.27 & 1.12 & 1.03 & 0.90** \\
DNSS-RF & 1.11 & 1.15 & 1.19 & 1.21 & 1.22 & 1.24 & 1.25 & 1.23 & 1.28 & 1.17 & 1.12 & 1.13 & 1.13 \\
DNSS-AR-PSO & 0.98 & 0.99 & 0.99 & 0.99 & 1.01 & 1.04 & 1.10 & 1.22 & 1.34 & 1.37 & 1.03 & 0.97 & 0.95 \\
DNSS-RF-PSO & 1.17 & 1.17 & 1.17 & 1.18 & 1.19 & 1.21 & 1.23 & 1.15 & 1.05 & 1.02 & 1.24 & 1.33 & 1.50 \\
\addlinespace
\multicolumn{14}{l}{\textit{Panel B: Macro-augmented models}} \\
Direct-RF-X & 1.01 & 1.00 & 1.02 & 1.03 & 1.04 & 1.06 & 1.10 & 1.14 & 1.10 & 1.12 & 0.96 & 0.97 & 0.96 \\
DNS-RF-X & 1.03 & 1.05 & 1.07 & 1.08 & 1.09 & 1.09 & 1.09 & 1.07 & 1.18 & 1.08 & 1.04 & 0.99 & 1.03 \\
DNSS-RF-X & 1.03 & 1.05 & 1.08 & 1.10 & 1.12 & 1.14 & 1.17 & 1.17 & 1.26 & 1.17 & 0.94 & 0.95 & 1.00 \\
DNSS-RF-PSO-X & 1.06 & 1.18 & 1.42 & 1.62 & 1.79 & 2.00 & 2.18 & 2.07 & 1.61 & 1.59 & 1.60 & 1.90 & 2.75 \\
\bottomrule
\end{tabular}
\begin{tablenotes}
\footnotesize
\item Notes: This table reports the RMSE at the 12-month horizon relative to that of the random walk benchmark. Values below 1 indicate an improvement over the random walk. Results are grouped by curve segment into the short end, belly, long end, and slope. Panel A reports yields-only specifications, while Panel B reports macro-augmented specifications. GR results are omitted. Statistical significance relative to the random walk benchmark is denoted by stars: *** \(1\%\), ** \(5\%\), and * \(10\%\).
\end{tablenotes}
\end{threeparttable}
}
\end{table}


\begin{table}[H]
\centering
\singlespacing
\caption{Absolute RMSE at the 1-month forecast horizon}
\label{tab:armse_h1}
\small
\setlength{\tabcolsep}{4pt}
\renewcommand{\arraystretch}{1.15}
\makebox[\textwidth][c]{%
\begin{threeparttable}
\begin{tabular}{lccccc@{\hspace{0.6cm}}cc@{\hspace{0.6cm}}ccc@{\hspace{0.6cm}}ccc}
\toprule
& \multicolumn{5}{c}{Short-end} & \multicolumn{2}{c}{Belly} & \multicolumn{3}{c}{Long-end} & \multicolumn{3}{c}{Slope} \\
\cmidrule(lr){2-6} \cmidrule(lr){7-8} \cmidrule(lr){9-11} \cmidrule(lr){12-14}
Model & 3 & 6 & 12 & 18 & 24 & 36 & 60 & 120 & 240 & 360 & 24--120 & 24--360 & 60--360 \\
\midrule
\multicolumn{14}{l}{\textit{Panel A: Yields-only models}} \\
Random Walk & 0.20 & 0.20 & 0.22 & 0.24 & 0.25 & 0.26 & 0.26 & 0.25 & 0.23 & 0.21 & 0.15 & 0.20 & 0.15 \\
Direct-RF & 0.17** & 0.18** & 0.20* & 0.23 & 0.24 & 0.26 & 0.27 & 0.26 & 0.24 & 0.22 & 0.15 & 0.20 & 0.16 \\
DNS-AR & 0.19 & 0.21 & 0.27 & 0.27 & 0.25 & 0.26 & 0.28 & 0.28 & 0.31 & 0.25 & 0.27 & 0.22 & 0.27 \\
DNS-RF & 0.18 & 0.18* & 0.25 & 0.25 & 0.25 & 0.26 & 0.29 & 0.30 & 0.30 & 0.25 & 0.28 & 0.22 & 0.27 \\
DNSS-AR & 0.21 & 0.21 & 0.25 & 0.25 & 0.25 & 0.28 & 0.30 & 0.28 & 0.31 & 0.24 & 0.20 & 0.22 & 0.16 \\
DNSS-RF & 0.17 & 0.18* & 0.24 & 0.24 & 0.25 & 0.27 & 0.28 & 0.27 & 0.31 & 0.23 & 0.19 & 0.22 & 0.17 \\
DNSS-AR-PSO & 0.22 & 0.21 & 0.22 & 0.23 & 0.25 & 0.27 & 0.27 & 0.27 & 0.27 & 0.24 & 0.16 & 0.21 & 0.16 \\
DNSS-RF-PSO & 0.40 & 0.40 & 0.41 & 0.44 & 0.47 & 0.49 & 0.49 & 0.50 & 0.48 & 0.47 & 0.18 & 0.23 & 0.20 \\
\addlinespace
\multicolumn{14}{l}{\textit{Panel B: Macro-augmented models}} \\
Direct-RF-X & 0.17* & 0.17* & 0.20* & 0.23 & 0.24 & 0.25 & 0.26 & 0.25 & 0.23 & 0.22 & 0.15 & 0.19 & 0.15 \\
DNS-RF-X & 0.18 & 0.18 & 0.24 & 0.25 & 0.25 & 0.26 & 0.28 & 0.29 & 0.30 & 0.25 & 0.26 & 0.22 & 0.27 \\
DNSS-RF-X & 0.17 & 0.18* & 0.24 & 0.25 & 0.24 & 0.26 & 0.28 & 0.26 & 0.32 & 0.22 & 0.19 & 0.20 & 0.16 \\
DNSS-RF-PSO-X & 0.33 & 0.33 & 0.35 & 0.37 & 0.39 & 0.44 & 0.50 & 0.49 & 0.39 & 0.32 & 0.24 & 0.26 & 0.28 \\
\bottomrule
\end{tabular}
\begin{tablenotes}
\footnotesize
\item Notes: This table reports absolute RMSE at the 1-month forecast horizon. Results are grouped by curve segment into the short end, belly, long end, and slope. Panel A reports yields-only specifications, while Panel B reports macro-augmented specifications. GR results are omitted. Statistical significance relative to the random walk benchmark is denoted by stars: *** \(1\%\), ** \(5\%\), and * \(10\%\).
\end{tablenotes}
\end{threeparttable}
}
\end{table}

\begin{table}[H]
\centering
\singlespacing
\caption{Absolute RMSE at the 3-month forecast horizon}
\label{tab:armse_h3}
\small
\setlength{\tabcolsep}{4pt}
\renewcommand{\arraystretch}{1.15}
\makebox[\textwidth][c]{%
\begin{threeparttable}
\begin{tabular}{lccccc@{\hspace{0.6cm}}cc@{\hspace{0.6cm}}ccc@{\hspace{0.6cm}}ccc}
\toprule
& \multicolumn{5}{c}{Short-end} & \multicolumn{2}{c}{Belly} & \multicolumn{3}{c}{Long-end} & \multicolumn{3}{c}{Slope} \\
\cmidrule(lr){2-6} \cmidrule(lr){7-8} \cmidrule(lr){9-11} \cmidrule(lr){12-14}
Model & 3 & 6 & 12 & 18 & 24 & 36 & 60 & 120 & 240 & 360 & 24--120 & 24--360 & 60--360 \\
\midrule
\multicolumn{14}{l}{\textit{Panel A: Yields-only models}} \\
Random Walk & 0.48 & 0.47 & 0.49 & 0.50 & 0.50 & 0.49 & 0.48 & 0.45 & 0.43 & 0.39 & 0.29 & 0.36 & 0.26 \\
Direct-RF & 0.42** & 0.42** & 0.45** & 0.47 & 0.48 & 0.49 & 0.49 & 0.50 & 0.47 & 0.43 & 0.30 & 0.38 & 0.27 \\
DNS-AR & 0.46 & 0.50 & 0.55 & 0.54 & 0.52 & 0.49 & 0.48 & 0.45 & 0.50 & 0.40 & 0.39 & 0.38 & 0.32 \\
DNS-RF & 0.43 & 0.46 & 0.52 & 0.53 & 0.52 & 0.52 & 0.52 & 0.51 & 0.53 & 0.44 & 0.38 & 0.39 & 0.34 \\
DNSS-AR & 0.48 & 0.49 & 0.50 & 0.50 & 0.50 & 0.52 & 0.53 & 0.50 & 0.52 & 0.45 & 0.34 & 0.39 & 0.26 \\
DNSS-RF & 0.39* & 0.42* & 0.47 & 0.48 & 0.48 & 0.50 & 0.50 & 0.48 & 0.51 & 0.42 & 0.32 & 0.38 & 0.27 \\
DNSS-AR-PSO & 0.51 & 0.51 & 0.53 & 0.54 & 0.55 & 0.57 & 0.56 & 0.56 & 0.56 & 0.50 & 0.30 & 0.37 & 0.28 \\
DNSS-RF-PSO & 0.49 & 0.59 & 0.76 & 0.87 & 0.94 & 0.99 & 0.98 & 0.89 & 0.84 & 0.81 & 0.58 & 0.86 & 0.71 \\
\addlinespace
\multicolumn{14}{l}{\textit{Panel B: Macro-augmented models}} \\
Direct-RF-X & 0.39** & 0.39** & 0.41** & 0.45* & 0.46* & 0.48 & 0.47 & 0.46 & 0.44 & 0.41 & 0.27 & 0.34 & 0.25 \\
DNS-RF-X & 0.38* & 0.41* & 0.46 & 0.48 & 0.48 & 0.49 & 0.49 & 0.48 & 0.51 & 0.43 & 0.35 & 0.35 & 0.34 \\
DNSS-RF-X & 0.36* & 0.39* & 0.44 & 0.45* & 0.46* & 0.48 & 0.48 & 0.46 & 0.50 & 0.40 & 0.31 & 0.34 & 0.26 \\
DNSS-RF-PSO-X & 0.56 & 0.64 & 0.80 & 0.92 & 1.00 & 1.09 & 1.13 & 1.03 & 0.84 & 0.75 & 0.52 & 0.75 & 0.73 \\
\bottomrule
\end{tabular}
\begin{tablenotes}
\footnotesize
\item Notes: This table reports absolute RMSE at the 3-month forecast horizon. Results are grouped by curve segment into the short end, belly, long end, and slope. Panel A reports yields-only specifications, while Panel B reports macro-augmented specifications. GR results are omitted. Statistical significance relative to the random walk benchmark is denoted by stars: *** \(1\%\), ** \(5\%\), and * \(10\%\).
\end{tablenotes}
\end{threeparttable}
}
\end{table}

\begin{table}[H]
\centering
\singlespacing
\caption{Absolute RMSE at the 6-month forecast horizon}
\label{tab:armse_h6}
\small
\setlength{\tabcolsep}{4pt}
\renewcommand{\arraystretch}{1.15}
\makebox[\textwidth][c]{%
\begin{threeparttable}
\begin{tabular}{lccccc@{\hspace{0.6cm}}cc@{\hspace{0.6cm}}ccc@{\hspace{0.6cm}}ccc}
\toprule
& \multicolumn{5}{c}{Short-end} & \multicolumn{2}{c}{Belly} & \multicolumn{3}{c}{Long-end} & \multicolumn{3}{c}{Slope} \\
\cmidrule(lr){2-6} \cmidrule(lr){7-8} \cmidrule(lr){9-11} \cmidrule(lr){12-14}
Model & 3 & 6 & 12 & 18 & 24 & 36 & 60 & 120 & 240 & 360 & 24--120 & 24--360 & 60--360 \\
\midrule
\multicolumn{14}{l}{\textit{Panel A: Yields-only models}} \\
Random Walk & 0.87 & 0.86 & 0.84 & 0.82 & 0.78 & 0.74 & 0.67 & 0.62 & 0.59 & 0.52 & 0.43 & 0.54 & 0.37 \\
Direct-RF & 0.85 & 0.84 & 0.83 & 0.83 & 0.81 & 0.77 & 0.72 & 0.71 & 0.67 & 0.61 & 0.43 & 0.56 & 0.40 \\
DNS-AR & 0.87 & 0.91 & 0.93 & 0.88 & 0.82 & 0.75 & 0.67 & 0.60 & 0.66 & 0.52 & 0.54 & 0.57 & 0.39 \\
DNS-RF & 0.79 & 0.84 & 0.90 & 0.89 & 0.86 & 0.82 & 0.78 & 0.72 & 0.74 & 0.61 & 0.50 & 0.55 & 0.42 \\
DNSS-AR & 0.85 & 0.85 & 0.84 & 0.81 & 0.78 & 0.76 & 0.75 & 0.71 & 0.72 & 0.65 & 0.51 & 0.59 & 0.36 \\
DNSS-RF & 0.83 & 0.87 & 0.90 & 0.88 & 0.85 & 0.81 & 0.77 & 0.69 & 0.71 & 0.58 & 0.47 & 0.58 & 0.40 \\
DNSS-AR-PSO & 0.93 & 0.93 & 0.92 & 0.90 & 0.89 & 0.87 & 0.83 & 0.80 & 0.81 & 0.71 & 0.44 & 0.53 & 0.38 \\
DNSS-RF-PSO & 0.74 & 0.77 & 0.80 & 0.81 & 0.81 & 0.80 & 0.77 & 0.71 & 0.66 & 0.58 & 0.41 & 0.53 & 0.44 \\
\addlinespace
\multicolumn{14}{l}{\textit{Panel B: Macro-augmented models}} \\
Direct-RF-X & 0.76* & 0.76* & 0.75 & 0.75 & 0.74 & 0.72 & 0.69 & 0.66 & 0.63 & 0.57 & 0.38* & 0.48** & 0.36 \\
DNS-RF-X & 0.73** & 0.77* & 0.81 & 0.79 & 0.76 & 0.73 & 0.70 & 0.65 & 0.71 & 0.58 & 0.45 & 0.49* & 0.40 \\
DNSS-RF-X & 0.82 & 0.85 & 0.89 & 0.87 & 0.85 & 0.82 & 0.78 & 0.71 & 0.76 & 0.61 & 0.42 & 0.50 & 0.36 \\
DNSS-RF-PSO-X & 0.82 & 0.96 & 1.23 & 1.44 & 1.60 & 1.79 & 1.89 & 1.65 & 1.16 & 0.93 & 0.77 & 1.22 & 1.34 \\
\bottomrule
\end{tabular}
\begin{tablenotes}
\footnotesize
\item Notes: This table reports absolute RMSE at the 6-month forecast horizon. Results are grouped by curve segment into the short end, belly, long end, and slope. Panel A reports yields-only specifications, while Panel B reports macro-augmented specifications. GR results are omitted. Statistical significance relative to the random walk benchmark is denoted by stars: *** \(1\%\), ** \(5\%\), and * \(10\%\).
\end{tablenotes}
\end{threeparttable}
}
\end{table}

\begin{table}[H]
\centering
\singlespacing
\caption{Absolute RMSE at the 12-month forecast horizon}
\label{tab:armse_h12}
\small
\setlength{\tabcolsep}{4pt}
\renewcommand{\arraystretch}{1.15}
\begin{threeparttable}
\begin{tabular}{lccccc@{\hspace{0.6cm}}cc@{\hspace{0.6cm}}ccc@{\hspace{0.6cm}}ccc}
\toprule
& \multicolumn{5}{c}{Short-end} & \multicolumn{2}{c}{Belly} & \multicolumn{3}{c}{Long-end} & \multicolumn{3}{c}{Slope} \\
\cmidrule(lr){2-6} \cmidrule(lr){7-8} \cmidrule(lr){9-11} \cmidrule(lr){12-14}
Model & 3 & 6 & 12 & 18 & 24 & 36 & 60 & 120 & 240 & 360 & 24--120 & 24--360 & 60--360 \\
\midrule
\multicolumn{14}{l}{\textit{Panel A: Yields-only models}} \\
Random Walk & 1.57 & 1.55 & 1.51 & 1.44 & 1.37 & 1.26 & 1.10 & 0.93 & 0.85 & 0.75 & 0.71 & 0.91 & 0.60 \\
Direct-RF & 1.81 & 1.78 & 1.72 & 1.64 & 1.57 & 1.45 & 1.27 & 1.09 & 0.96 & 0.83 & 0.77 & 1.08 & 0.70 \\
DNS-AR & 1.62 & 1.64 & 1.63 & 1.54 & 1.42 & 1.27 & 1.08 & 0.88* & 0.94 & 0.74 & 0.84 & 0.92 & 0.52** \\
DNS-RF & 1.68 & 1.71 & 1.72 & 1.66 & 1.59 & 1.46 & 1.28 & 1.04 & 1.01 & 0.81 & 0.83 & 1.03 & 0.68 \\
DNSS-AR & 1.45 & 1.45 & 1.42 & 1.35 & 1.28 & 1.20 & 1.13 & 1.05 & 1.05 & 0.95 & 0.79 & 0.94 & 0.54** \\
DNSS-RF & 1.75 & 1.78 & 1.80 & 1.74 & 1.67 & 1.56 & 1.38 & 1.14 & 1.10 & 0.88 & 0.79 & 1.03 & 0.67 \\
DNSS-AR-PSO & 1.54 & 1.53 & 1.49 & 1.43 & 1.38 & 1.31 & 1.22 & 1.13 & 1.15 & 1.02 & 0.73 & 0.89 & 0.57 \\
DNSS-RF-PSO & 1.84 & 1.82 & 1.77 & 1.70 & 1.63 & 1.53 & 1.36 & 1.06 & 0.89 & 0.77 & 0.88 & 1.22 & 0.89 \\
\addlinespace
\multicolumn{14}{l}{\textit{Panel B: Macro-augmented models}} \\
Direct-RF-X & 1.58 & 1.56 & 1.55 & 1.49 & 1.43 & 1.34 & 1.21 & 1.06 & 0.94 & 0.84 & 0.68 & 0.89 & 0.57 \\
DNS-RF-X & 1.62 & 1.63 & 1.62 & 1.56 & 1.49 & 1.37 & 1.21 & 1.00 & 1.01 & 0.81 & 0.74 & 0.90 & 0.61 \\
DNSS-RF-X & 1.61 & 1.63 & 1.64 & 1.59 & 1.53 & 1.44 & 1.30 & 1.09 & 1.07 & 0.88 & 0.66 & 0.86 & 0.59 \\
DNSS-RF-PSO-X & 1.67 & 1.84 & 2.14 & 2.34 & 2.45 & 2.52 & 2.41 & 1.92 & 1.38 & 1.19 & 1.13 & 1.74 & 1.64 \\
\bottomrule
\end{tabular}
\begin{tablenotes}
\footnotesize
\item Notes: This table reports absolute RMSE at the 12-month forecast horizon. Results are grouped by curve segment into the short end, belly, long end, and slope. Panel A reports yields-only specifications, while Panel B reports macro-augmented specifications. GR results are omitted. Statistical significance relative to the random walk benchmark is denoted by stars: *** \(1\%\), ** \(5\%\), and * \(10\%\).
\end{tablenotes}
\end{threeparttable}
\end{table}

\clearpage
\newgeometry{top=0.1cm,bottom=0.1cm,left=0.1cm,right=0.1cm}
\begin{landscape}

\begin{table}[p]
\centering
\caption{Relative RMSE of macro-augmented models against their non-augmented benchmarks}
\label{tab:relative_rmse_macro_models}
\small
\renewcommand{\arraystretch}{0.8}
\begin{threeparttable}
\begin{tabular}{llccccccccccccc}
\toprule
& & \multicolumn{5}{c}{Short-end} & \multicolumn{2}{c}{Belly} & \multicolumn{3}{c}{Long-end} & \multicolumn{3}{c}{Slope} \\
\cmidrule(lr){3-7} \cmidrule(lr){8-9} \cmidrule(lr){10-12} \cmidrule(lr){13-15}
Model & $h$ & 3 & 6 & 12 & 18 & 24 & 36 & 60 & 120 & 240 & 360 & 24\_120 & 24\_360 & 60\_360 \\
\midrule

\multirow{4}{*}{Direct-RF-X}
& 1  & 0.99 & 0.97 & 0.97 & 0.98 & 0.99 & 0.97 & 0.97** & 0.98 & 0.98 & 0.98 & 0.97* & 0.94** & 0.95*** \\
& 3  & 0.93 & 0.92* & 0.93* & 0.94 & 0.96 & 0.98 & 0.95 & 0.93*** & 0.94*** & 0.95** & 0.91** & 0.89*** & 0.94* \\
& 6  & 0.89** & 0.90** & 0.91* & 0.91** & 0.92* & 0.94* & 0.95 & 0.93** & 0.94** & 0.93** & 0.89*** & 0.85*** & 0.90*** \\
& 12 & 0.87*** & 0.88*** & 0.90*** & 0.91*** & 0.91*** & 0.92*** & 0.95* & 0.97 & 0.98 & 1.00 & 0.88*** & 0.82*** & 0.82*** \\
\addlinespace

\multirow{4}{*}{DNS-RF-X}
& 1  & 1.00 & 0.98 & 0.98 & 0.98 & 0.99 & 0.99 & 0.98 & 0.97** & 1.02 & 1.00 & 0.96*** & 0.97* & 1.02 \\
& 3  & 0.90** & 0.88** & 0.89** & 0.90** & 0.91** & 0.93** & 0.94** & 0.95*** & 0.97 & 0.97* & 0.90*** & 0.91*** & 1.00 \\
& 6  & 0.93*** & 0.91*** & 0.90*** & 0.89*** & 0.89*** & 0.89*** & 0.89*** & 0.91*** & 0.95** & 0.95** & 0.89*** & 0.88*** & 0.94** \\
& 12 & 0.96 & 0.95 & 0.94* & 0.94* & 0.94** & 0.94** & 0.94** & 0.96* & 1.00 & 1.00 & 0.89*** & 0.88*** & 0.90*** \\
\addlinespace

\multirow{4}{*}{DNSS-RF-X}
& 1  & 1.00 & 1.03 & 1.02 & 1.00 & 0.99 & 0.98 & 0.97* & 0.96** & 1.03 & 0.95** & 0.96** & 0.92*** & 0.99 \\
& 3  & 0.91** & 0.91* & 0.93 & 0.94 & 0.95 & 0.96 & 0.97 & 0.97 & 1.00 & 0.97 & 0.96 & 0.91** & 0.95 \\
& 6  & 0.98 & 0.98 & 0.98 & 0.99 & 1.00 & 1.01 & 1.02 & 1.02 & 1.06 & 1.06 & 0.91** & 0.87*** & 0.92** \\
& 12 & 0.92*** & 0.91*** & 0.91*** & 0.91*** & 0.92*** & 0.93*** & 0.94*** & 0.96** & 0.98 & 1.00 & 0.84*** & 0.84*** & 0.89*** \\

\bottomrule
\end{tabular}
\begin{tablenotes}
\footnotesize
\item Notes: Entries report RMSE relative to the corresponding non-macro benchmark, so values below one indicate that the macro-augmented model has lower RMSE. Results are grouped by model and forecast horizon. Stars denote statistical significance from the Diebold--Mariano test: *** \(1\%\), ** \(5\%\), and * \(10\%\).
\end{tablenotes}
\end{threeparttable}
\end{table}



\begin{table}[p]
\centering
\caption{Relative RMSE of PSO-based models against their corresponding non-PSO benchmarks}
\label{tab:relative_rmse_pso_models}
\small
\renewcommand{\arraystretch}{0.8}
\begin{threeparttable}
\begin{tabular}{llccccccccccccc}
\toprule
& & \multicolumn{5}{c}{Short-end} & \multicolumn{2}{c}{Belly} & \multicolumn{3}{c}{Long-end} & \multicolumn{3}{c}{Slope} \\
\cmidrule(lr){3-7} \cmidrule(lr){8-9} \cmidrule(lr){10-12} \cmidrule(lr){13-15}
Model & $h$ & 3 & 6 & 12 & 18 & 24 & 36 & 60 & 120 & 240 & 360 & 24\_120 & 24\_360 & 60\_360 \\
\midrule

\multirow{4}{*}{DNSS-AR-PSO}
& 1  & 1.04 & 0.98 & 0.89 & 0.95 & 1.02 & 0.98 & 0.91* & 0.98 & 0.87* & 1.00 & 0.76** & 0.93 & 1.01 \\
& 3  & 1.08 & 1.06 & 1.04 & 1.07 & 1.10 & 1.09 & 1.06 & 1.11 & 1.09 & 1.11 & 0.86 & 0.95 & 1.08 \\
& 6  & 1.10 & 1.09 & 1.09 & 1.12 & 1.14 & 1.14 & 1.11 & 1.13 & 1.12 & 1.09 & 0.86* & 0.90 & 1.05 \\
& 12 & 1.06 & 1.06 & 1.05 & 1.06 & 1.08 & 1.09 & 1.07 & 1.07 & 1.09 & 1.07 & 0.92 & 0.94 & 1.06 \\
\addlinespace

\multirow{4}{*}{DNSS-RF-PSO}
& 1  & 2.34 & 2.28 & 1.75 & 1.81 & 1.92 & 1.85 & 1.75 & 1.88 & 1.56 & 2.03 & 0.92 & 1.07 & 1.18 \\
& 3  & 1.25 & 1.38 & 1.60 & 1.80 & 1.94 & 2.01 & 1.96 & 1.88 & 1.67 & 1.95 & 1.83 & 2.28 & 2.66 \\
& 6  & 0.89* & 0.88* & 0.89* & 0.91 & 0.95 & 0.99 & 1.01 & 1.02 & 0.93 & 1.00 & 0.89** & 0.92 & 1.11 \\
& 12 & 1.05 & 1.02 & 0.98 & 0.97 & 0.98 & 0.98 & 0.98 & 0.93 & 0.82*** & 0.87*** & 1.11 & 1.18 & 1.33 \\
\addlinespace

\bottomrule
\end{tabular}
\begin{tablenotes}
\footnotesize
\item Notes: Entries report RMSE relative to the corresponding non-PSO benchmark, so values below one indicate that the PSO-based model has lower RMSE. Results are grouped by model and forecast horizon. Stars denote statistical significance from the Diebold--Mariano test: *** \(1\%\), ** \(5\%\), and * \(10\%\).
\end{tablenotes}
\end{threeparttable}
\end{table}

\end{landscape}
\restoregeometry
\clearpage

\clearpage
\newgeometry{top=0.1cm,bottom=0.1cm,left=0.1cm,right=0.1cm}
\begin{landscape}
\begin{table}[htbp]
\centering
\caption{Relative RMSE of factor-based models against their corresponding direct-yield benchmark}
\label{tab:approach_relative_rmse}
\small
\begin{tabular}{llccccccccccccc}
\toprule
 & & \multicolumn{5}{c}{Short-end} & \multicolumn{2}{c}{Belly} & \multicolumn{3}{c}{Long-end} & \multicolumn{3}{c}{Slope} \\
\cmidrule(lr){3-7} \cmidrule(lr){8-9} \cmidrule(lr){10-12} \cmidrule(lr){13-15}
Model & $h$ & 3 & 6 & 12 & 18 & 24 & 36 & 60 & 120 & 240 & 360 & 24--120 & 24--360 & 60--360 \\
\midrule
\multicolumn{15}{l}{\textit{Panel A: Yields-only models}} \\

DNS-AR & 1  & 1.12*** & 1.17 & 1.35** & 1.17** & 1.04 & 1.00 & 1.04 & 1.08** & 1.29** & 1.14** & 1.80*** & 1.10* & 1.69*** \\
       & 3  & 1.10**  & 1.19* & 1.22*  & 1.15*  & 1.08 & 1.00 & 0.98 & 0.90** & 1.06 & 0.93 & 1.30** & 1.00 & 1.19** \\
       & 6  & 1.02    & 1.08  & 1.12   & 1.06   & 1.01 & 0.97 & 0.93 & 0.85*** & 0.99 & 0.85** & 1.26* & 1.02 & 0.97 \\
       & 12 & 0.90    & 0.92  & 0.95   & 0.94   & 0.90 & 0.88* & 0.85** & 0.81*** & 0.98 & 0.89 & 1.09 & 0.85 & 0.74*** \\

DNS-RF & 1  & 1.06 & 1.00* & 1.25** & 1.09** & 1.04 & 1.00 & 1.07** & 1.15*** & 1.25** & 1.14*** & 1.87*** & 1.10*** & 1.69*** \\
       & 3  & 1.02 & 1.10** & 1.16** & 1.13** & 1.08** & 1.06* & 1.06 & 1.02 & 1.13** & 1.02 & 1.27*** & 1.03 & 1.26*** \\
       & 6  & 0.93 & 1.00 & 1.08** & 1.07** & 1.06** & 1.06** & 1.08* & 1.01 & 1.10** & 1.00 & 1.16** & 0.98 & 1.05 \\
       & 12 & 0.93** & 0.96* & 1.00 & 1.01 & 1.01 & 1.01 & 1.01 & 0.95* & 1.05 & 0.98* & 1.08** & 0.95** & 0.97 \\

DNSS-AR & 1  & 1.24*** & 1.17** & 1.25** & 1.09 & 1.04 & 1.08* & 1.11* & 1.08 & 1.29*** & 1.09* & 1.33** & 1.10* & 1.00 \\
        & 3  & 1.14**  & 1.17** & 1.11*  & 1.06 & 1.04 & 1.06 & 1.08 & 1.00 & 1.11* & 1.05 & 1.13 & 1.03 & 0.96 \\
        & 6  & 1.00    & 1.01 & 1.01 & 0.98 & 0.96 & 0.99 & 1.04 & 1.00 & 1.07 & 1.07 & 1.19 & 1.05 & 0.90 \\
        & 12 & 0.80*** & 0.81*** & 0.83*** & 0.82*** & 0.82*** & 0.83*** & 0.89* & 0.96 & 1.09* & 1.14 & 1.03 & 0.87 & 0.77*** \\

DNSS-RF & 1  & 1.00 & 1.00 & 1.20*** & 1.04 & 1.04 & 1.04* & 1.04* & 1.04 & 1.29*** & 1.05 & 1.27*** & 1.10* & 1.06** \\
        & 3  & 0.93 & 1.00 & 1.04* & 1.02 & 1.00 & 1.02 & 1.02 & 0.96 & 1.09 & 0.98 & 1.07 & 1.00 & 1.00 \\
        & 6  & 0.98 & 1.04 & 1.08*** & 1.06* & 1.05 & 1.05 & 1.07 & 0.97 & 1.06 & 0.95 & 1.09* & 1.04 & 1.00 \\
        & 12 & 0.97 & 1.00 & 1.05* & 1.06** & 1.06** & 1.08** & 1.09** & 1.05 & 1.15** & 1.06 & 1.03 & 0.95 & 0.96 \\



\midrule
\multicolumn{15}{l}{\textit{Panel B: Macro-augmented models}} \\

DNS-RF-X & 1  & 1.06 & 1.06** & 1.20*** & 1.09** & 1.04 & 1.04** & 1.08** & 1.16*** & 1.30*** & 1.14*** & 1.73*** & 1.16*** & 1.80*** \\
         & 3  & 0.97 & 1.05** & 1.12**  & 1.07   & 1.04 & 1.02   & 1.04*  & 1.04    & 1.16**  & 1.05    & 1.30*** & 1.03    & 1.36*** \\
         & 6  & 0.96 & 1.01   & 1.08    & 1.05   & 1.03 & 1.01   & 1.01   & 0.98    & 1.13**  & 1.02    & 1.18**  & 1.02    & 1.11 \\
         & 12 & 1.03 & 1.04   & 1.05    & 1.05   & 1.04 & 1.02   & 1.00   & 0.94**  & 1.07    & 0.96*   & 1.09    & 1.01    & 1.07* \\

DNSS-RF-X & 1  & 1.00 & 1.06 & 1.20*** & 1.09 & 1.00 & 1.04** & 1.08** & 1.04 & 1.39*** & 1.00 & 1.27*** & 1.05 & 1.07*** \\
          & 3  & 0.92 & 1.00 & 1.07*   & 1.00 & 1.00 & 1.00   & 1.02   & 1.00 & 1.14**  & 0.98 & 1.15**  & 1.00 & 1.04 \\
          & 6  & 1.08 & 1.12*** & 1.19*** & 1.16*** & 1.15*** & 1.14** & 1.13** & 1.08 & 1.21** & 1.07 & 1.11 & 1.04 & 1.00 \\
          & 12 & 1.02 & 1.04 & 1.06*   & 1.07** & 1.07** & 1.07*  & 1.07*  & 1.03 & 1.14*** & 1.05 & 0.97 & 0.97 & 1.04 \\


\bottomrule
\end{tabular}

\vspace{0.5em}
\begin{minipage}{0.95\linewidth}
\footnotesize
\textit{Notes:} Entries report RMSE relative to the corresponding direct-yield benchmark, so values below one indicate that the direct-yield model has lower RMSE. Results are grouped by model and forecast horizon. Stars denote statistical significance from the Diebold--Mariano test: *** 1\%, ** 5\%, and * 10\%.
\end{minipage}
\end{table}
\end{landscape}
\restoregeometry
\clearpage

\clearpage
\newgeometry{top=0.1cm,bottom=0.1cm,left=0.1cm,right=0.1cm}
\begin{landscape}

\begin{table}[p]
\centering
\caption{Relative RMSE of macro-augmented models against their non-augmented benchmarks}
\label{tab:relative_rmse_macro_models_appendix}
\small
\renewcommand{\arraystretch}{0.8}
\begin{threeparttable}
\begin{tabular}{llccccccccccccc}
\toprule
& & \multicolumn{5}{c}{Short-end} & \multicolumn{2}{c}{Belly} & \multicolumn{3}{c}{Long-end} & \multicolumn{3}{c}{Slope} \\
\cmidrule(lr){3-7} \cmidrule(lr){8-9} \cmidrule(lr){10-12} \cmidrule(lr){13-15}
Model & $h$ & 3 & 6 & 12 & 18 & 24 & 36 & 60 & 120 & 240 & 360 & 24\_120 & 24\_360 & 60\_360 \\
\midrule

\multirow{4}{*}{Direct-RF-X}
& 1  & 0.99 & 0.97 & 0.97 & 0.98 & 0.99 & 0.97 & 0.97** & 0.98 & 0.98 & 0.98 & 0.97* & 0.94** & 0.95*** \\
& 3  & 0.93 & 0.92* & 0.93* & 0.94 & 0.96 & 0.98 & 0.95 & 0.93*** & 0.94*** & 0.95** & 0.91** & 0.89*** & 0.94* \\
& 6  & 0.89** & 0.90** & 0.91* & 0.91** & 0.92* & 0.94* & 0.95 & 0.93** & 0.94** & 0.93** & 0.89*** & 0.85*** & 0.90*** \\
& 12 & 0.87*** & 0.88*** & 0.90*** & 0.91*** & 0.91*** & 0.92*** & 0.95* & 0.97 & 0.98 & 1.00 & 0.88*** & 0.82*** & 0.82*** \\
\addlinespace

\multirow{4}{*}{DNS-RF-X}
& 1  & 1.00 & 0.98 & 0.98 & 0.98 & 0.99 & 0.99 & 0.98 & 0.97** & 1.02 & 1.00 & 0.96*** & 0.97* & 1.02 \\
& 3  & 0.90** & 0.88** & 0.89** & 0.90** & 0.91** & 0.93** & 0.94** & 0.95*** & 0.97 & 0.97* & 0.90*** & 0.91*** & 1.00 \\
& 6  & 0.93*** & 0.91*** & 0.90*** & 0.89*** & 0.89*** & 0.89*** & 0.89*** & 0.91*** & 0.95** & 0.95** & 0.89*** & 0.88*** & 0.94** \\
& 12 & 0.96 & 0.95 & 0.94* & 0.94* & 0.94** & 0.94** & 0.94** & 0.96* & 1.00 & 1.00 & 0.89*** & 0.88*** & 0.90*** \\
\addlinespace

\multirow{4}{*}{DNSS-RF-X}
& 1  & 1.00 & 1.03 & 1.02 & 1.00 & 0.99 & 0.98 & 0.97* & 0.96** & 1.03 & 0.95** & 0.96** & 0.92*** & 0.99 \\
& 3  & 0.91** & 0.91* & 0.93 & 0.94 & 0.95 & 0.96 & 0.97 & 0.97 & 1.00 & 0.97 & 0.96 & 0.91** & 0.95 \\
& 6  & 0.98 & 0.98 & 0.98 & 0.99 & 1.00 & 1.01 & 1.02 & 1.02 & 1.06 & 1.06 & 0.91** & 0.87*** & 0.92** \\
& 12 & 0.92*** & 0.91*** & 0.91*** & 0.91*** & 0.92*** & 0.93*** & 0.94*** & 0.96** & 0.98 & 1.00 & 0.84*** & 0.84*** & 0.89*** \\
\addlinespace

\multirow{4}{*}{DNSS-RF-PSO-X}
& 1  & 0.81 & 0.81 & 0.86 & 0.84 & 0.84 & 0.89 & 1.00 & 0.96 & 0.80 & 0.69* & 1.36 & 1.13 & 1.42 \\
& 3  & 1.15 & 1.10 & 1.05 & 1.05 & 1.06 & 1.10 & 1.15 & 1.15 & 1.00 & 0.92 & 0.89 & 0.87* & 1.02 \\
& 6  & 1.11 & 1.24 & 1.54 & 1.79 & 1.99 & 2.24 & 2.45 & 2.32 & 1.75 & 1.60 & 1.87 & 2.29 & 3.06 \\
& 12 & 0.91 & 1.01 & 1.21 & 1.37 & 1.50 & 1.65 & 1.78 & 1.80 & 1.54 & 1.55 & 1.29 & 1.43 & 1.83 \\
\bottomrule
\end{tabular}
\begin{tablenotes}
\footnotesize
\item Notes: Entries report RMSE relative to the corresponding non-macro benchmark, so values below one indicate that the macro-augmented model has lower RMSE. Results are grouped by model and forecast horizon. Stars denote statistical significance from the Diebold--Mariano test: *** \(1\%\), ** \(5\%\), and * \(10\%\).
\end{tablenotes}
\end{threeparttable}
\end{table}

\end{landscape}
\restoregeometry

\subsection{Variable Importance}

\begin{figure}[H]
\centering
\begin{tabular}{ccc}
\includegraphics[width=0.32\textwidth]{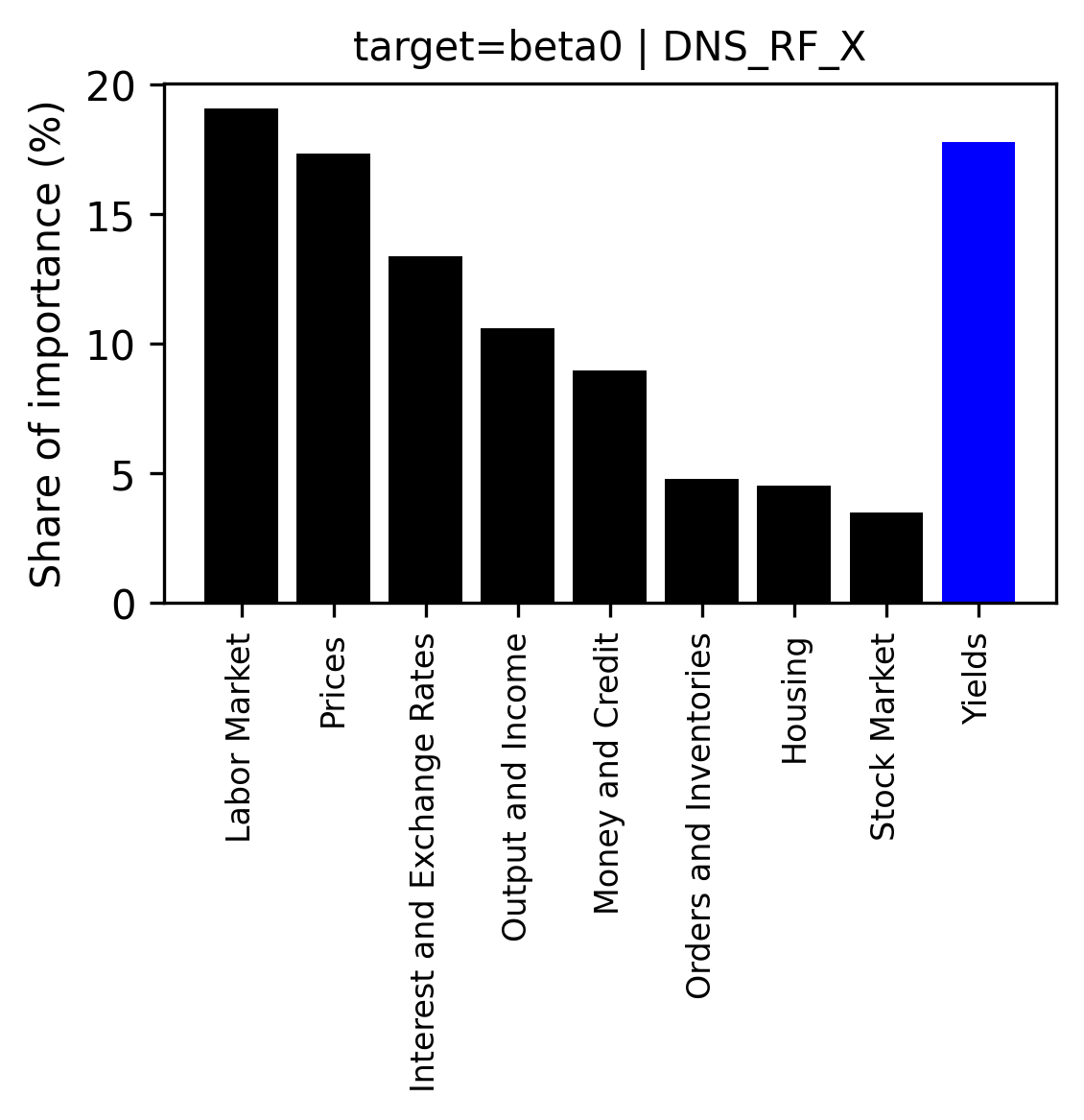} &
\includegraphics[width=0.32\textwidth]{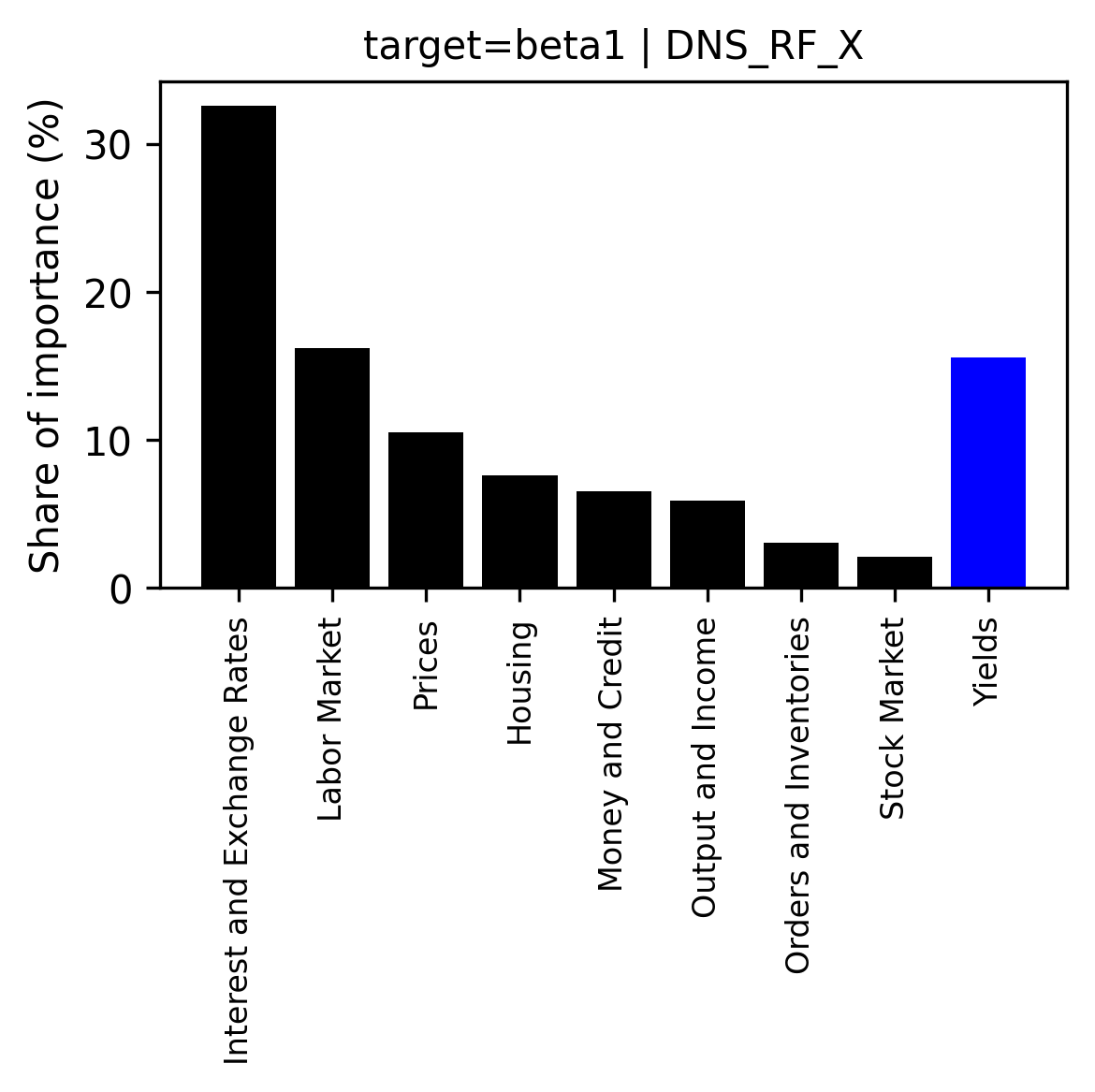} &
\includegraphics[width=0.32\textwidth]{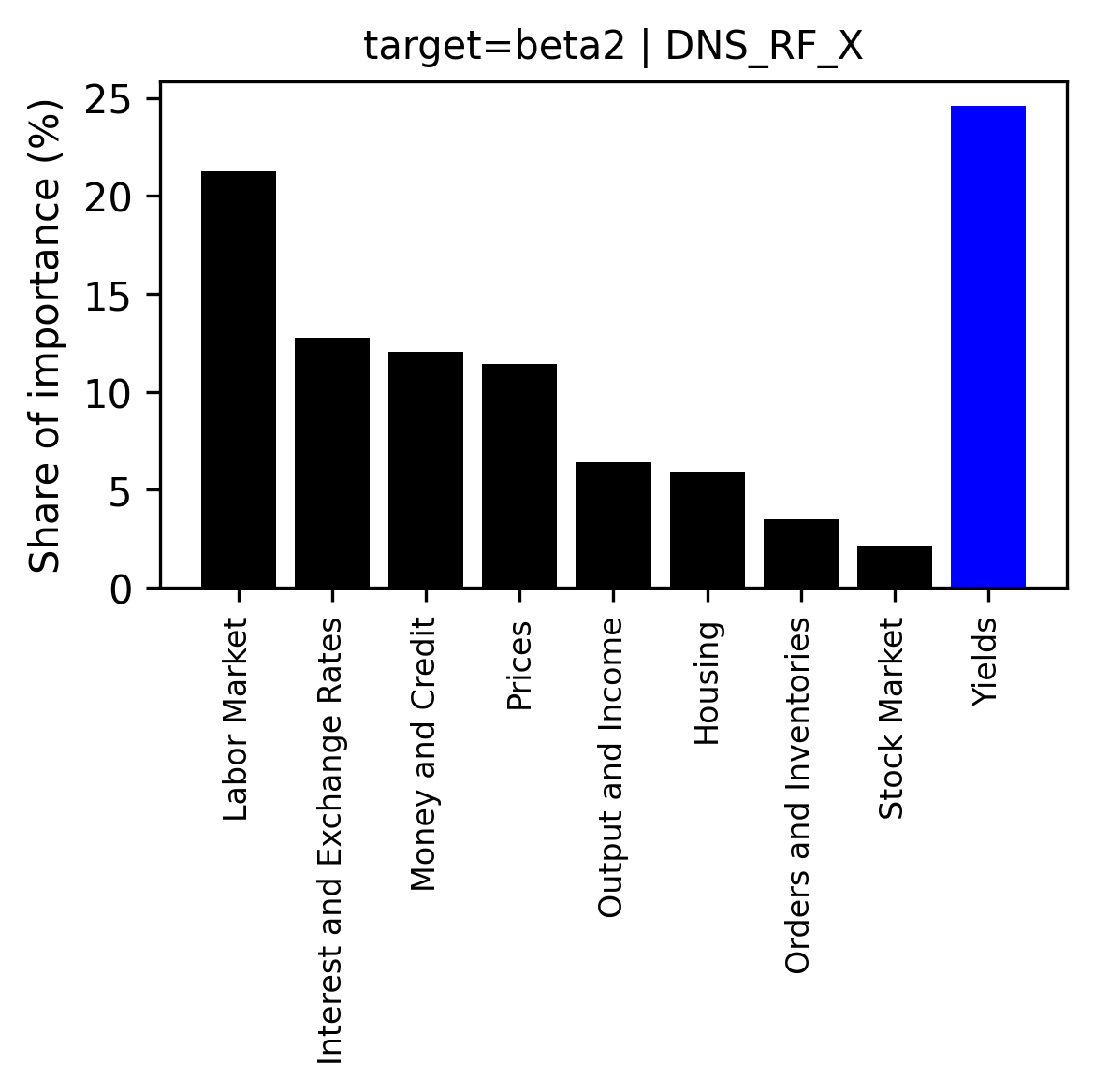} \\
\end{tabular}
\caption{Predictor importance shares, averaged across forecast horizons, for DNS-RF-X across factor targets.}
\label{fig:varimp_dns_rf_x}
\end{figure}

\begin{figure}[H]
\centering
\begin{tabular}{ccc}
\includegraphics[width=0.31\textwidth]{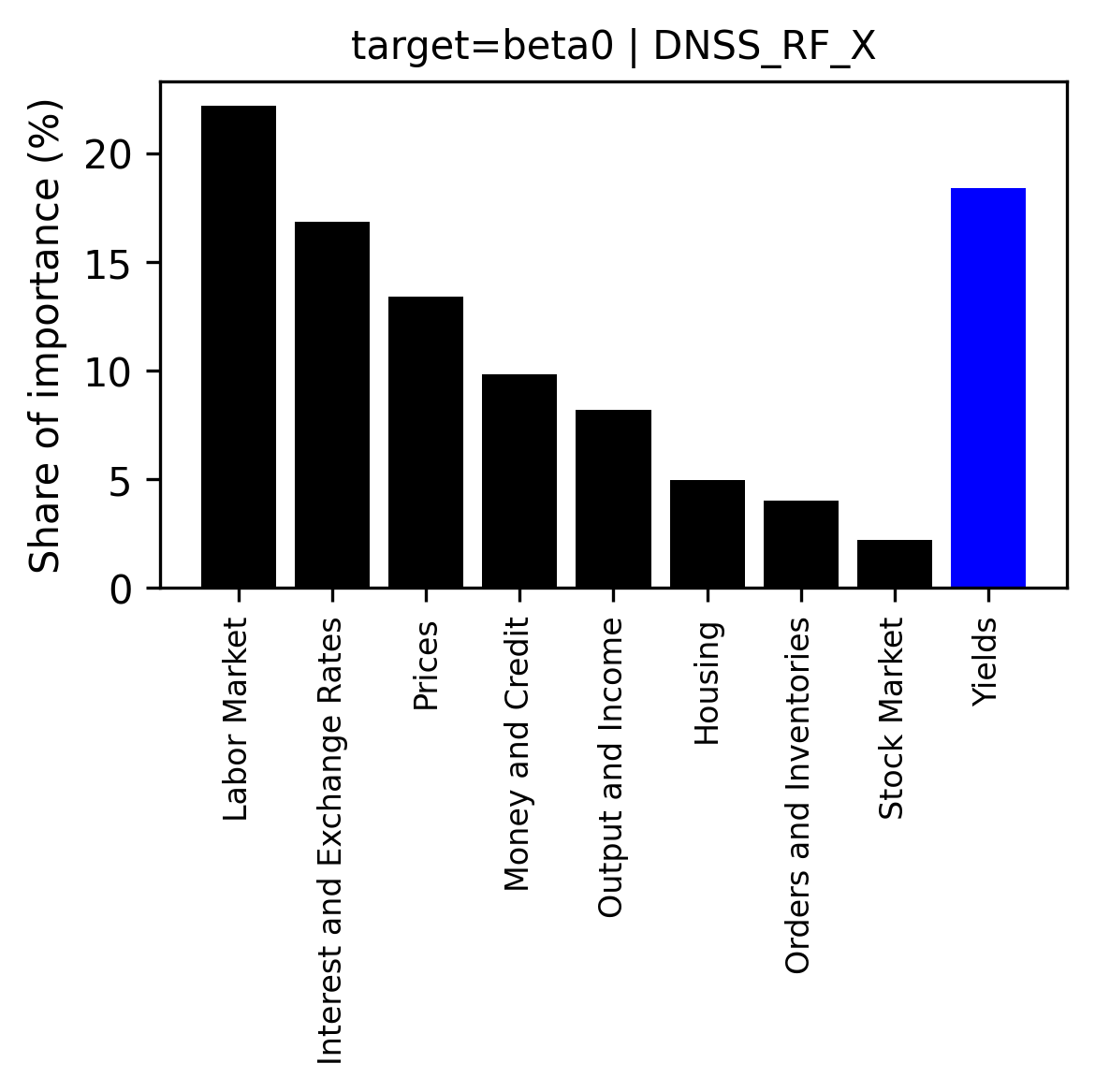} &
\includegraphics[width=0.31\textwidth]{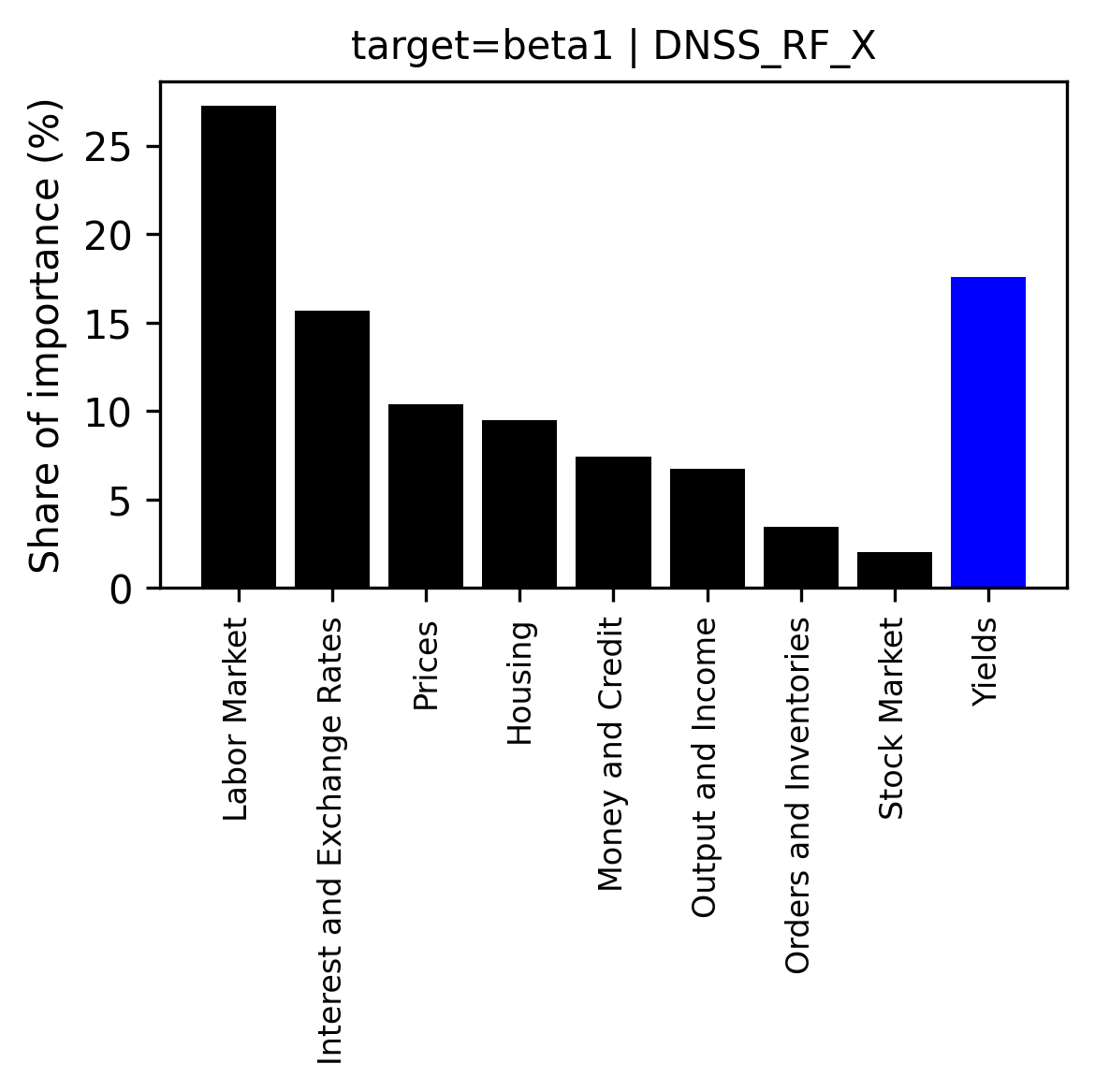} &
\includegraphics[width=0.31\textwidth]{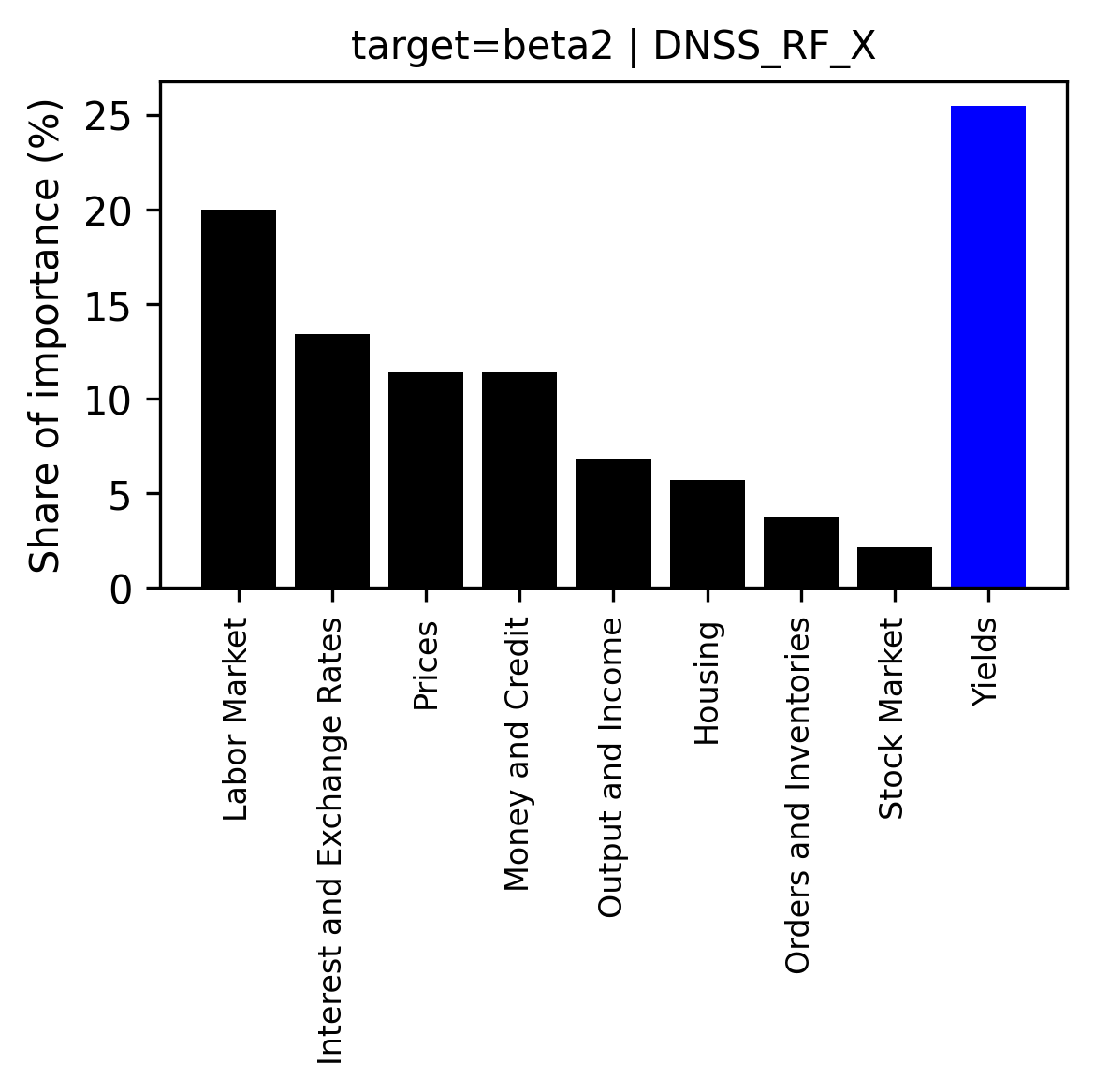} \\[0.6em]
\multicolumn{3}{c}{
\includegraphics[width=0.31\textwidth]{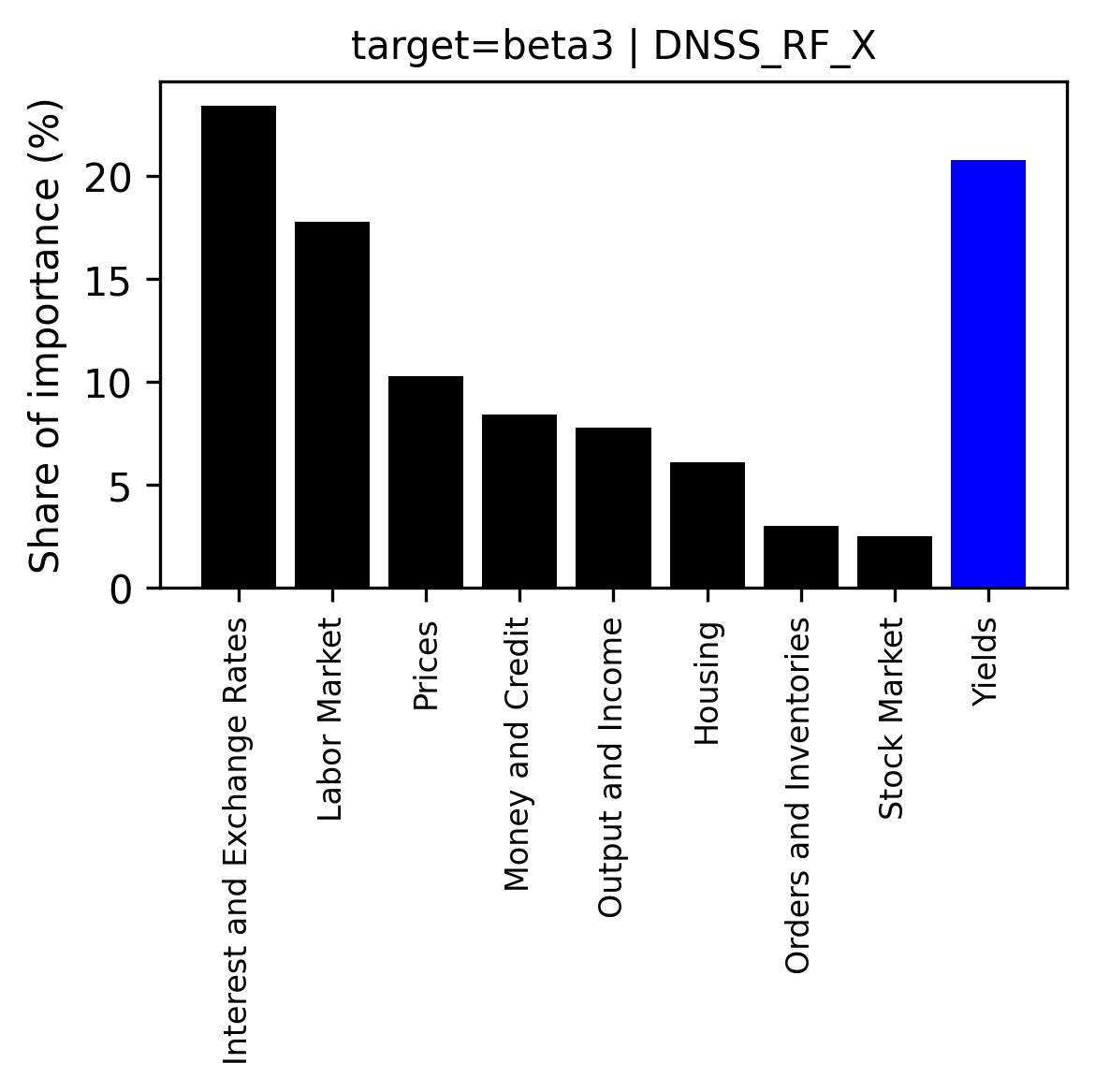}
}
\end{tabular}
\caption{Predictor importance shares, averaged across forecast horizons, for DNSS-RF-X across factor targets.}
\label{fig:varimp_dnss_rf_x}
\end{figure}

\begin{figure}[H]
\centering
\begin{tabular}{ccc}
\includegraphics[width=0.30\textwidth]{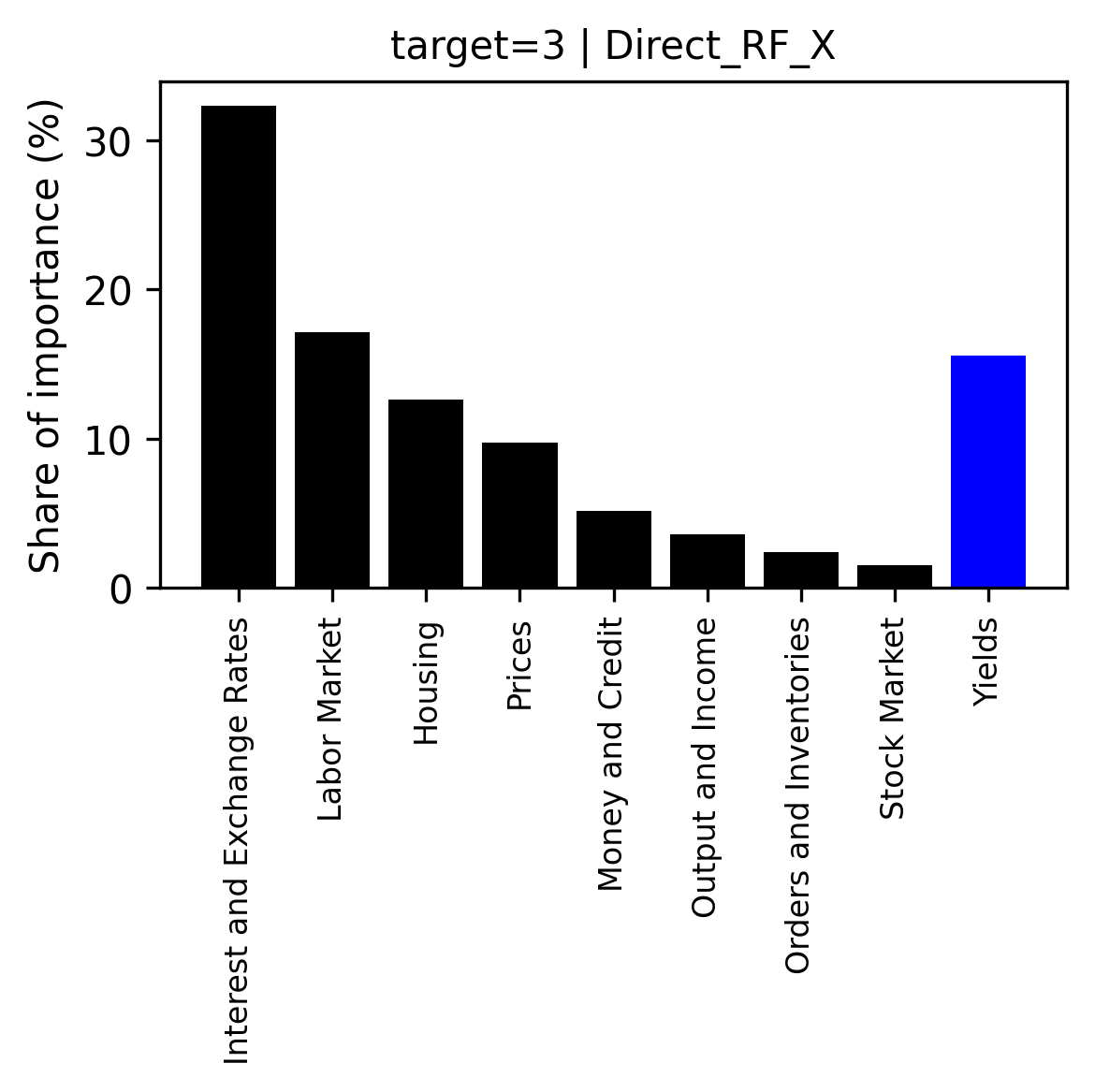} &
\includegraphics[width=0.30\textwidth]{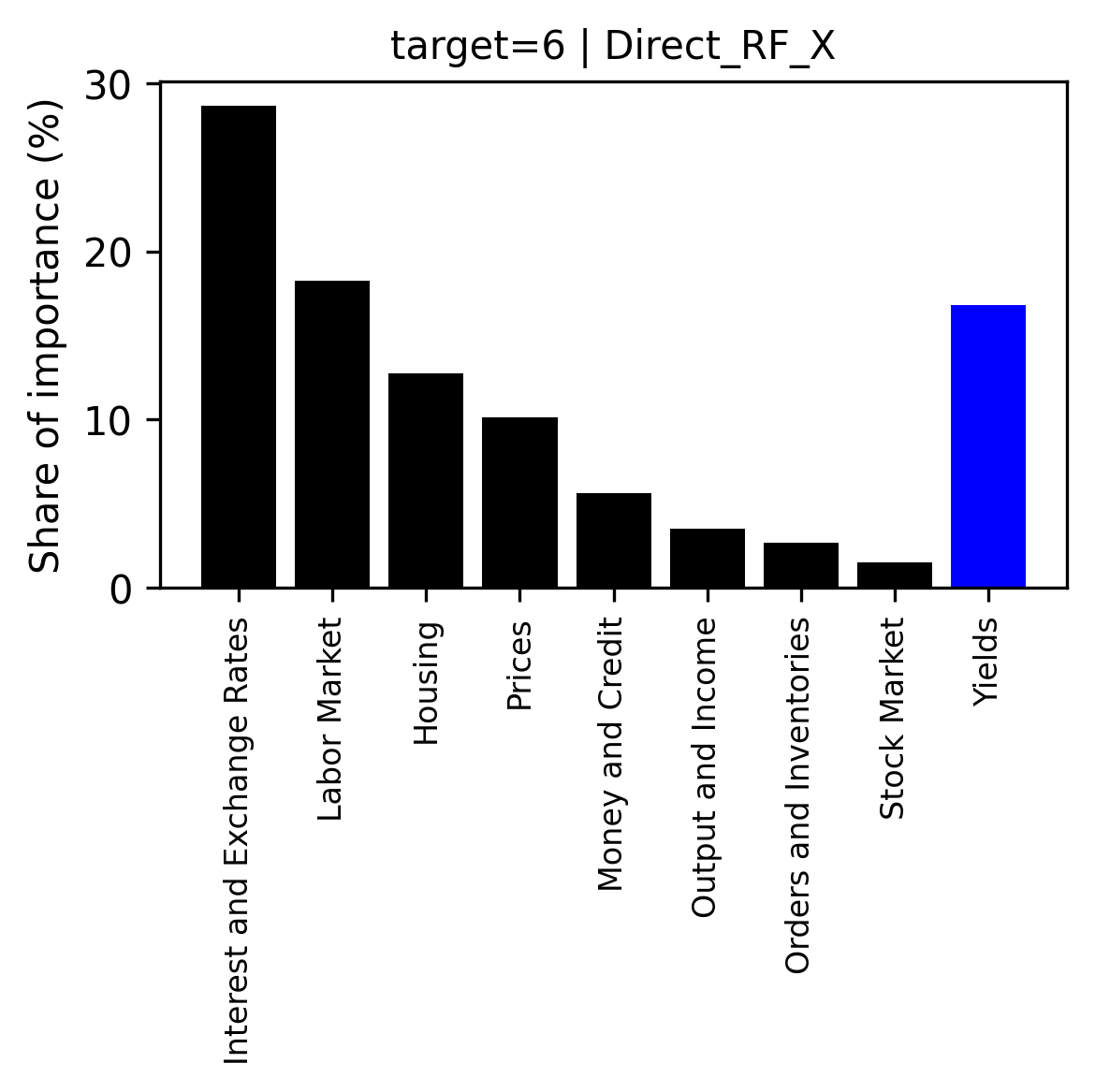} &
\includegraphics[width=0.30\textwidth]{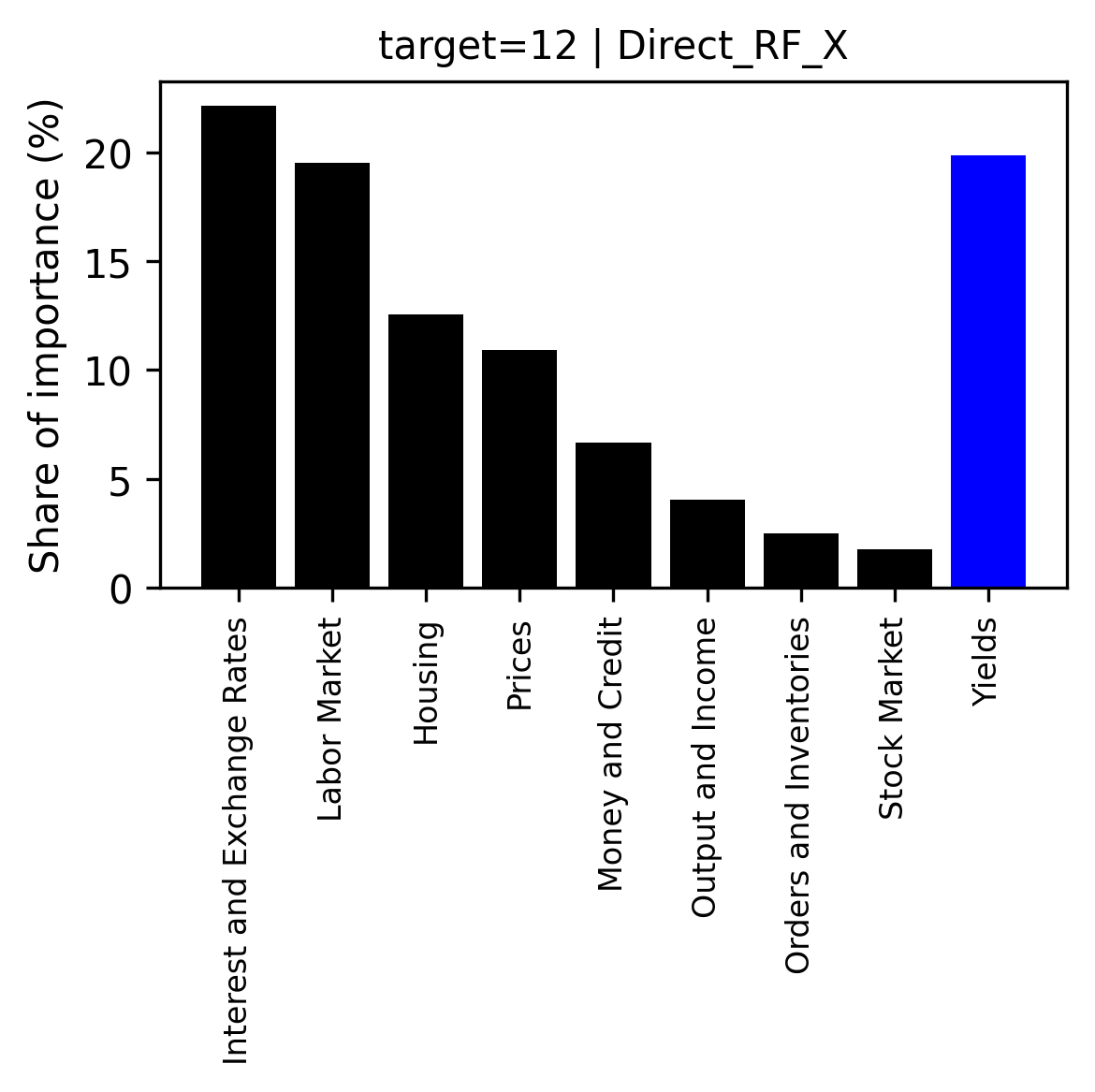} \\[0.5em]

\includegraphics[width=0.30\textwidth]{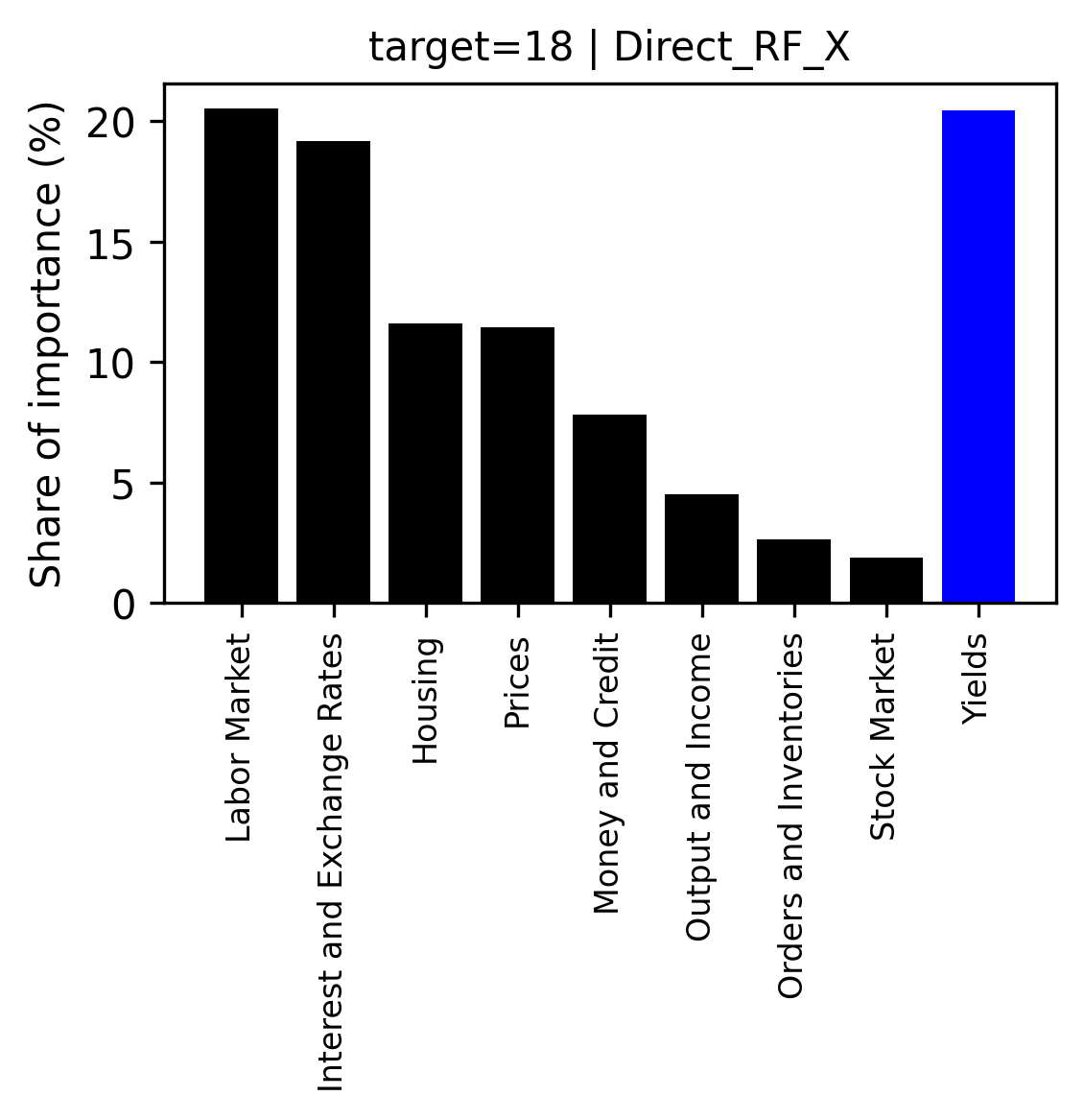} &
\includegraphics[width=0.30\textwidth]{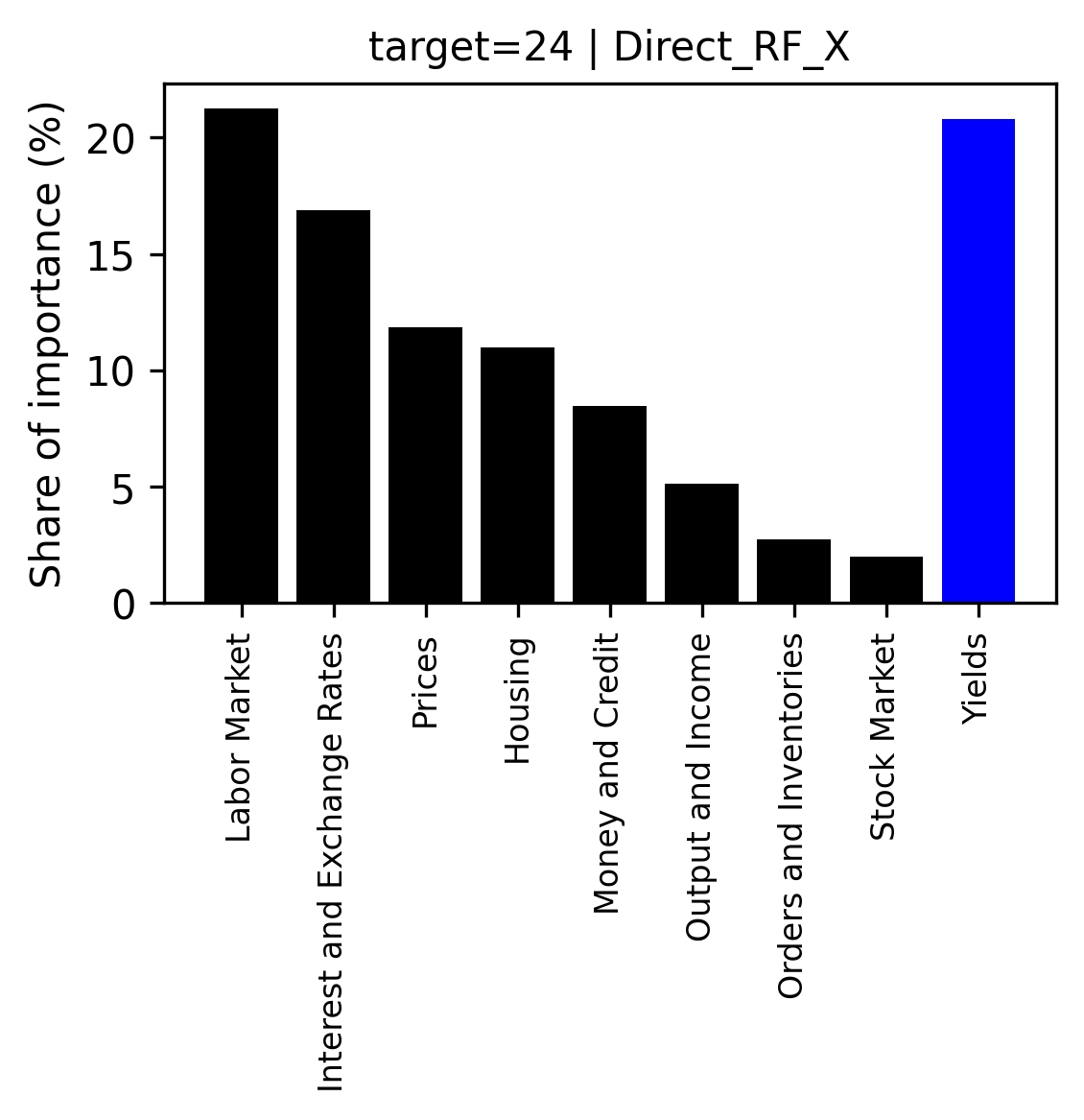} &
\includegraphics[width=0.30\textwidth]{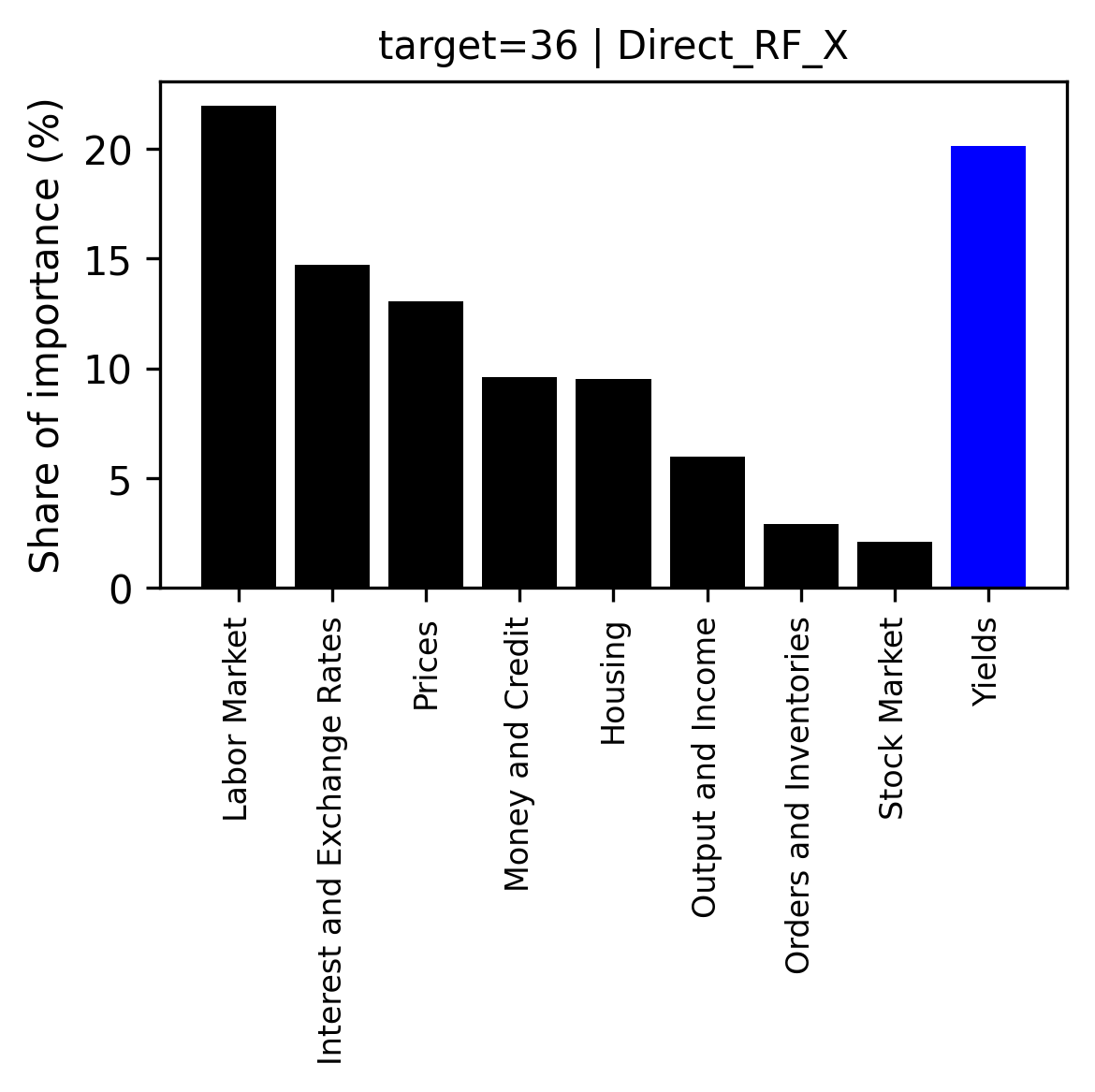} \\[0.5em]

\includegraphics[width=0.30\textwidth]{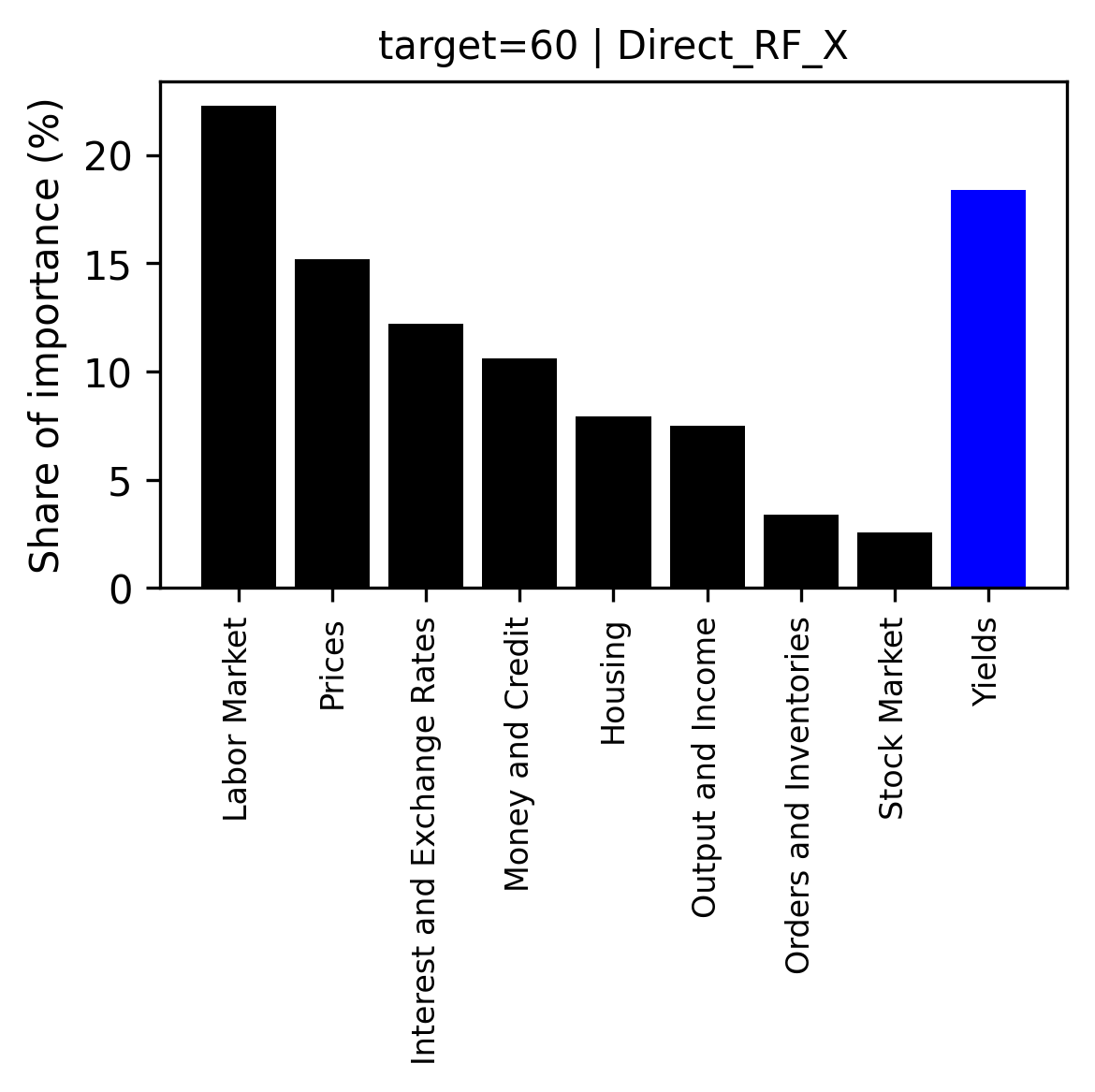} &
\includegraphics[width=0.30\textwidth]{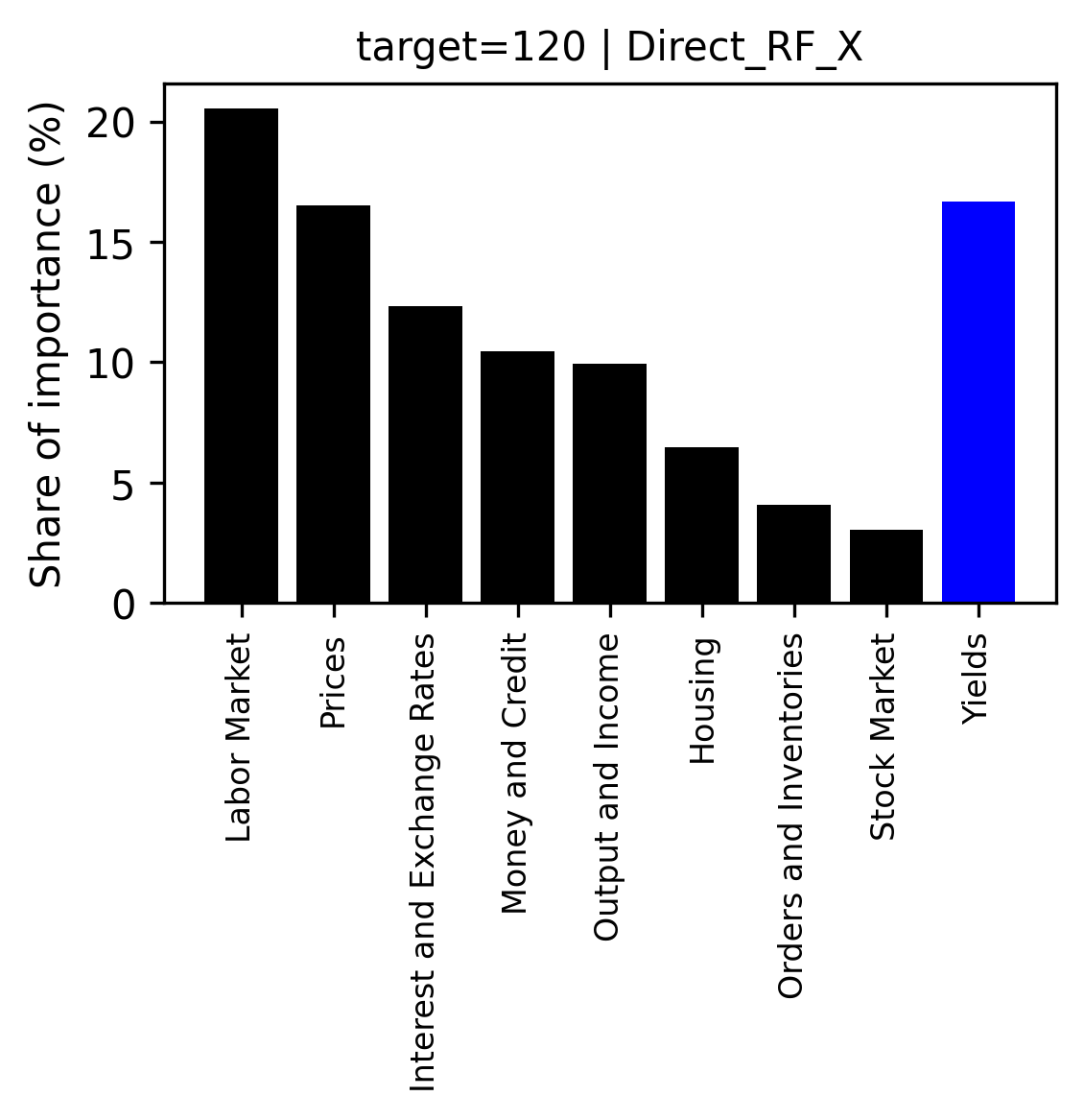} &
\includegraphics[width=0.30\textwidth]{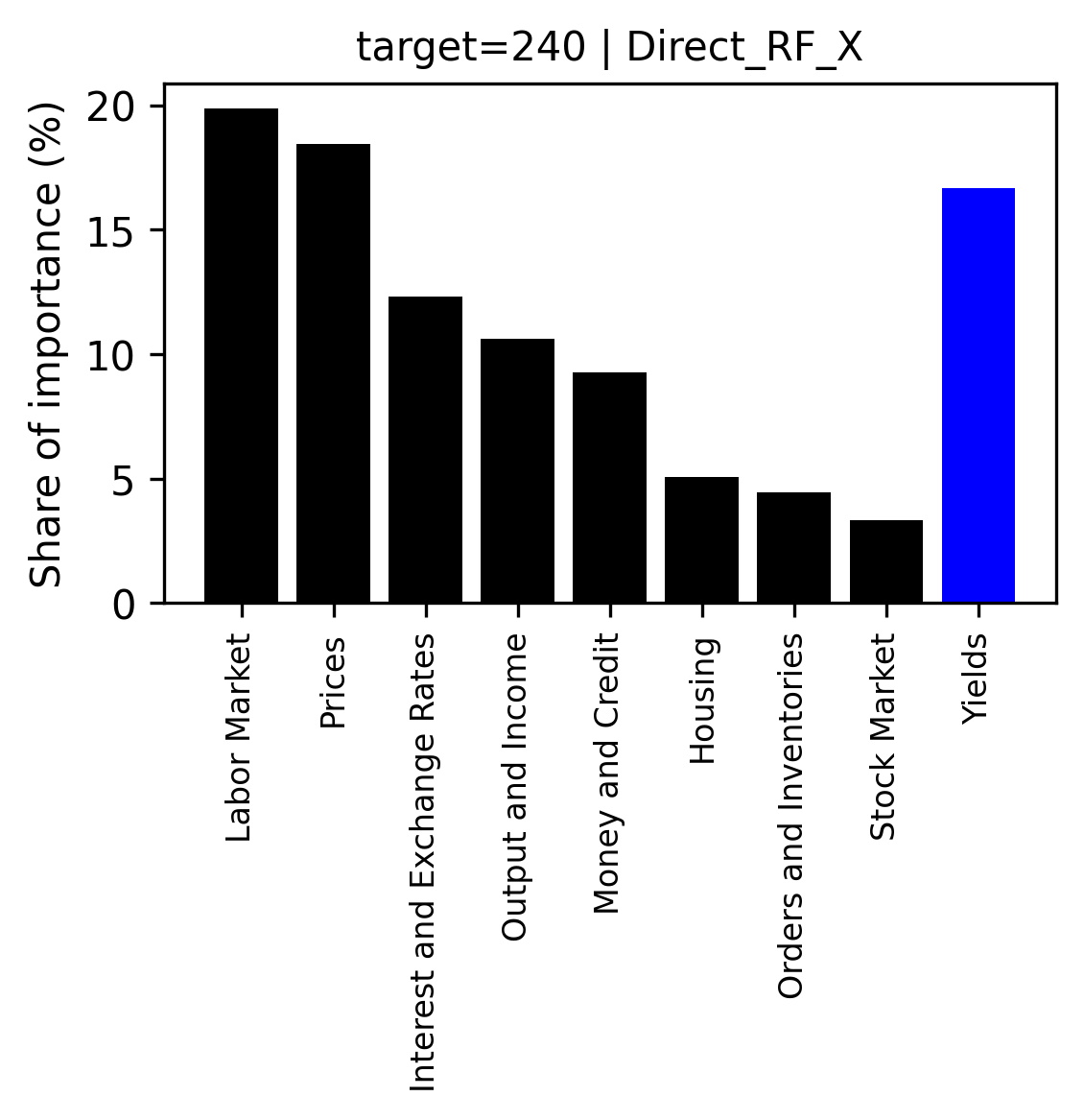} \\[0.5em]

\includegraphics[width=0.30\textwidth]{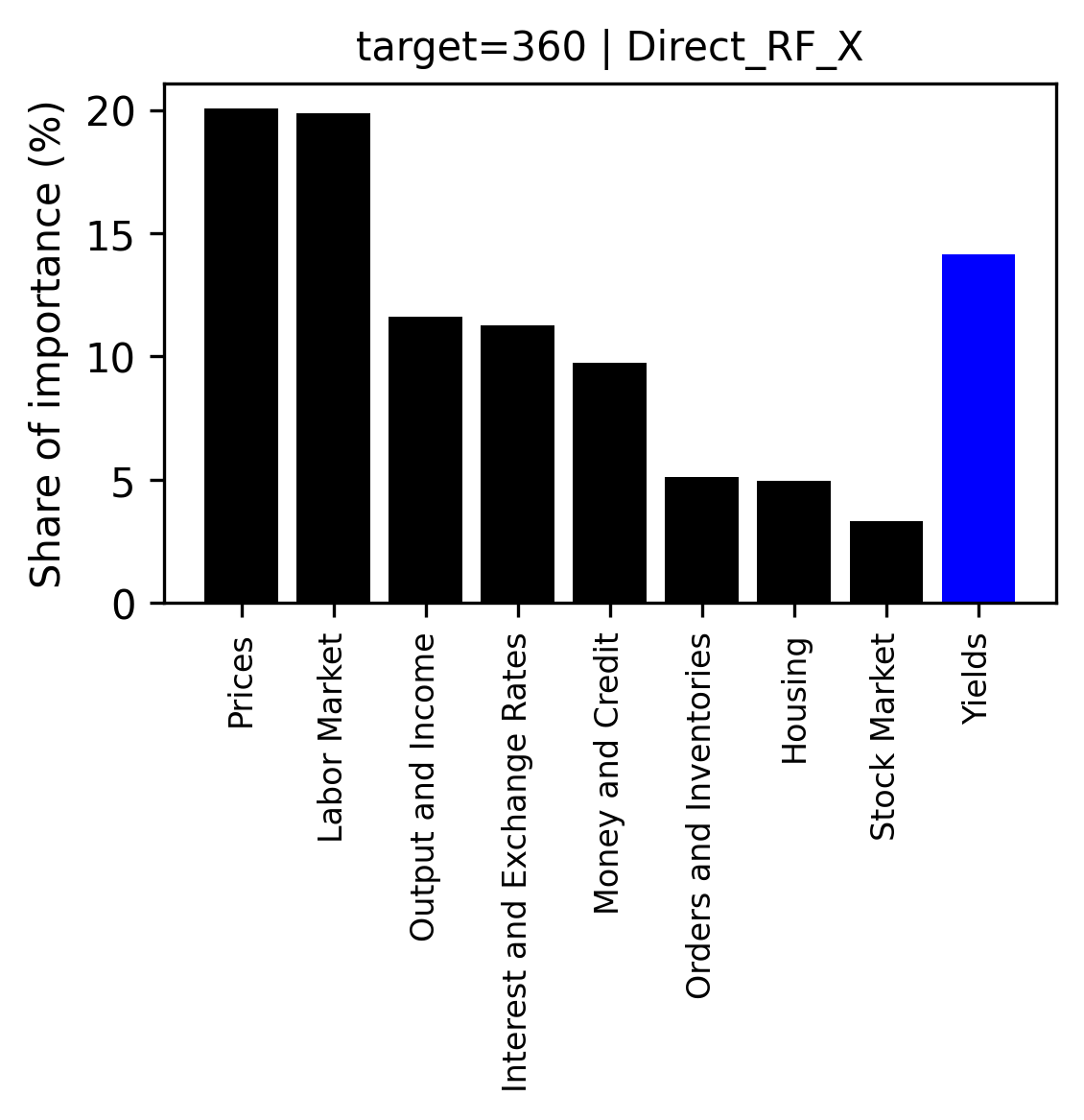} &
\includegraphics[width=0.30\textwidth]{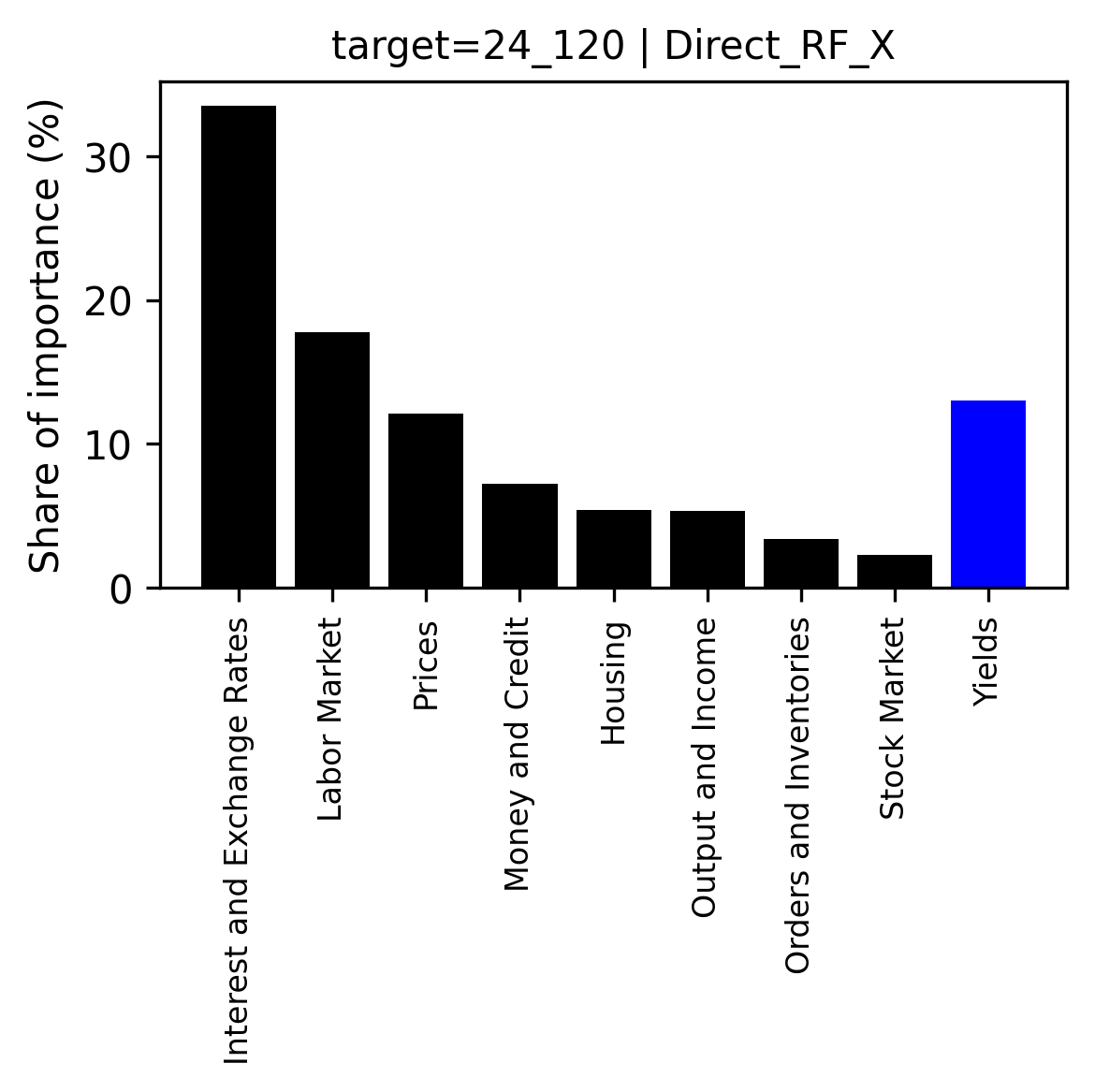} &
\includegraphics[width=0.30\textwidth]{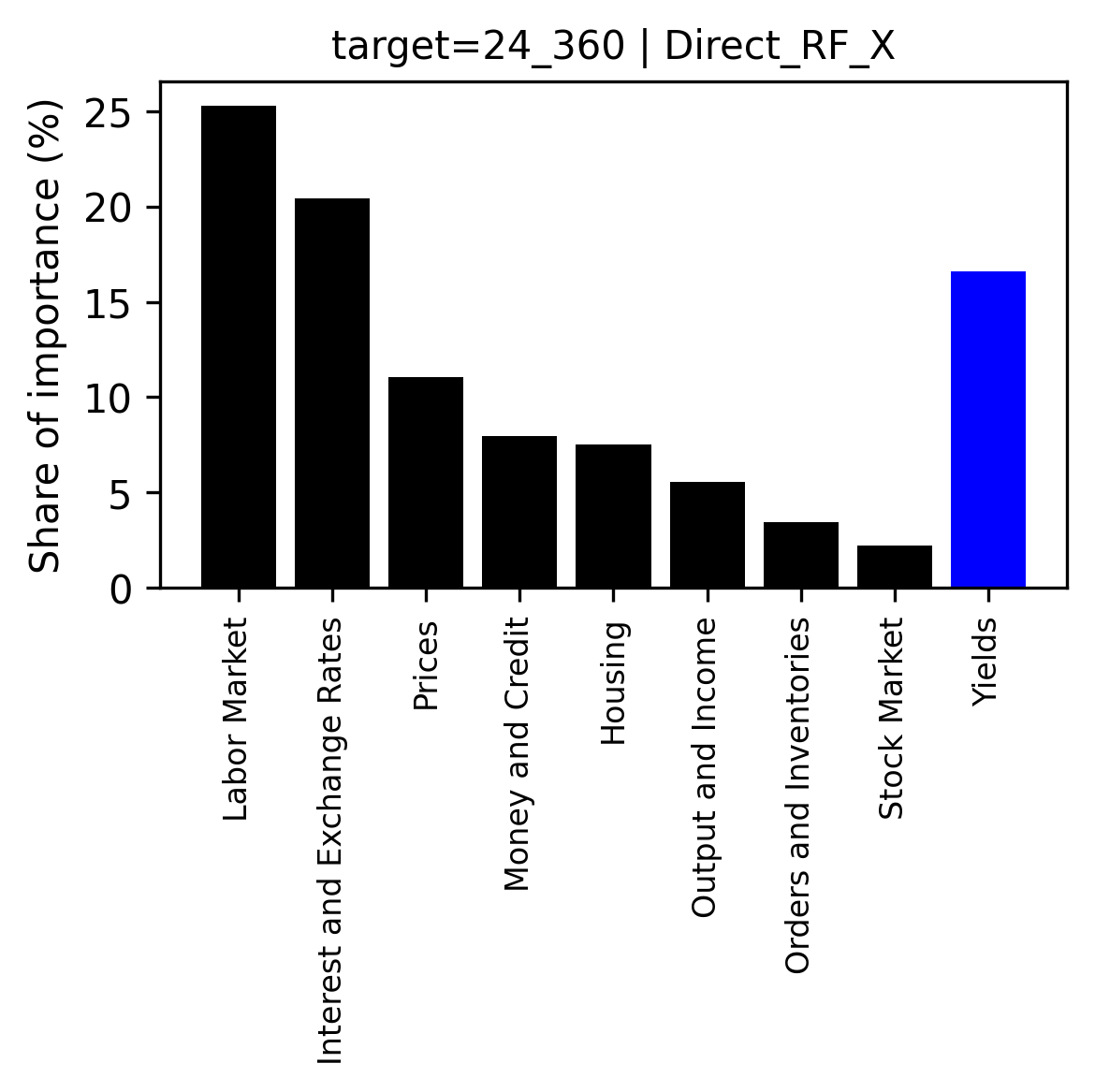} \\[0.5em]

\multicolumn{3}{c}{
\includegraphics[width=0.30\textwidth]{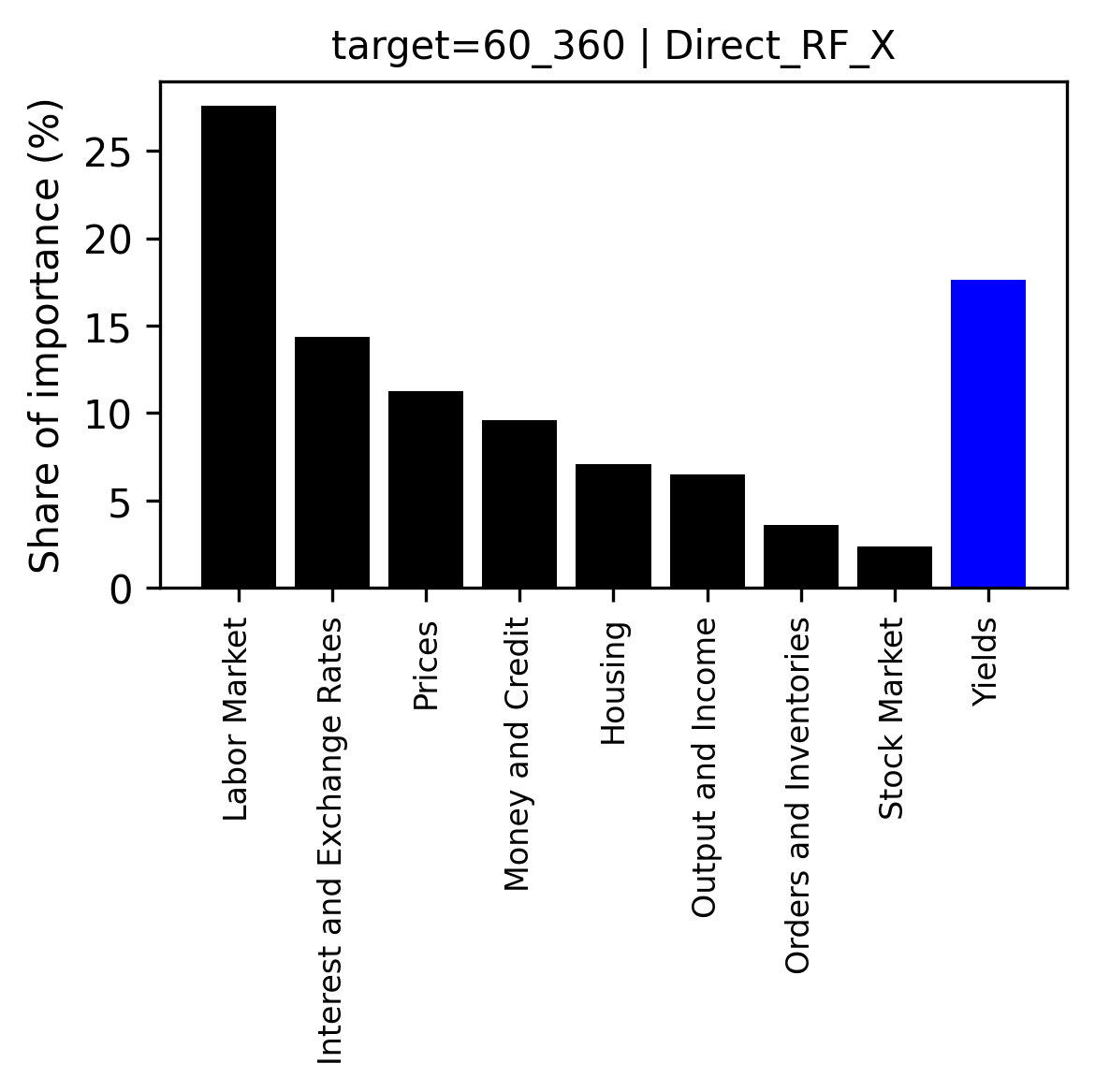}
}
\end{tabular}
\caption{Predictor importance shares, averaged across forecast horizons, for the Direct-RF-X model across yield and slope targets.}
\label{fig:varimp_direct_rf_x}
\end{figure}

\begin{landscape}
\begin{figure}[p]
\centering
\includegraphics[width=\linewidth]{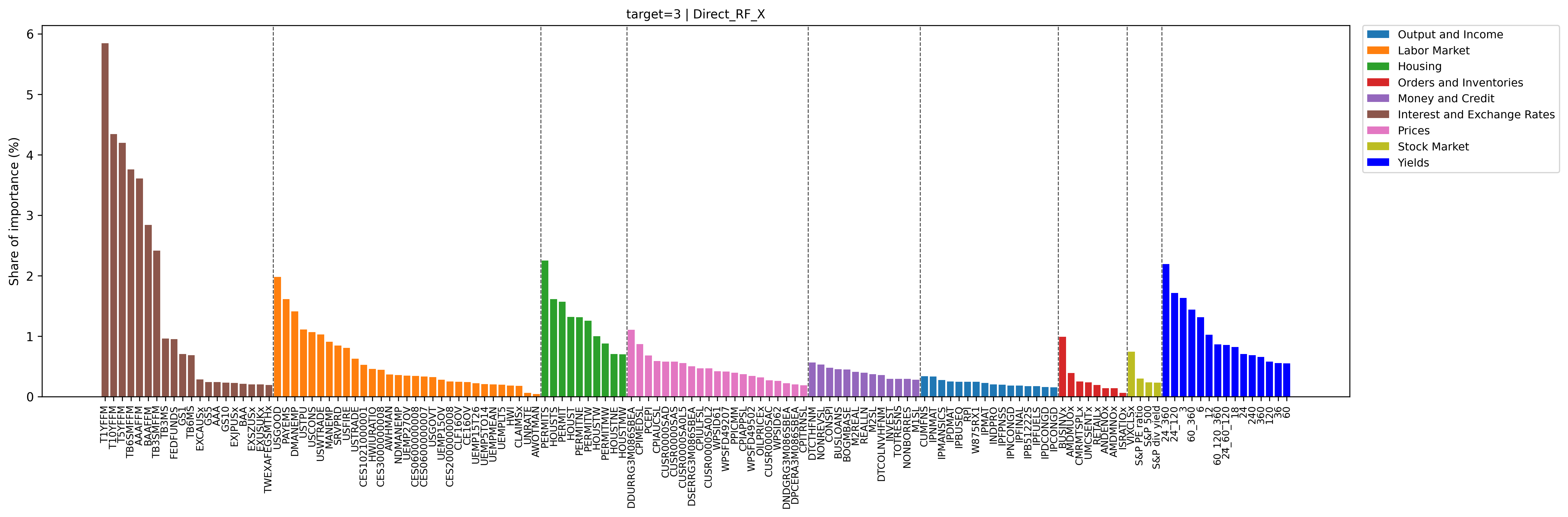}\\[0.5em]
\includegraphics[width=\linewidth]{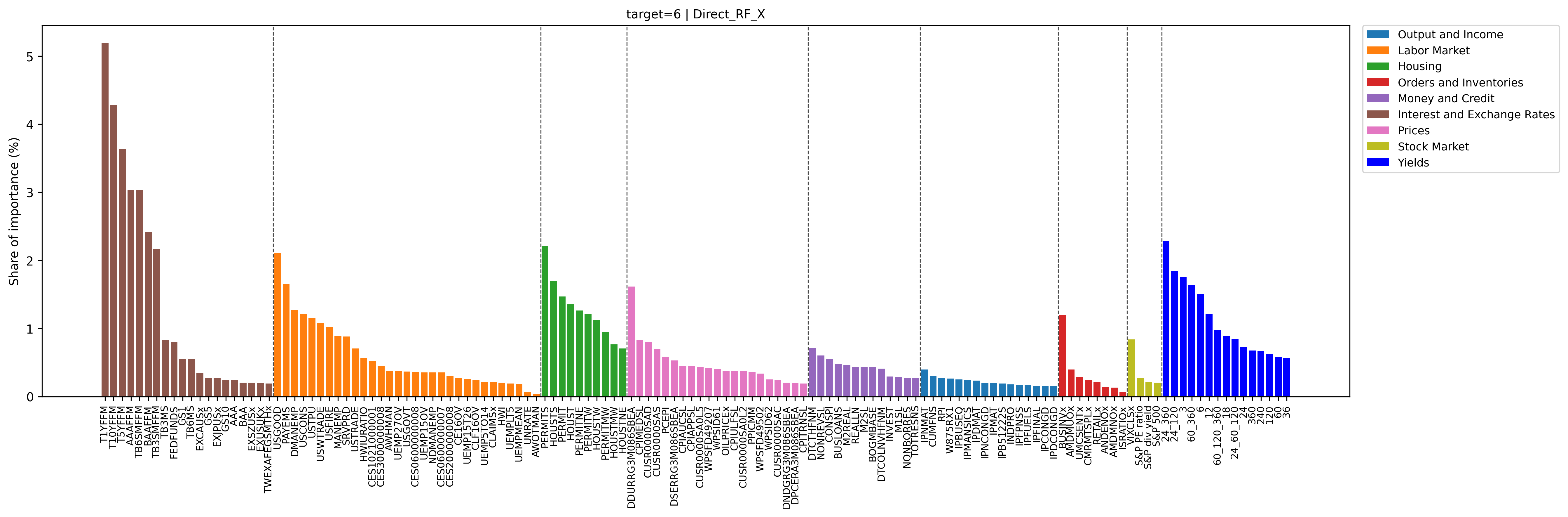}
\caption{Macro-variable importance profiles for the Direct-RF-X model across maturity and slope targets.}
\label{fig:macro_indiv_direct_rf_x}
\end{figure}
\end{landscape}

\begin{landscape}
\begin{figure}[p]\ContinuedFloat
\centering
\includegraphics[width=\linewidth]{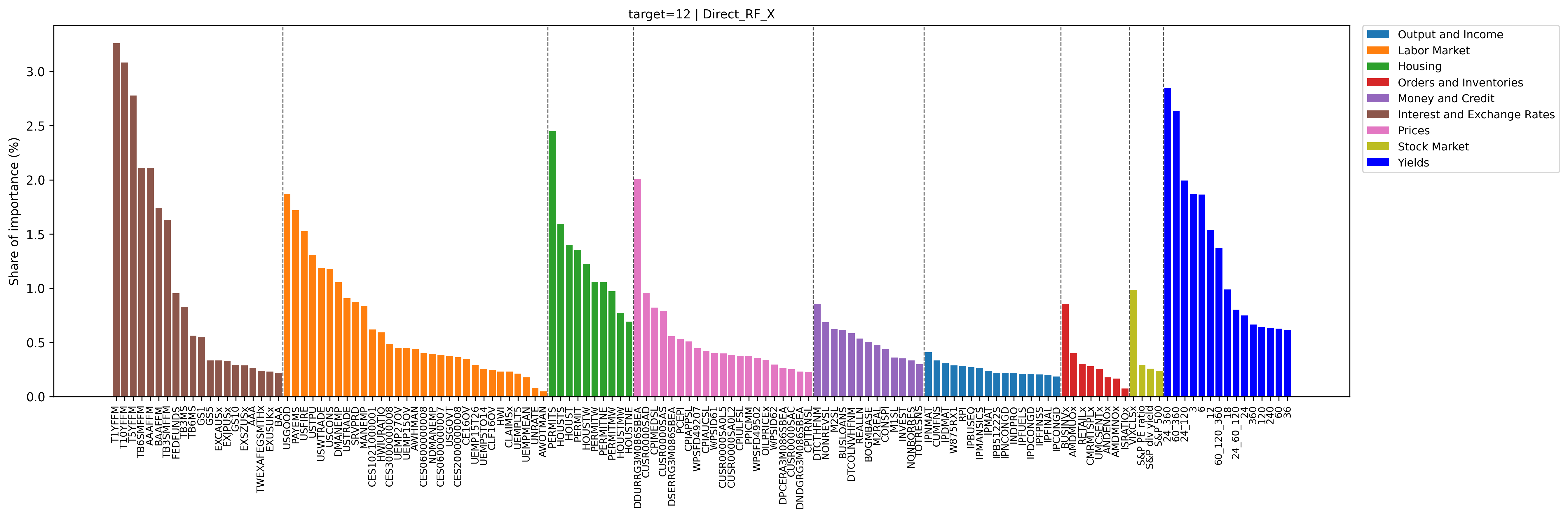}\\[0.5em]
\includegraphics[width=\linewidth]{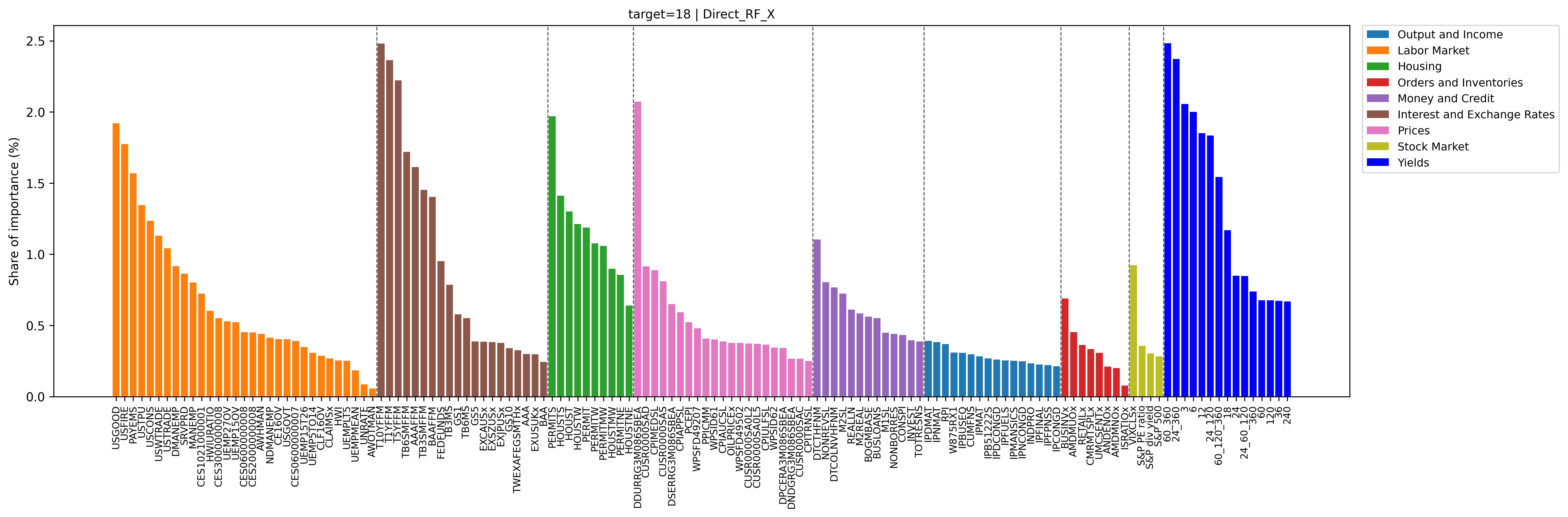}
\caption{Macro-variable importance profiles for the Direct-RF-X model across maturity and slope targets (continued).}
\end{figure}
\end{landscape}

\begin{landscape}
\begin{figure}[p]\ContinuedFloat
\centering
\includegraphics[width=\linewidth]{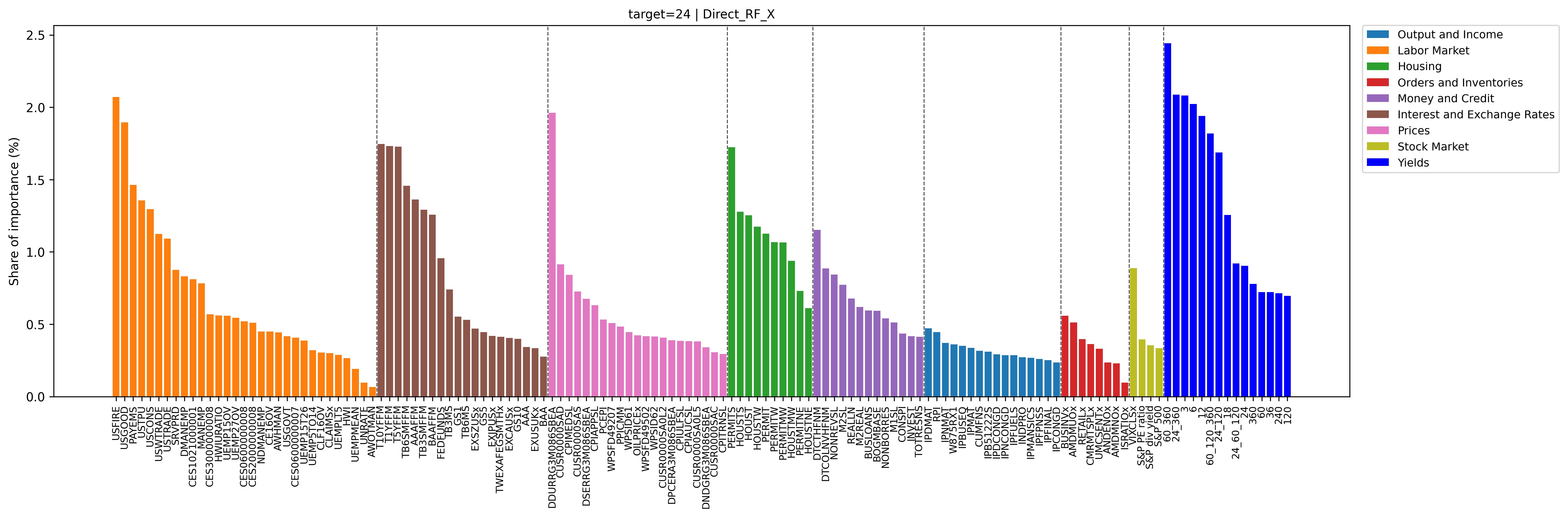}\\[0.5em]
\includegraphics[width=\linewidth]{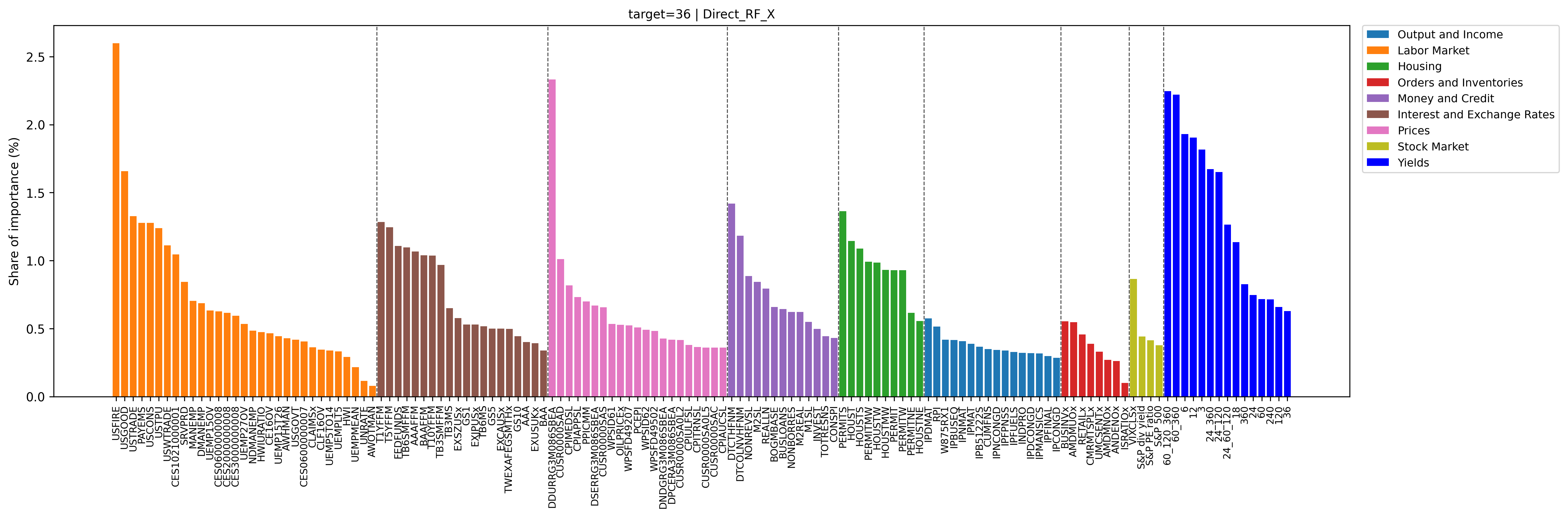}
\caption{Macro-variable importance profiles for the Direct-RF-X model across maturity and slope targets (continued).}
\end{figure}
\end{landscape}

\begin{landscape}
\begin{figure}[p]\ContinuedFloat
\centering
\includegraphics[width=\linewidth]{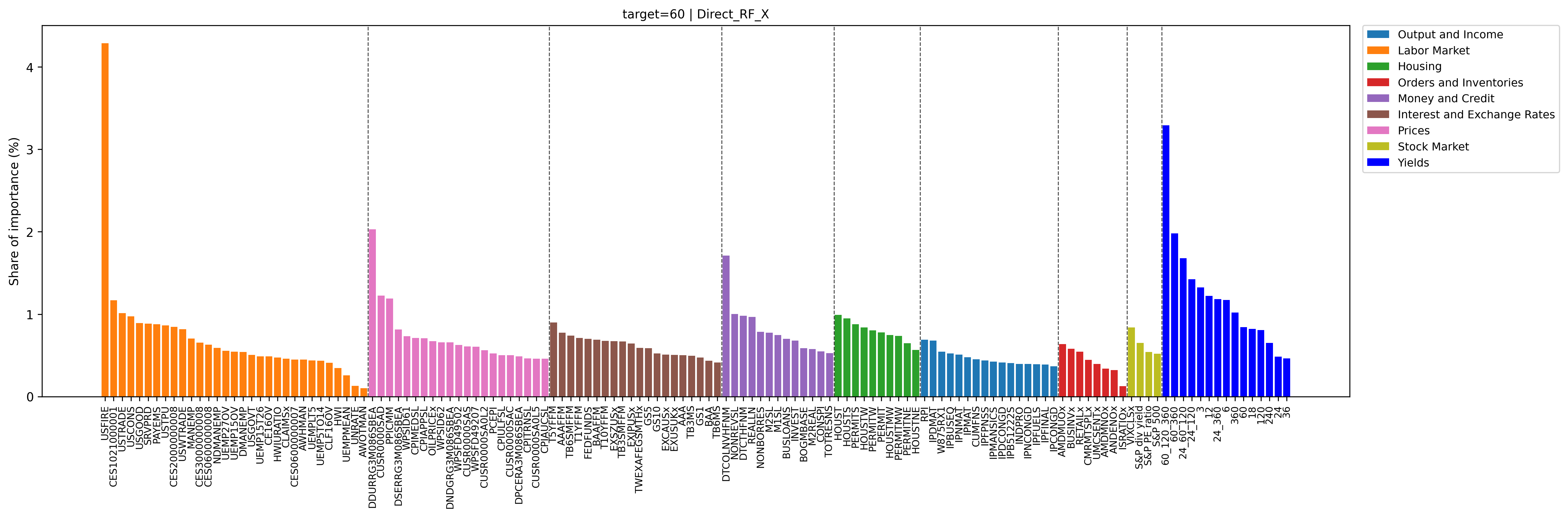}\\[0.5em]
\includegraphics[width=\linewidth]{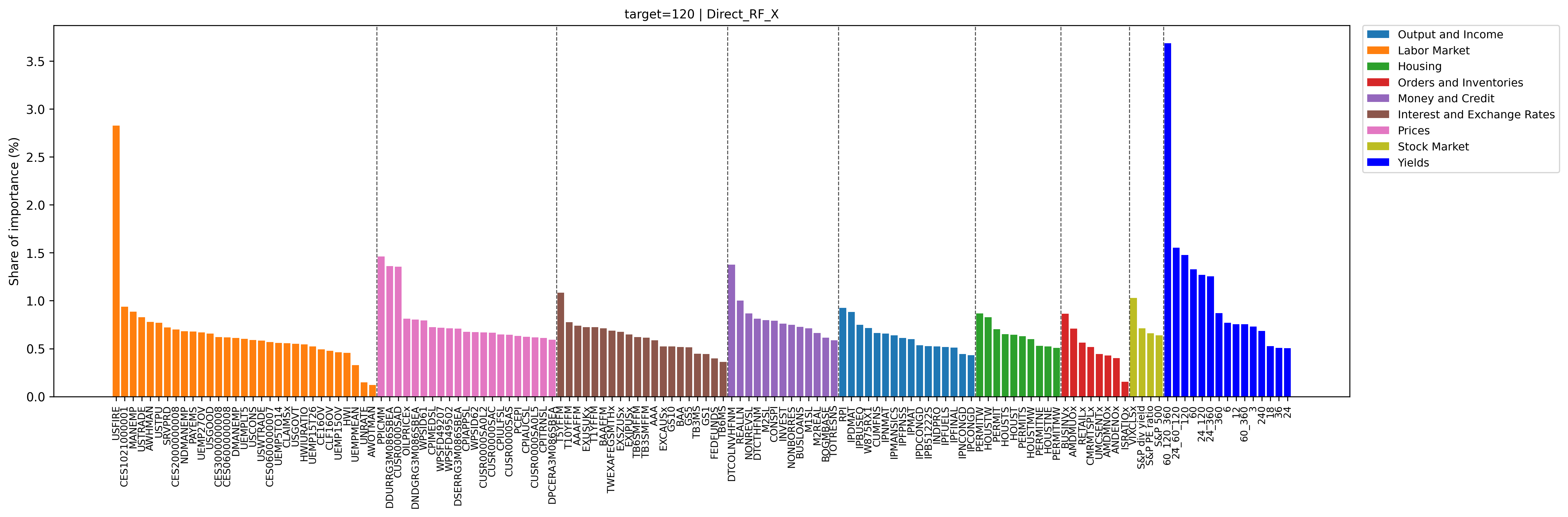}
\caption{Macro-variable importance profiles for the Direct-RF-X model across maturity and slope targets (continued).}
\end{figure}
\end{landscape}

\begin{landscape}
\begin{figure}[p]\ContinuedFloat
\centering
\includegraphics[width=\linewidth]{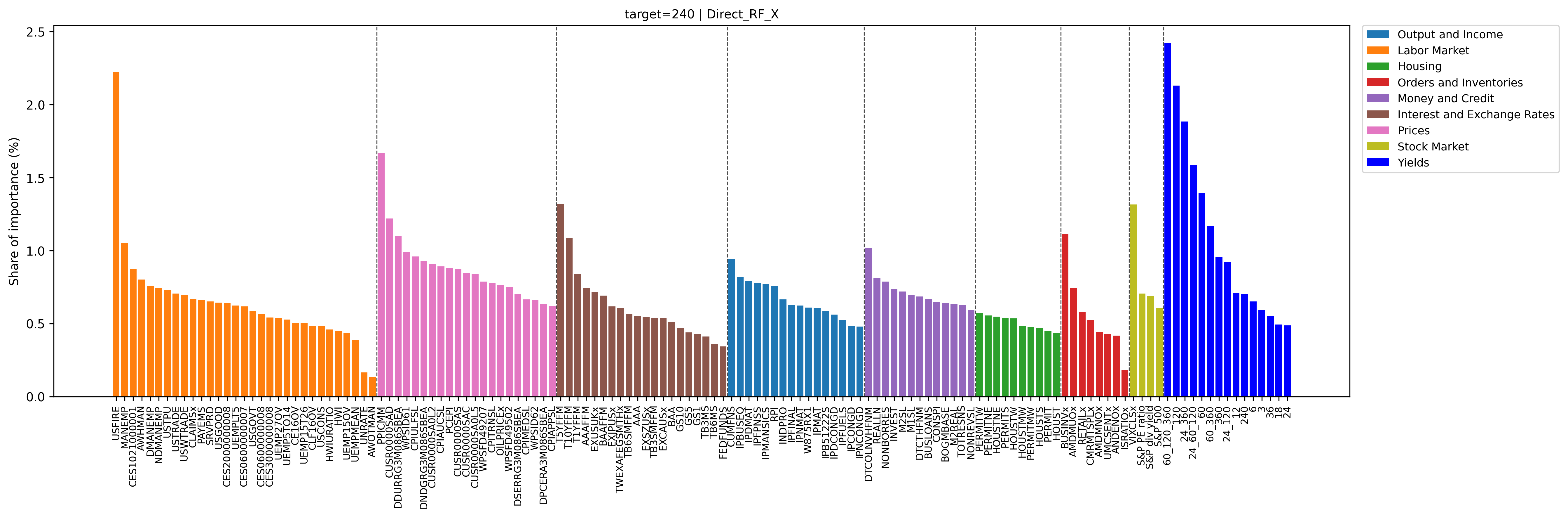}\\[0.5em]
\includegraphics[width=\linewidth]{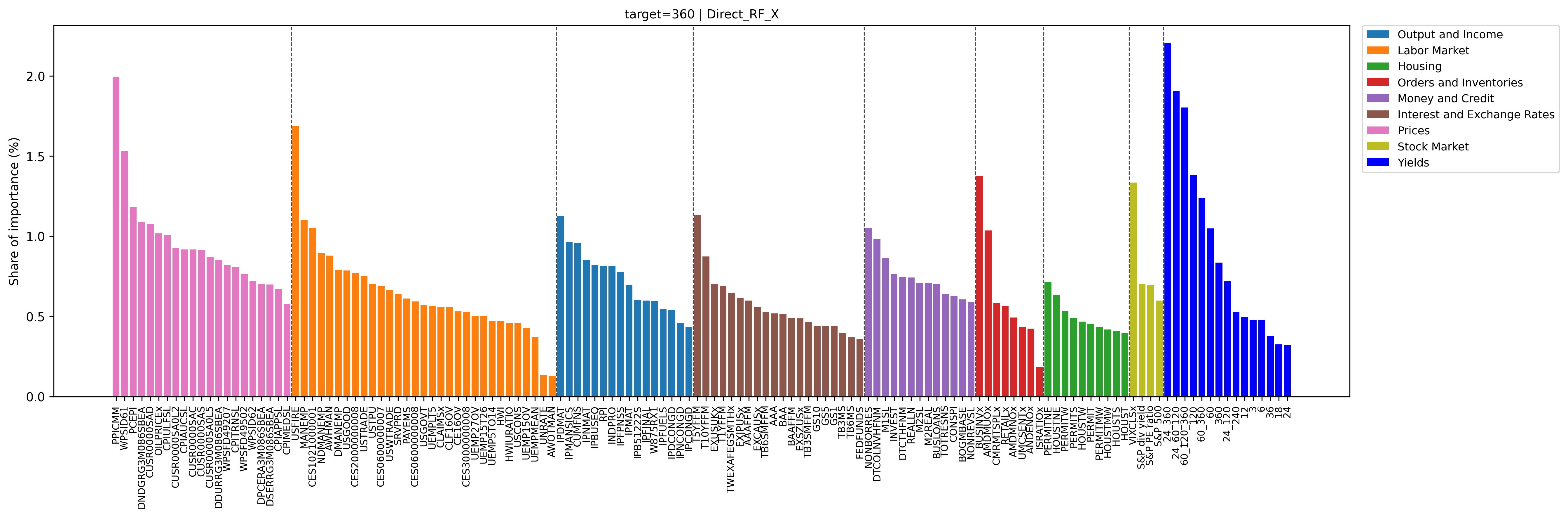}
\caption{Macro-variable importance profiles for the Direct-RF-X model across maturity and slope targets (continued).}
\end{figure}
\end{landscape}

\begin{landscape}
\begin{figure}[p]\ContinuedFloat
\centering
\includegraphics[width=\linewidth]{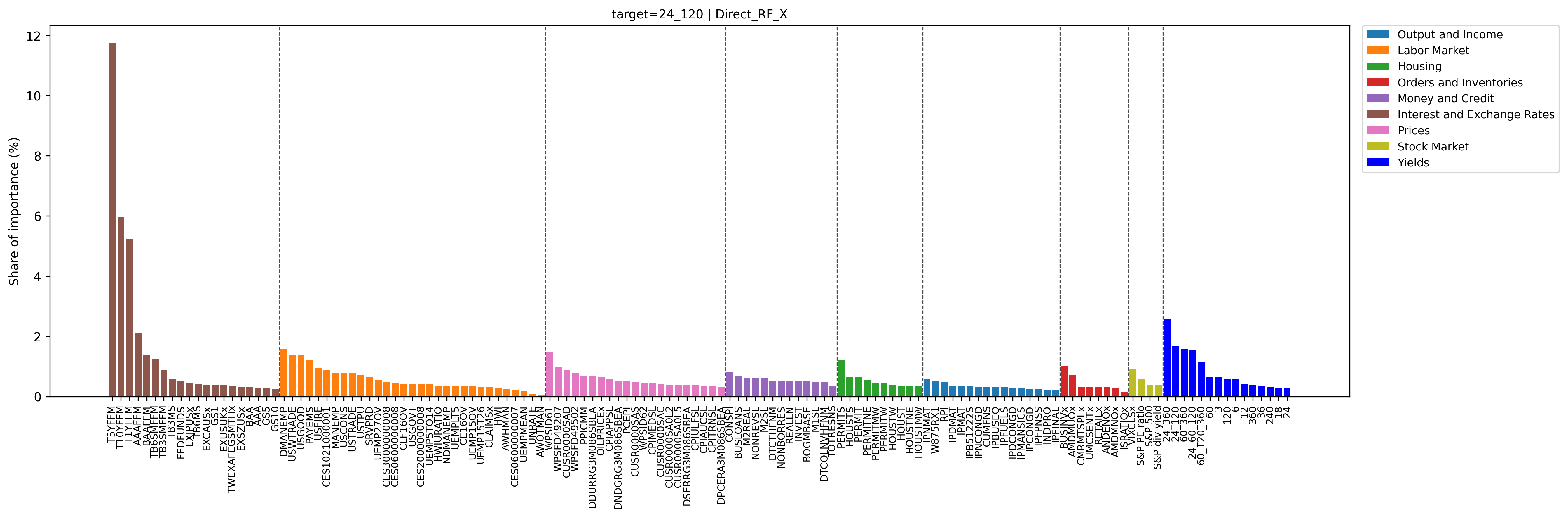}\\[0.5em]
\includegraphics[width=\linewidth]{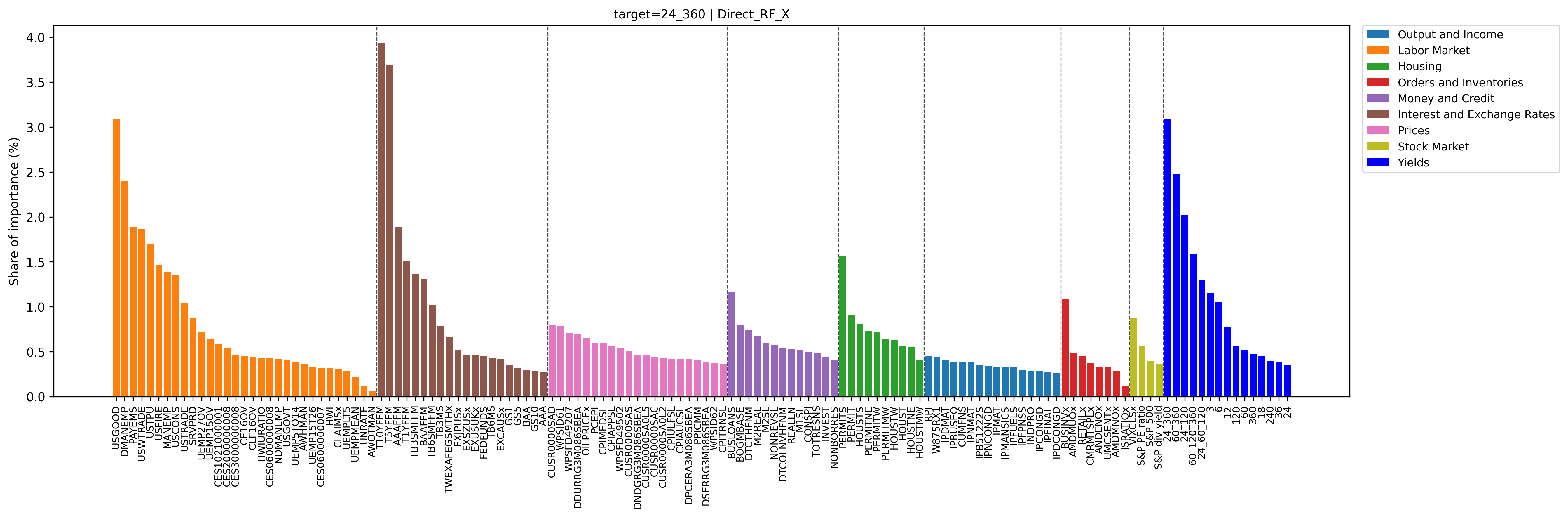}
\caption{Macro-variable importance profiles for the Direct-RF-X model across maturity and slope targets (continued).}
\end{figure}
\end{landscape}

\begin{landscape}
\begin{figure}[p]\ContinuedFloat
\centering
\includegraphics[width=\linewidth]{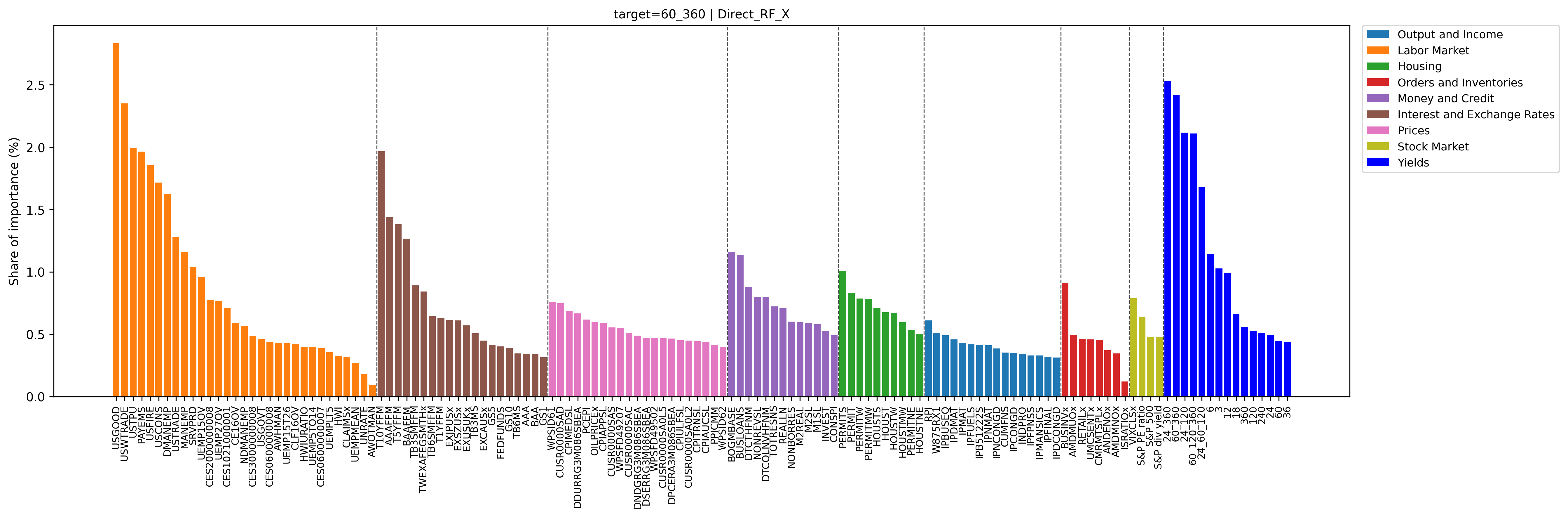}
\caption{Macro-variable importance profiles for the Direct-RF-X model across maturity and slope targets (continued).}
\end{figure}
\end{landscape}

\begin{landscape}
\subsection{Trading Summary Tables}
\begin{table}[htbp]
\centering
\caption{League table of duration trading performance across all tasks, ranked by average annualized return}
\label{tab:duration_summary_total_return}
\begin{tabular}{lccccccccc}
\toprule
Model & Avg. & Median & 1st-Place & Top-3 & Avg. Ann. & Avg. Ann. & Avg. Ann. & Avg. & Avg. Hit \\
      & Rank & Rank   & Finishes  & Finishes & Return (\%) & Roll-Down (\%) & Curve Move (\%) & Sharpe & Rate (\%) \\
\midrule
DNSS\_AR\_PSO           & 2.67  & 2.0  & 5 & 8 & 2.433  & -0.786 & 3.318  & 0.235  & 57.803 \\
DNS\_AR                 & 3.08  & 2.0  & 5 & 9 & 2.277  & 0.399  & 1.956  & 0.247  & 58.125 \\
Carry-Roll              & 4.08  & 3.0  & 2 & 7 & 0.465  & 2.492  & -1.995 & 0.170  & 56.978 \\
Random-Walk             & 5.92  & 4.0  & 0 & 3 & -0.197 & 2.550  & -2.448 & 0.067  & 52.267 \\
DNSS\_RF\_PSO\_X        & 6.50  & 5.0  & 0 & 2 & 0.755  & -0.238 & 1.043  & 0.094  & 50.483 \\
DNSS\_AR                & 6.75  & 6.5  & 0 & 2 & 0.681  & -1.490 & 2.296  & 0.061  & 50.392 \\
DNS\_RF\_X              & 7.58  & 7.5  & 0 & 1 & -0.447 & -0.335 & -0.028 & 0.044  & 50.374 \\
DNSS\_RF\_X             & 8.42  & 9.0  & 0 & 1 & -0.701 & -0.763 & 0.135  & -0.009 & 48.908 \\
Time-Series-Momentum    & 8.67  & 9.0  & 0 & 3 & -0.274 & -0.448 & 0.183  & 0.006  & 50.271 \\
DNSS\_RF                & 8.67  & 8.0  & 0 & 0 & -0.264 & -1.037 & 0.858  & -0.007 & 49.643 \\
DNSS\_RF\_PSO           & 9.08  & 8.5  & 0 & 0 & -0.029 & -0.605 & 0.636  & -0.002 & 49.011 \\
DNS\_RF                 & 10.67 & 11.0 & 0 & 0 & -1.564 & -0.589 & -0.894 & -0.066 & 46.847 \\
Direct\_RF\_X           & 11.17 & 12.0 & 0 & 0 & -2.510 & -0.605 & -1.839 & -0.069 & 46.357 \\
Direct\_RF              & 11.75 & 13.0 & 0 & 0 & -2.929 & -1.048 & -1.822 & -0.110 & 44.912 \\
\bottomrule
\end{tabular}
\end{table}
\end{landscape}

\begin{landscape}
\begin{table}[htbp]
\centering
\caption{League table of slope trading performance across all tasks, ranked by average annualized curve-move return}
\label{tab:slope_summary_total_return}
\begin{tabular}{lccccccccc}
\toprule
Model & Avg. & Median & 1st-Place & Top-3 & Avg. Ann. & Avg. Ann. & Avg. Ann. & Avg. & Avg. Hit \\
      & Rank & Rank   & Finishes  & Finishes & Return (\%) & Roll-Down (\%) & Curve Move (\%) & Sharpe & Rate (\%) \\
\midrule
DNSS\_AR\_PSO        & 1.67  & 1.0  & 8 & 11 & 2.433  & -0.786 & 3.318  & 0.235  & 57.803 \\
DNSS\_AR             & 4.83  & 3.0  & 1 & 7  & 0.681  & -1.490 & 2.296  & 0.061  & 50.392 \\
DNS\_AR              & 5.75  & 3.5  & 1 & 6  & 2.277  & 0.399  & 1.956  & 0.247  & 58.125 \\
DNS\_RF\_X           & 6.25  & 6.5  & 1 & 2  & -0.447 & -0.335 & -0.028 & 0.044  & 50.374 \\
Time-Series-Momentum & 6.50  & 5.5  & 0 & 2  & -0.274 & -0.448 & 0.183  & 0.006  & 50.271 \\
DNSS\_RF\_PSO\_X     & 6.58  & 7.0  & 0 & 0  & 0.755  & -0.238 & 1.043  & 0.094  & 50.483 \\
DNSS\_RF\_X          & 6.75  & 6.5  & 0 & 3  & -0.701 & -0.763 & 0.135  & -0.009 & 48.908 \\
DNSS\_RF\_PSO        & 6.75  & 7.0  & 0 & 1  & -0.029 & -0.605 & 0.636  & -0.002 & 49.011 \\
DNSS\_RF             & 7.00  & 6.5  & 0 & 1  & -0.264 & -1.037 & 0.858  & -0.007 & 49.643 \\
Direct\_RF\_X        & 8.50  & 10.0 & 1 & 3  & -2.510 & -0.605 & -1.839 & -0.069 & 46.357 \\
DNS\_RF              & 9.25  & 9.5  & 0 & 0  & -1.564 & -0.589 & -0.894 & -0.066 & 46.847 \\
Direct\_RF           & 9.75  & 10.5 & 0 & 0  & -2.929 & -1.048 & -1.822 & -0.110 & 44.912 \\
Carry-Roll           & 12.17 & 13.0 & 0 & 0  & 0.465  & 2.492  & -1.995 & 0.170  & 56.978 \\
Random-Walk          & 13.00 & 13.5 & 0 & 0  & -0.197 & 2.550  & -2.448 & 0.067  & 52.267 \\
\bottomrule
\end{tabular}
\end{table}
\end{landscape}

\begin{landscape}
\begin{table}[htbp]
\centering
\caption{League table of duration trading performance across all tasks, ranked by average annualized return}
\label{tab:duration_summary_curve_return}
\begin{tabular}{lccccccccc}
\toprule
Model & Avg. & Median & 1st-Place & Top-3 & Avg. Ann. & Avg. Ann. & Avg. Ann. & Avg. & Avg. Hit \\
      & Rank & Rank   & Finishes  & Finishes & Return (\%) & Roll-Down (\%) & Curve Move (\%) & Sharpe & Rate (\%) \\
\midrule
DNSS\_AR              & 2.00  & 1.0  & 7 & 10 & 0.813 & 1.091  & 0.104 & 0.404 & 63.227 \\
DNSS\_AR\_PSO         & 3.17  & 3.0  & 2 & 8  & 0.849 & 0.833  & 0.350 & 0.367 & 65.877 \\
Carry-Roll            & 3.92  & 3.5  & 1 & 6  & 0.768 & 1.617  & -0.326 & 0.314 & 61.663 \\
DNS\_AR               & 4.92  & 3.5  & 1 & 6  & 0.315 & 0.583  & 0.182 & 0.262 & 58.211 \\
DNSS\_RF\_X           & 6.17  & 6.0  & 0 & 1  & 0.428 & 0.019  & 0.767 & 0.176 & 51.899 \\
DNS\_RF\_X            & 6.83  & 7.5  & 0 & 2  & 0.138 & -0.012 & 0.585 & 0.137 & 49.067 \\
DNSS\_RF              & 7.25  & 7.0  & 0 & 1  & 0.115 & 0.311  & 0.162 & 0.124 & 49.589 \\
Direct\_RF\_X         & 7.33  & 8.5  & 1 & 1  & 0.377 & -0.093 & 0.897 & 0.134 & 49.660 \\
DNS\_RF               & 7.75  & 7.5  & 0 & 0  & 0.024 & 0.076  & 0.377 & 0.105 & 47.113 \\
Direct\_RF            & 9.08  & 9.0  & 0 & 0  & 0.064 & 0.163  & 0.297 & 0.075 & 47.469 \\
DNSS\_RF\_PSO\_X      & 9.25  & 10.5 & 0 & 1  & 0.139 & -0.099 & 0.571 & 0.042 & 48.517 \\
DNSS\_RF\_PSO         & 10.33 & 11.0 & 0 & 0  & -0.205 & -0.140 & 0.283 & -0.038 & 43.989 \\
\bottomrule
\end{tabular}
\end{table}
\end{landscape}

\begin{landscape}
\begin{table}[htbp]
\centering
\caption{League table of slope trading performance across all tasks, ranked by average annualized curve-move return}
\label{tab:slope_summary_curve_return}
\begin{tabular}{lccccccccc}
\toprule
Model & Avg. & Median & 1st-Place & Top-3 & Avg. Ann. & Avg. Ann. & Avg. Ann. & Avg. & Avg. Hit \\
      & Rank & Rank   & Finishes  & Finishes & Return (\%) & Roll-Down (\%) & Curve Move (\%) & Sharpe & Rate (\%) \\
\midrule
Direct\_RF\_X         & 1.58  & 1.0  & 8 & 11 & 0.377 & -0.093 & 0.897 & 0.134 & 49.660 \\
DNSS\_RF\_X           & 2.83  & 3.0  & 3 & 8  & 0.428 & 0.019  & 0.767 & 0.176 & 51.899 \\
DNS\_RF\_X            & 3.58  & 2.5  & 3 & 8  & 0.138 & -0.012 & 0.585 & 0.137 & 49.067 \\
DNSS\_RF\_PSO\_X      & 4.75  & 5.0  & 0 & 5  & 0.139 & -0.099 & 0.571 & 0.042 & 48.517 \\
DNS\_RF               & 6.75  & 6.5  & 0 & 1  & 0.024 & 0.076  & 0.377 & 0.105 & 47.113 \\
DNSS\_AR\_PSO         & 6.75  & 6.5  & 0 & 0  & 0.849 & 0.833  & 0.350 & 0.367 & 65.877 \\
DNSS\_RF\_PSO         & 6.75  & 6.0  & 0 & 0  & -0.205 & -0.140 & 0.283 & -0.038 & 43.989 \\
Direct\_RF            & 7.25  & 8.0  & 0 & 2  & 0.064 & 0.163  & 0.297 & 0.075 & 47.469 \\
DNSS\_RF              & 7.58  & 7.0  & 0 & 1  & 0.115 & 0.311  & 0.162 & 0.124 & 49.589 \\
DNS\_AR               & 8.92  & 10.0 & 0 & 0  & 0.315 & 0.583  & 0.182 & 0.262 & 58.211 \\
DNSS\_AR              & 9.08  & 10.0 & 0 & 0  & 0.813 & 1.091  & 0.104 & 0.404 & 63.227 \\
Carry-Roll            & 11.83 & 12.0 & 0 & 0  & 0.768 & 1.617  & -0.326 & 0.314 & 61.663 \\
\bottomrule
\end{tabular}
\end{table}
\end{landscape}

\begin{landscape}
\subsection{Individual Trading Performance Metrics}
\begin{table}[htbp]
\centering
\caption{Performance metrics for 24--120 slope trades at the one-month-ahead forecast horizon}
\label{tab:slope_table_24_120_h1}
\small
\begin{tabular}{lccccccccccc}
\toprule
Strategy & Ann. Return & Ann. Curve & Ann. Roll- & Ann. TC & Hit Rate & Vol. & Sharpe & Trade & Trade & Trade & Trade \\
      & (\%) & Move (\%) & down (\%) & (\%) & (\%) & Period & Period & Return (\%) & Roll-down (\%) & Curve Move (\%) & TC (\%) \\
\midrule
DNSS\_AR                & 0.118 & -0.153 & 1.257 & 0.983 & 48.8 & 0.257 & 0.038 & 0.010 & 0.104 & -0.013 & 0.082 \\
DNS\_RF\_X              & 0.040 & -0.048 & 1.157 & 1.068 & 47.2 & 0.262 & 0.013 & 0.003 & 0.096 & -0.004 & 0.089 \\
DNS\_AR                 & -0.016 & -0.152 & 1.270 & 1.133 & 47.2 & 0.261 & -0.005 & -0.001 & 0.105 & -0.013 & 0.094 \\
DNS\_RF                 & -0.041 & -0.049 & 1.075 & 1.067 & 44.8 & 0.263 & -0.013 & -0.003 & 0.089 & -0.004 & 0.088 \\
DNSS\_RF                & -0.060 & 0.296 & 0.624 & 0.982 & 45.6 & 0.266 & -0.019 & -0.005 & 0.052 & 0.025 & 0.082 \\
DNSS\_RF\_X             & -0.195 & 0.175 & 0.559 & 0.931 & 42.4 & 0.264 & -0.061 & -0.016 & 0.046 & 0.015 & 0.077 \\
DNSS\_AR\_PSO           & -0.242 & 0.181 & 0.462 & 0.888 & 47.2 & 0.274 & -0.074 & -0.020 & 0.038 & 0.015 & 0.074 \\
Direct\_RF\_X           & -0.248 & 0.736 & 0.014 & 1.001 & 44.8 & 0.269 & -0.077 & -0.021 & 0.001 & 0.061 & 0.083 \\
Direct\_RF              & -0.299 & 0.624 & 0.041 & 0.968 & 40.0 & 0.275 & -0.091 & -0.025 & 0.003 & 0.052 & 0.080 \\
Roll-Down             & -0.324 & -0.395 & 1.292 & 1.220 & 42.4 & 0.269 & -0.100 & -0.027 & 0.107 & -0.033 & 0.101 \\
DNSS\_RF\_PSO\_X        & -0.358 & 0.166 & 0.243 & 0.770 & 41.6 & 0.277 & -0.108 & -0.030 & 0.020 & 0.014 & 0.064 \\
DNSS\_RF\_PSO           & -0.776 & 0.093 & -0.048 & 0.827 & 37.6 & 0.282 & -0.230 & -0.065 & -0.004 & 0.008 & 0.069 \\
\bottomrule
\end{tabular}
\end{table}
\end{landscape}

\subsection{Trading Exercise Strategy Diagnostic Plots}

\begin{figure}[!htbp]
    \centering
    \makebox[\textwidth][c]{%
    \begin{minipage}[t]{0.72\textwidth}
        \centering
        \begin{subfigure}[t]{\textwidth}
            \centering
            \includegraphics[width=\textwidth]{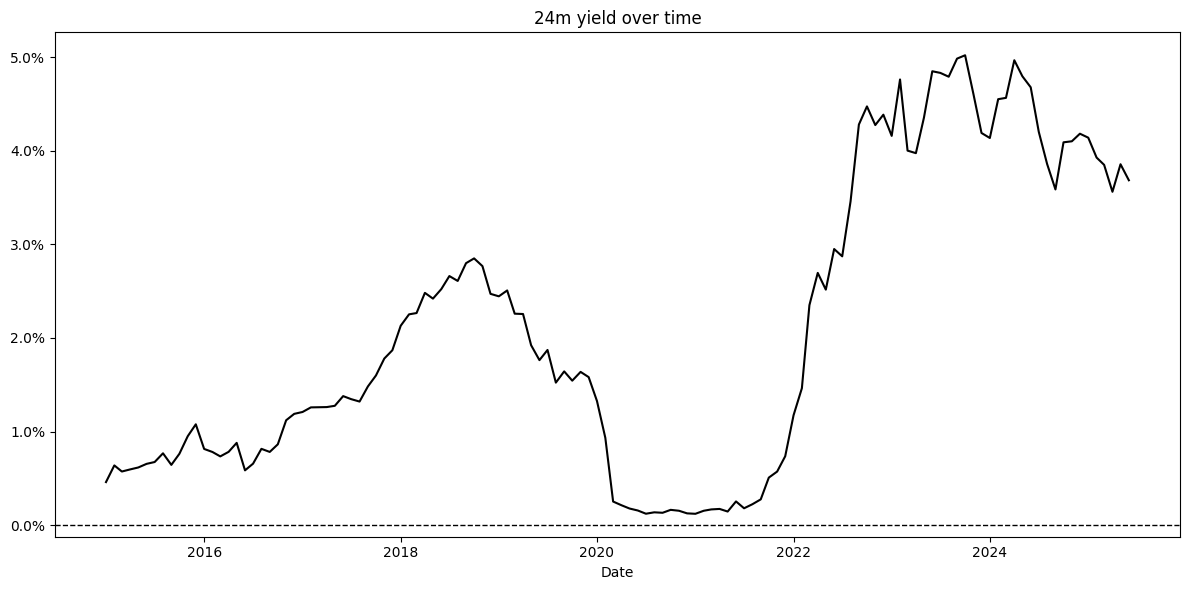}
            \caption{24-month yield over time}
            \label{fig:24m_overtime}
        \end{subfigure}

        \vspace{0.5cm}

        \begin{subfigure}[t]{\textwidth}
            \centering
            \includegraphics[width=\textwidth]{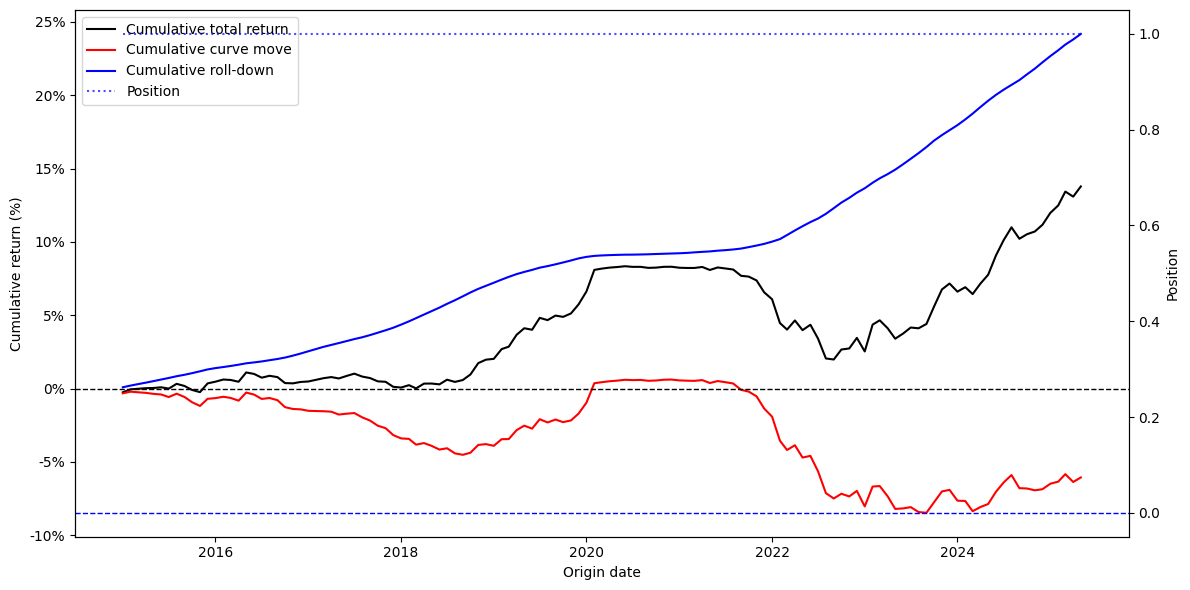}
            \caption{Random Walk PnL attribution and signal position}
            \label{fig:rw_24_h1}
        \end{subfigure}

        \vspace{0.5cm}

        \begin{subfigure}[t]{\textwidth}
            \centering
            \includegraphics[width=\textwidth]{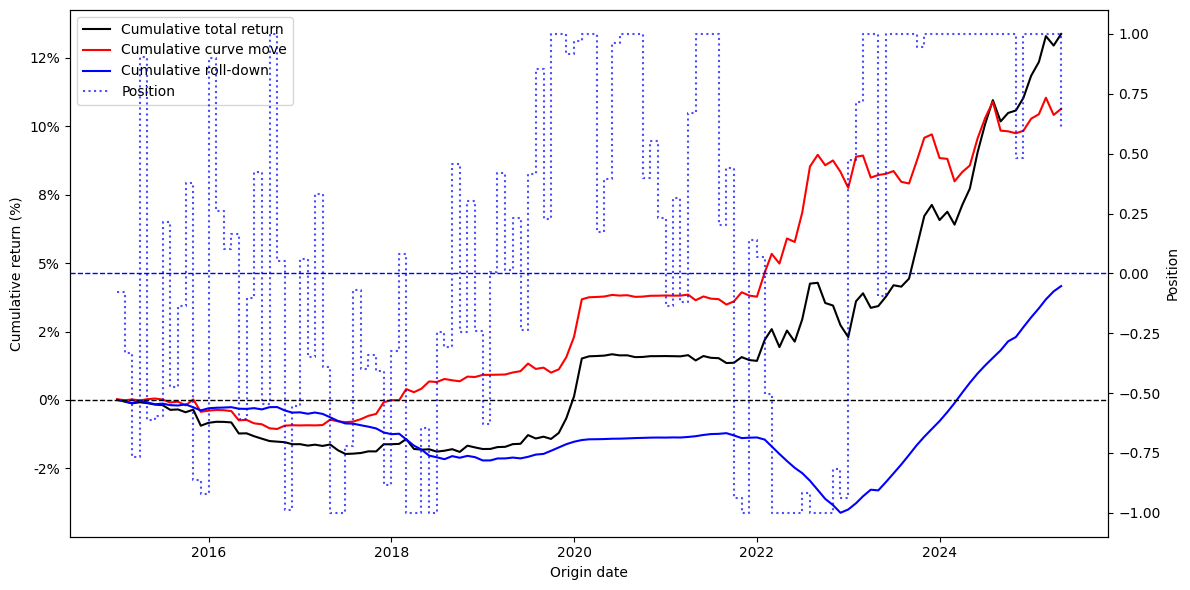}
            \caption{DNSS-RF-X PnL attribution and signal position}
            \label{fig:dnss_rf_x_24_h1}
        \end{subfigure}

        \caption{Strategy diagnostic plots for the 24 month yield duration trade at horizon $h=1$.}
        \label{fig:duration_trading_diagnostics_24_h1}
    \end{minipage}}
\end{figure}

\begin{figure}[!htbp]
    \centering
    \makebox[\textwidth][c]{%
    \begin{minipage}[t]{0.72\textwidth}
        \centering
        \begin{subfigure}[t]{\textwidth}
            \centering
            \includegraphics[width=\textwidth]{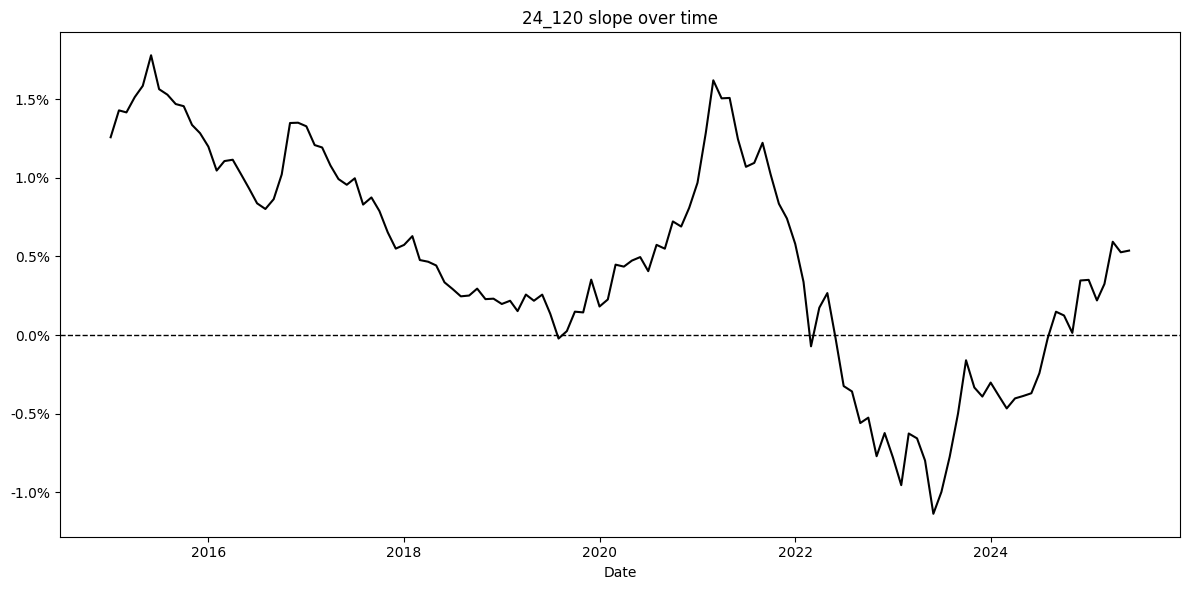}
            \caption{24--120 slope over time}
            \label{fig:24_120_overtime}
        \end{subfigure}

        \vspace{0.5cm}

        \begin{subfigure}[t]{\textwidth}
            \centering
            \includegraphics[width=\textwidth]{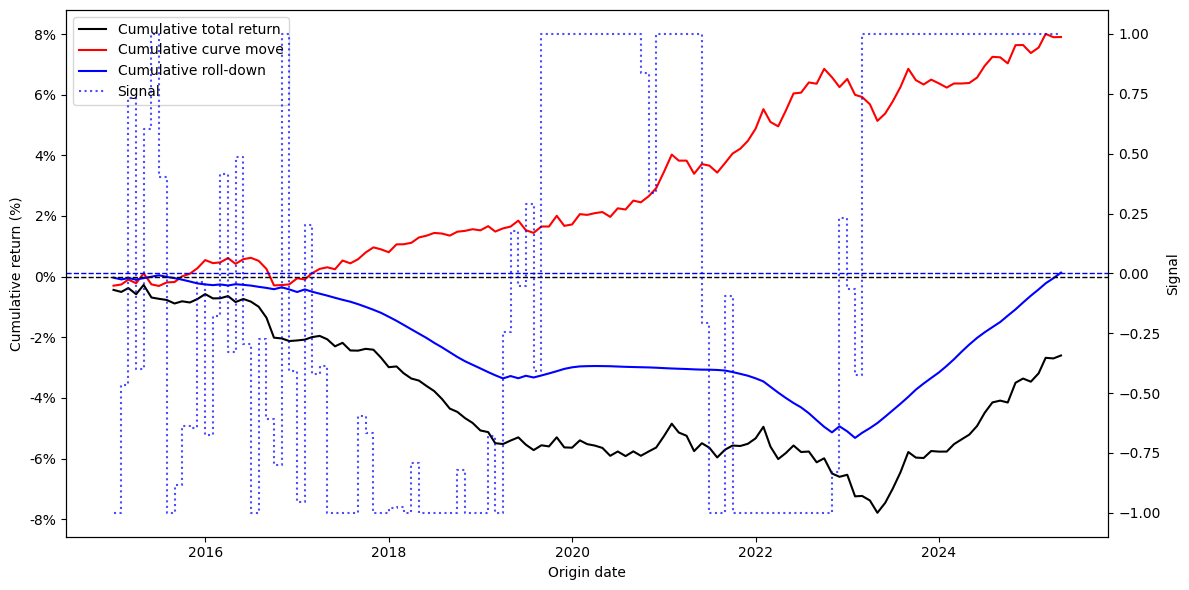}
            \caption{Direct-RF-X PnL attribution and signal position}
            \label{fig:direct_rf_x_24_120_h1}
        \end{subfigure}

        \vspace{0.5cm}

        \begin{subfigure}[t]{\textwidth}
            \centering
            \includegraphics[width=\textwidth]{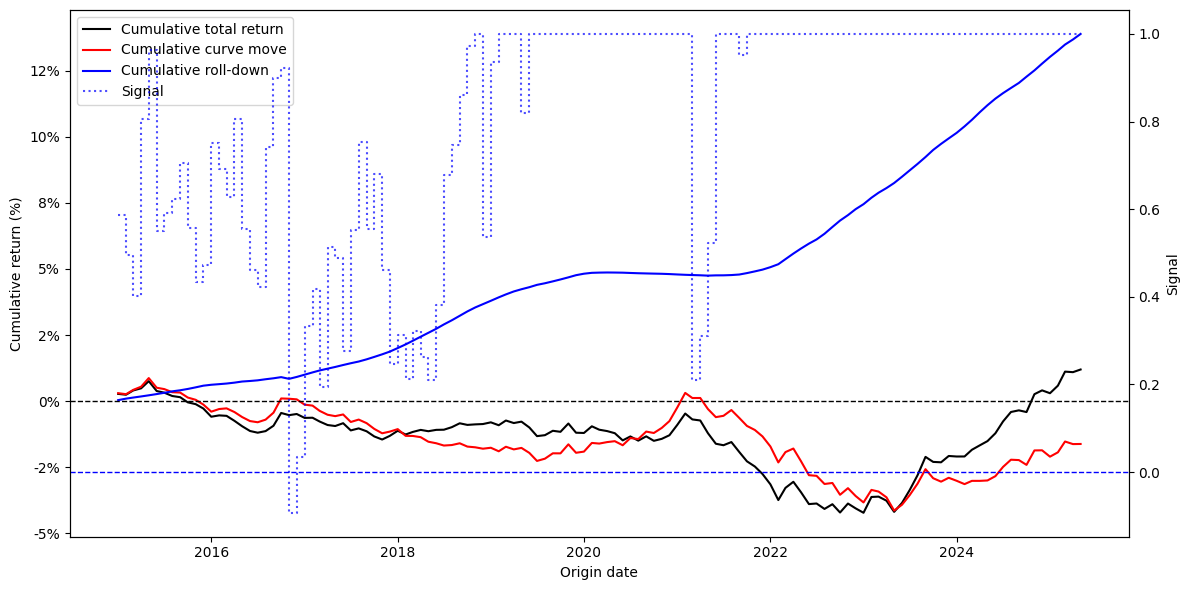}
            \caption{DNSS-AR PnL attribution and signal position}
            \label{fig:dnss_ar_24_120_h1}
        \end{subfigure}

        \caption{Strategy diagnostic plots for the 24--120 slope trade at horizon $h=1$.}
        \label{fig:slope_trading_diagnostics_24_120_h1}
    \end{minipage}}
\end{figure}

\end{document}